\UseRawInputEncoding
\documentclass[%
reprint,
superscriptaddress,
aps,
prl,
sort&compress,
square,
numbers,
]{revtex4-2}

\usepackage{natbib}
\usepackage{xr-hyper}
\makeatletter
\newcommand*{\addFileDependency}[1]{
\typeout{(#1)}
\@addtofilelist{#1}
\IfFileExists{#1}{}{\typeout{No file #1.}}
}\makeatother

\usepackage{hyperref}
\hypersetup{
    colorlinks=false,
    hidelinks=true,
    linkcolor=black,
    filecolor=magenta, 
    citecolor=black,
    urlcolor=black,
    pdfpagemode=FullScreen,
    }

\usepackage{graphicx}
\usepackage{float}
\usepackage{braket}
\usepackage{subcaption}
\usepackage{comment}
\usepackage{color, ulem}
\usepackage{gensymb}
\usepackage{xcolor}
\usepackage{wrapfig}
\usepackage{amsmath,amsfonts,amsthm,bm,bigints,amssymb}
\usepackage{mathtools}
\usepackage{bbm}
\usepackage{gensymb}
\usepackage{relsize}
\usepackage[version=4]{mhchem}
\usepackage{chemformula}

\newcommand{\vc}[1]{\mathbf{#1}}
\newcommand{\ang}{\mathrm{\AA}}

\begin{document}

\title{Coexisting charge density waves in twisted bilayer \ch{NbSe2}}

\author{Christopher T. S. Cheung}
\affiliation{Departments of Physics and Materials and the Thomas Young center for Theory and Simulation of Materials, Imperial College London, South Kensington Campus, London SW7 2AZ, UK\\}
\author{Zachary A. H. Goodwin}
\affiliation{John A. Paulson School of Engineering and  Applied Sciences, Harvard  University, Cambridge, MA 02138, USA}
\author{Yixuan Han}
\affiliation{Institute for Functional Intelligent Materials, National University of Singapore, Singapore 117544, Singapore}
\author{Jiong Lu}
\affiliation{Institute for Functional Intelligent Materials, National University of Singapore, Singapore 117544, Singapore}
\author{Arash A. Mostofi}
\affiliation{Departments of Physics and Materials and the Thomas Young center for Theory and Simulation of Materials, Imperial College London, South Kensington Campus, London SW7 2AZ, UK\\}
\author{Johannes Lischner}
\affiliation{Departments of Physics and Materials and the Thomas Young center for Theory and Simulation of Materials, Imperial College London, South Kensington Campus, London SW7 2AZ, UK\\}

\date{\today}

\begin{abstract}
Twisted bilayers of two-dimensional materials have emerged as a highly tunable platform for studying broken symmetry phases. While most interest has been focused on emergent states in systems whose constituent monolayers do not feature broken symmetry states, assembling monolayers that exhibit ordered states  into twisted bilayers can also give rise to interesting phenomena. Here, we use large-scale first-principles density-functional theory calculations to study the atomic structure of twisted bilayer \ch{NbSe2} whose constituent monolayers feature a charge density wave. We find that different charge density wave states coexist in the ground state of the twisted bilayer: monolayer-like $3\times 3$ triangular and hexagonal charge density waves are observed in low-energy stacking regions, while stripe charge density waves are found in the domain walls surrounding the low-energy stacking regions. These predictions, which can be tested by scanning tunneling microscopy experiments, highlight the potential to create complex charge density wave ground states in twisted bilayer systems and can serve as a starting point for understanding superconductivity occurring at low temperatures.
\end{abstract}

\maketitle

\section{Introduction}
\label{sec:intro}
Twisted bilayers of two-dimensional (2D) materials have been intensely studied in recent years because of their novel electronic \cite{Bistritzer_MacDonald_2011,Cao_Fatemi_Demir_Fang_Tomarken_Luo_Sanchez-Yamagishi_Watanabe_Taniguchi_Kaxiras_2018,Cao_Fatemi_Fang_Watanabe_Taniguchi_Kaxiras_Jarillo-Herrero_2018,attrac_elel_goodwin,flat_mag_goodwin,sqi_tmd_vdw,Kennes_Claassen_Xian_Georges_Millis_Hone_Dean_Basov_Pasupathy_Rubio_2021}, optical \cite{Zheng_Wu_Li_Ding_He_Liu_Wang_Wang_Pan_Liu_2023,Lian_Meng_Ma_Maity_Yan_Wu_Huang_Chen_Chen_Chen_2023,Chen_Lian_Huang_Su_Rashetnia_Ma_Yan_Blei_Xiang_Taniguchi_et_2022,Brotons-Gisbert_Baek_Campbell_Watanabe_Taniguchi_Gerardot_2021,interlayer_x_transport} and vibrational properties \cite{phason_maity,chiral_phon_maity}. These materials are formed by vertically stacking two monolayers and then rotating them relative to each other. This gives rise to an emergent moir\'e pattern whose lattice constant depends on the twist angle. For small twist angles, the lattice constant of the moir\'e pattern can be thousands of times larger than that of the underlying atomic lattice \cite{local_dirac_el,commensuration_mele}. 

To date, many studies on twisted bilayer materials have focused on their emergent broken-symmetry states, i.e., states that are absent in the monolayer. For example, magic-angle twisted bilayer graphene exhibits correlated insulator states and superconductivity, which are not found in the constituent graphene sheets~\cite{Cao_Fatemi_Demir_Fang_Tomarken_Luo_Sanchez-Yamagishi_Watanabe_Taniguchi_Kaxiras_2018, Cao_Fatemi_Fang_Watanabe_Taniguchi_Kaxiras_Jarillo-Herrero_2018}. Recently, there has been increasing interest in twisted bilayer materials composed of monolayers with broken-symmetry states. In particular, bilayers composed of magnetic monolayers, such as CrI$_3$, have been found to exhibit coexisiting ferromagnetism and antiferromagnetism, non-collinear magnetic states and moir\'e magnon bands~\cite{xu2022coexisting, hejazi2020noncollinear, tong2018skyrmions, akram2021skyrmions, xiao2021magnetization}. 

Another class of broken-symmetry states that can be found in monolayers are charge-density waves (CDWs). CDWs are commonly found in monolayer or bulk transition metal dichalcogenides (TMDs) which typically also exhibit superconductivity at low temperatures. For example, monolayer \ch{NbSe2} adopts a $3 \times 3$ triangular CDW in which some Nb atoms move towards an interstitial site \cite{unveil_cdw_sc}. However, other types of CDWs have also been predicted and observed, such as $2\times 2$, $2\times 3$, hexagonal and chalcogen-centered triangular $3\times 3$ CDWs \cite{unveil_cdw_imp,elas_cdw_guster}, and stripe-like $4\times 4$ or $\sqrt{13} \times \sqrt{13}$ CDWs \cite{Ugeda_Bradley_Zhang_Onishi_Chen_Ruan_Ojeda-Aristizabal_Ryu_Edmonds_Tsai_etal._2015,Soumyanarayanan_Yee_He_van_wezel_Rahn_Rossnagel_Hudson_Norman_Hoffman_2013,uni_strain_cdw_flicker}. The various CDWs can be viewed as different linear combinations of the three soft phonon modes of the high-symmetry phase~\cite{mcmillan-cdw1,mcmillan-cdw2,mcmillan-cdw3,jw-cdw1,jw-cdw2,jw-cdw3}. 

Importantly, the condensation energies of the various CDWs in the monolayer differ only by a few meV and their relative stability is highly sensitive to external perturbations, such as strain and doping~\cite{elas_cdw_guster,strain_stripe_cdw,uni_strain_cdw_flicker}. It is therefore interesting to ask what happens to the CDW when a twisted bilayer is formed.


\begin{figure*}[htb!]
    \centering
    \begin{subfigure}[h]{0.495\textwidth}
        \centering
        \caption{\hspace{110pt}\ch{MoSe2}}
        \includegraphics[width=0.98\textwidth]{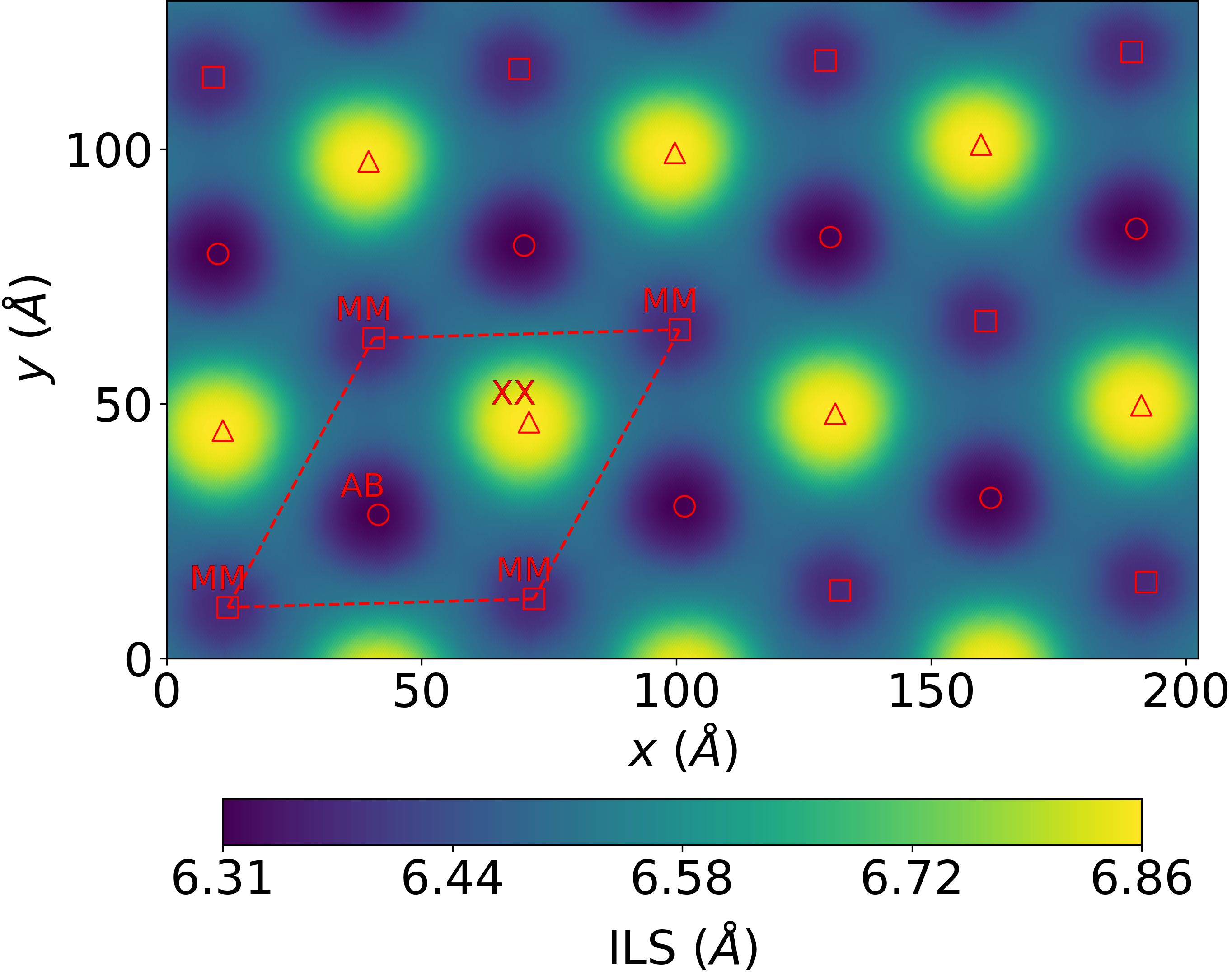}
    \end{subfigure}%
    ~
    \begin{subfigure}[h]{0.475\textwidth}
        \centering
        \caption{\hspace{110pt}\ch{NbSe2}}
        \includegraphics[width=\textwidth]{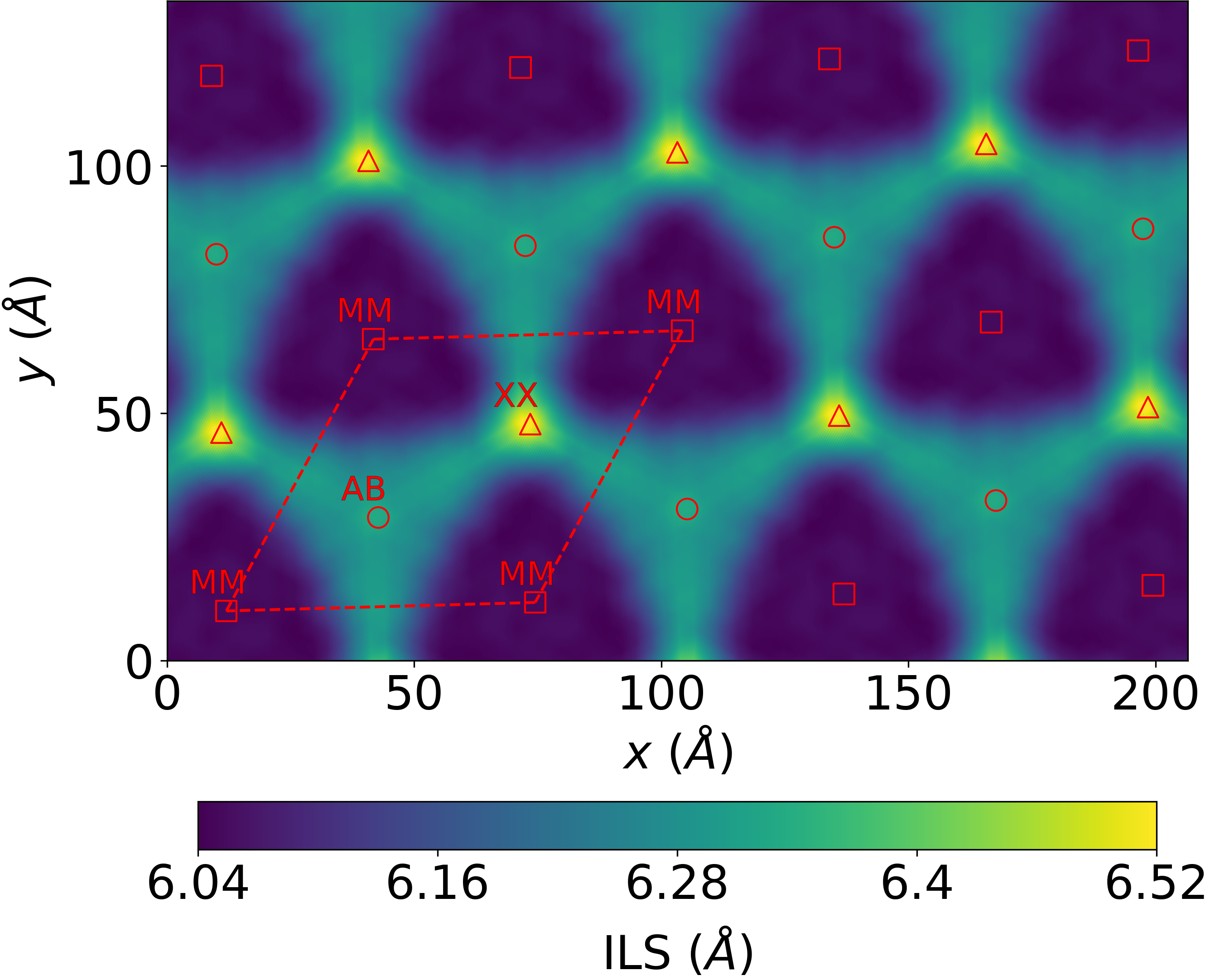}
        \captionsetup{justification=centering}
    \end{subfigure}
    ~
    \begin{subfigure}[h]{0.49\textwidth}
        \centering
        \caption{\hspace{110pt}\ch{MoSe2}}
        \includegraphics[width=\textwidth]{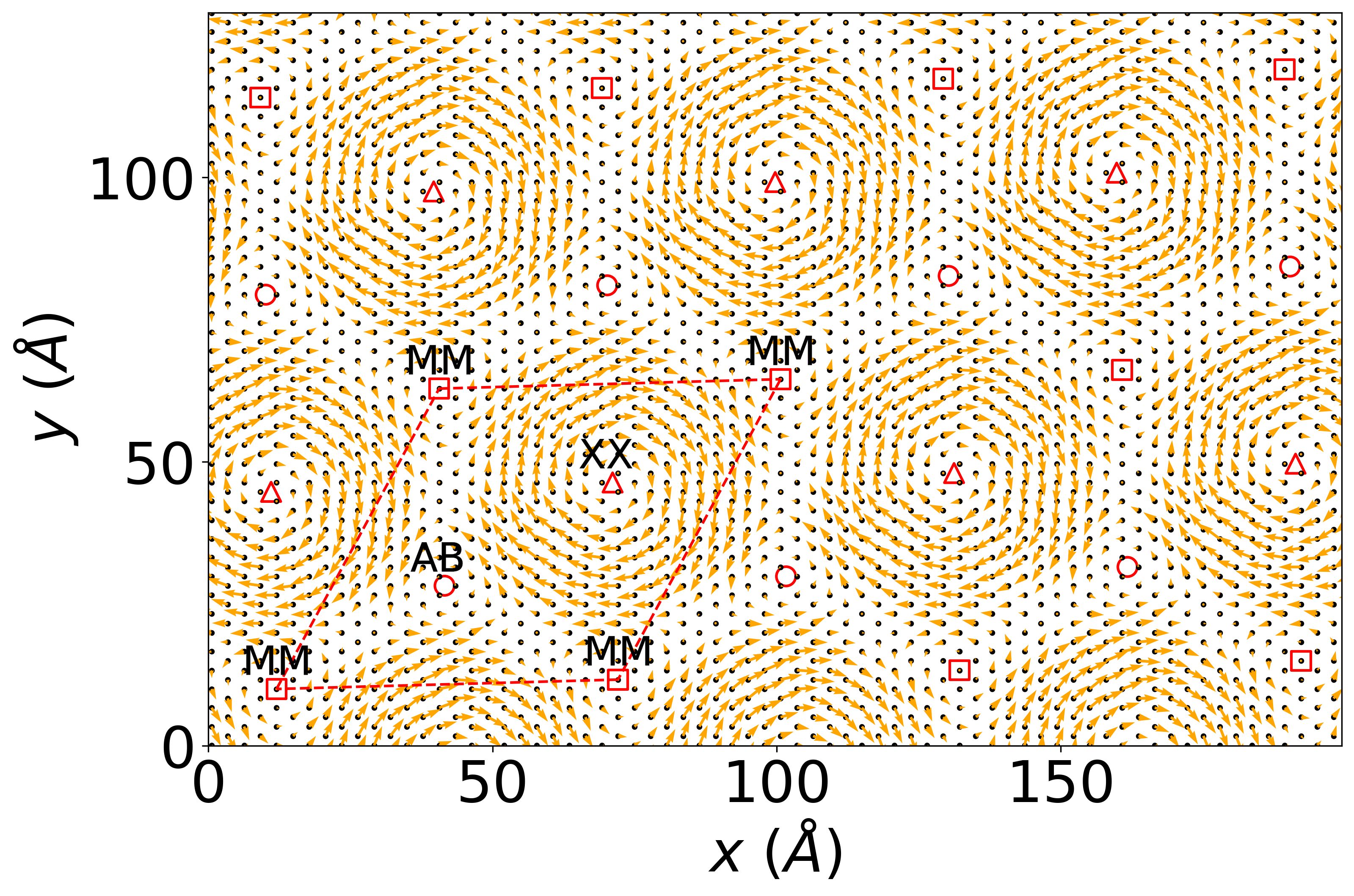}
        \captionsetup{justification=centering}
    \end{subfigure}
    ~
    \begin{subfigure}[h]{0.49\textwidth}
        \centering
        \caption{\hspace{110pt}\ch{NbSe2}}
        \includegraphics[width=1.01\textwidth]{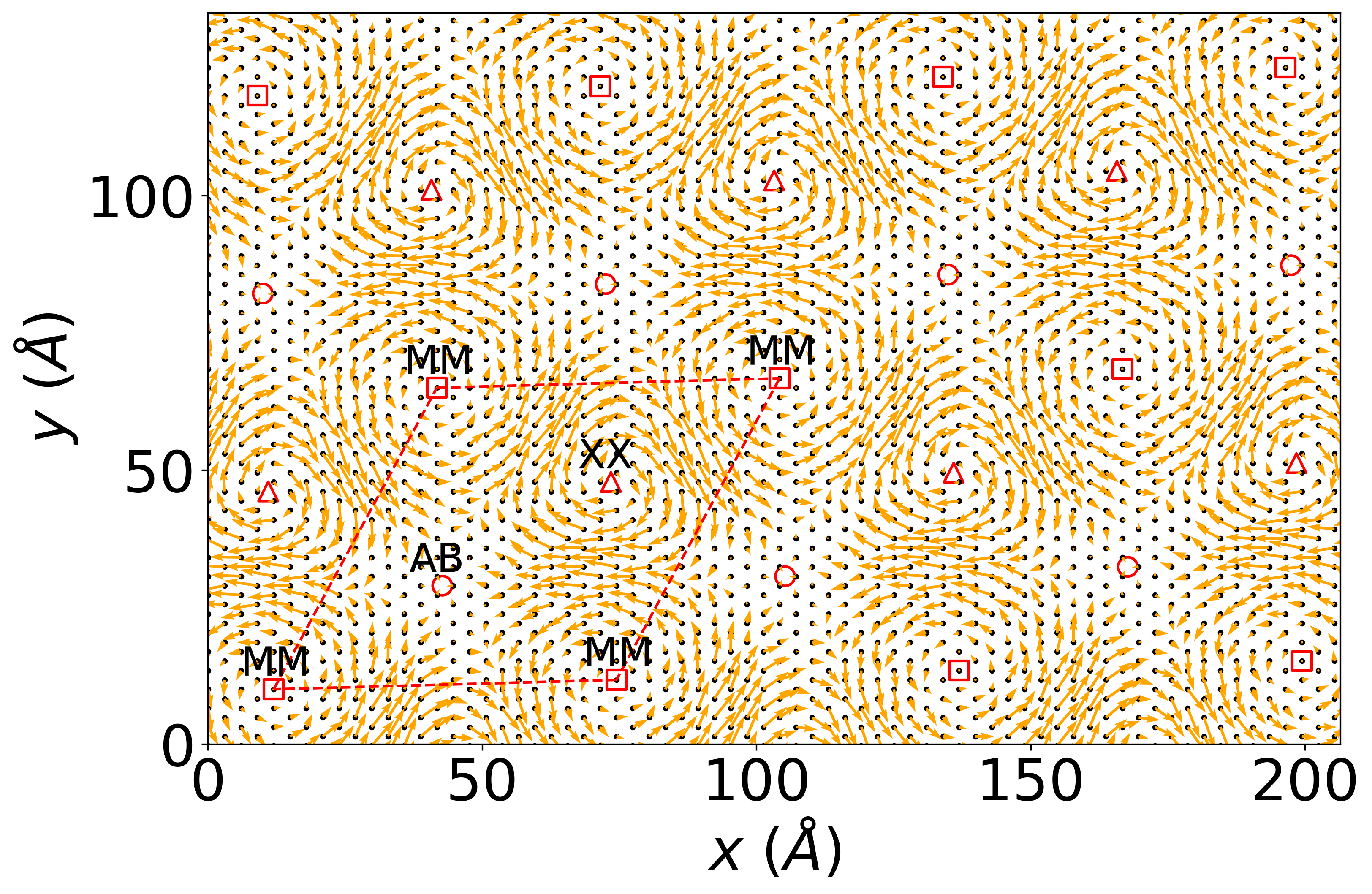}
        \captionsetup{justification=centering}
    \end{subfigure}
    \caption{In-plane (bottom panels) and out-of-plane (top panels) atomic relaxations of twisted bilayers \ch{MoSe2} (left panels) and \ch{NbSe2} (right panels). The interlayer separation (ILS) between the metal atom layers in (a) twisted bilayer \ch{MoSe2} and (b) twisted bilayer \ch{NbSe2} at a twist angle of 3.14$^\circ$. In-plane displacements of metal atoms relative to unrelaxed initial structure in the bottom layer for (c) twisted bilayer \ch{MoSe2} and (d) twisted bilayer \ch{NbSe2}. The red circles denote the centers of the AB stacking regions (metal on top of chalcogen and vice versa), the red squares denote the centers of the MM stacking regions (metal on top of metal), and the red triangles denote the centers of the XX stacking regions (chalcogen on top of chalcogen). The red parallelogram marks the repeating unit used in the DFT calculation.}
    \label{fig:ils}
\end{figure*}

Based on earlier models of CDWs in the monolayer~\cite{mcmillan-cdw1,mcmillan-cdw2,mcmillan-cdw3,jw-cdw1,jw-cdw2,jw-cdw3}, Goodwin and Fal'ko developed a Landau free energy approach based on a number of simplifying assumptions to understand CDWs in a twisted bilayer system at small twist angles, ~\cite{goodwin_falko_cdw_moire}. They analyzed the interplay of the periodicities of the moir\'e pattern and various CDWs and predicted that certain CDWs (such as $\sqrt{3} \times \sqrt{3}$) can propagate through the moir\'e system without geometric constraints. Other CDWs (such as $3 \times 3$ or $2 \times 2$) are modified by the presence of the moir\'e pattern with CDWs only occuring in specific regions of the moir\'e unit cell. Later, McHugh \textit{et al.} developed a multi-scale model based on first-principles DFT calculations of untwisted bilayers to predict the relaxed structure of marginally twisted bilayer \ch{NbSe2} and predicted the formation of domains surrounded by domain walls of specific stacking ~\cite{falko_tNbSe2}. However, no fully atomistic study of atomic relaxations and CDWs in twisted bilayers has been reported yet.

In this paper, we present results for the relaxed atomic structure of anti-parallel twisted bilayer \ch{NbSe2} for a twist angle of 3.14$^{\circ}$ (with a moir\'e unit cell containing 1986 atoms) using \textit{ab initio} density-functional theory \cite{kohn_dft}. We find that twisted bilayer \ch{NbSe2} exhibits significant in-plane and out-of-plane atomic relaxations. Comparing the relaxed atomic structure of twisted bilayer \ch{NbSe2} to that of twisted bilayer \ch{MoSe2}, we find striking qualitative differences that arise as a consequence of different stacking energetics in the two materials. We also analyze the CDWs in twisted bilayer \ch{NbSe2} and find that different CDW motifs coexist in different regions of the moir\'e unit cell: low-energy MM regions are dominated by filled-center, hollow-center and hexagonal $3\times 3$ CDWs while the surrounding domain wall regions exhibit a stripe CDW whose formation is a consequence of local strain in the twisted bilayer which suppresses contributions to the CDW from all but a single wavevector.  

\section{Results and Discussion}
We start by discussing the atomic structure of the relaxed twisted \ch{NbSe2} bilayer and compare our findings to twisted bilayer \ch{MoSe2}, which does not feature a CDW in its monolayer form. 

\subsection{Out-of-plane atomic relaxation}
Figures~\ref{fig:ils} (a) and (b) show the interlayer separation (ILS) between the metal atom layers with the high-symmetry stacking regions indicated by red symbols. In agreement with many previous studies \cite{kc_ff_tmd,Ultraflatbands_solitons_naik,recon_moire_tmd_maity,Vitale_Atalar_Mostofi_Lischner_2021,Weston_Zou_Enaldiev_Summerfield_Clark_Zólyomi_Graham_Yelgel_Magorrian_Zhou_etal,stack_domains_enaldiev}, we find that the relaxed twisted bilayers exhibit a corrugated structure. 

For twisted bilayer \ch{MoSe2} (shown in Figure~\ref{fig:ils}(a)), the ILS reaches a maximum of 6.86~$\ang$ in the XX stacking region (in which the chalcogen atoms of the top layer are directly above the chalcogen atoms of the bottom layer) and a minimum of 6.31~$\ang$ at the AB stacking region (in which the metal atoms of one layer have the same in-plane positions as the chalcogen atoms of the other layer). The ILS in the MM stacking region (in which the metal atoms of the top layer are directly above the metal atoms of the bottom layer) is 6.37~$\ang$, only slightly larger than the ILS of the AB stacking region. We note that the interlayer separations at the AB, MM, and XX points are close to those found in the corresponding untwisted bilayers \cite{kc_ff_tmd}. 

In twisted bilayer \ch{NbSe2} (shown in Figure~\ref{fig:ils}(b)), the ILS is also largest at the XX stacking region with a value of 6.57~$\ang$. However, the ILS of the AB point (6.31~$\ang$) is significantly larger than that of the MM point (6.03~$\ang$). Again, the interlayer separations at different stacking points are similar to those of the untwisted bilayers (6.57~$\ang$ for XX stacking, 6.25~$\ang$ for AB stacking, and 6.05~$\ang$ for MM stacking). 

The different relative ordering of the interlayer separation of the MM stacking regions and the AB stacking regions in \ch{NbSe2} compared to \ch{MoSe2} results in a qualitatively different interlayer separation landscape in the two twisted bilayer materials, as seen in Figs.~\ref{fig:ils}(a) and (b). In twisted bilayer \ch{NbSe2}, a triangular plateau of relatively constant interlayer separation emerges, which is centered on the AB point and has corners at the XX points. Another triangular region of small interlayer separation is centered on the MM points. Its size is significantly larger than the MM centered spherical region of small ILS in twisted \ch{MoSe2}. Our findings are consistent with the results of McHugh~\textit{et al.}, who used a continuum elasticity approach \cite{falko_tNbSe2}.

\subsection{In-plane atomic relaxations} 
Figures~\ref{fig:ils}(c) and (d) compare the in-plane atomic displacements (relative to the unrelaxed starting structure) of the bottom layers of twisted bilayer \ch{NbSe2} and twisted bilayer \ch{MoSe2} (the results for the top layers are shown in Supplementary Fig.~\ref{fig:displ_top}). In twisted bilayer \ch{MoSe2}, large vortex-like displacement patterns are observed which are centered at the points of XX stacking. In between these vortices are domain walls connecting the MM and AB stacking points where the displacements are small. These findings can be understood by considering the energetics of the different stackings: since XX stacking is energetically unfavorable, atoms move to minimize the area of XX stacking regions resulting in the formation of vortices. This, in turn, leads to shear strain solitons around the vortex, giving rise to domain walls \cite{kc_ff_tmd,Ultraflatbands_solitons_naik}. 

In twisted bilayer \ch{NbSe2}, we also find vortices centered on points of XX stacking. However, their shape is less circular than in twisted \ch{MoSe2} and the magnitude of the displacements along a circular path centered at the XX point is strongly modulated. In twisted bilayer \ch{MoSe2}, the magnitudes of the in-plane displacements along the vortices are up to 0.17~$\ang$ and relatively uniform along circular paths centered on the XX points. In contrast, in twisted bilayer \ch{NbSe2}, the magnitudes of the displacements are both larger and exhibit more variation, e.g., they can be as large as 0.44~$\ang$ where the circular displacement pattern intersects the lines connecting the vortex center to the MM points, and 0.25~$\ang$ on the lines connecting the vortex center to the AB points. 

The different behaviour of the in-plane displacements of twisted bilayer \ch{NbSe2} and \ch{MoSe2} is a consequence of their different out-of-plane relaxations. While the interlayer separation of twisted bilayer \ch{MoSe2} is approximately constant along a circle centerd at the XX point, this is not the case for twisted bilayer \ch{NbSe2}, as can be seen Figs.~\ref{fig:ils}(a) and (b). In particular, in \ch{NbSe2} the ILS is small along lines joining XX and MM points, which gives rise to a strong steric repulsion which in turn drives large in-plane displacements. In contrast, the ILS is larger along lines joining XX and AB points giving rise to weaker steric repulsion and smaller in-plane displacements.

\subsection{Charge density waves}

\begin{figure*}[htb!]
    \begin{subfigure}[t]{0.235\textwidth}
        \centering
        \caption{\hspace{12pt}Filled-center CDW}
        \includegraphics[width=\textwidth]{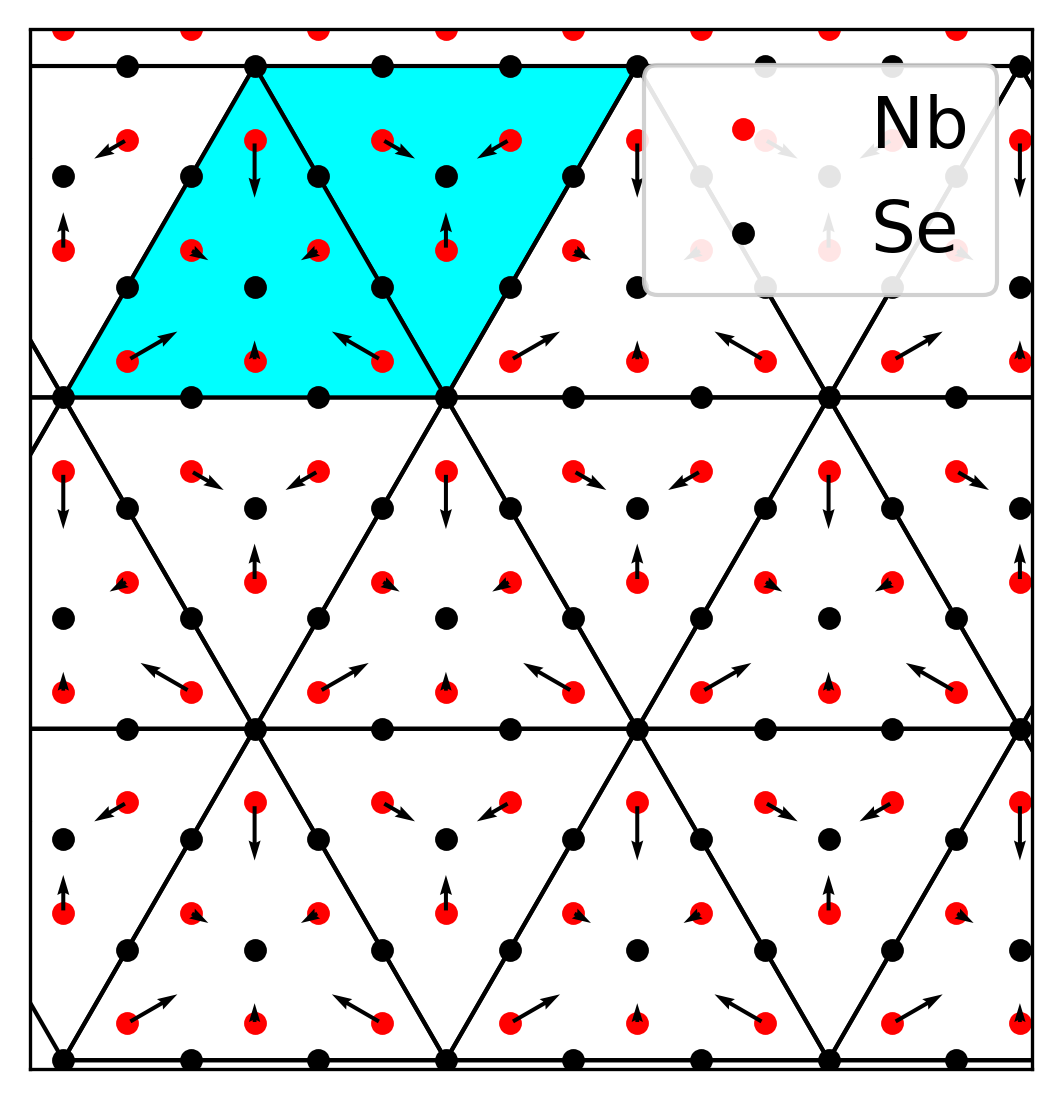}
    \end{subfigure}
    ~ 
    \begin{subfigure}[t]{0.235\textwidth}
        \centering
        \caption{\hspace{9pt}Hollow-center CDW}
        \includegraphics[width=\textwidth]{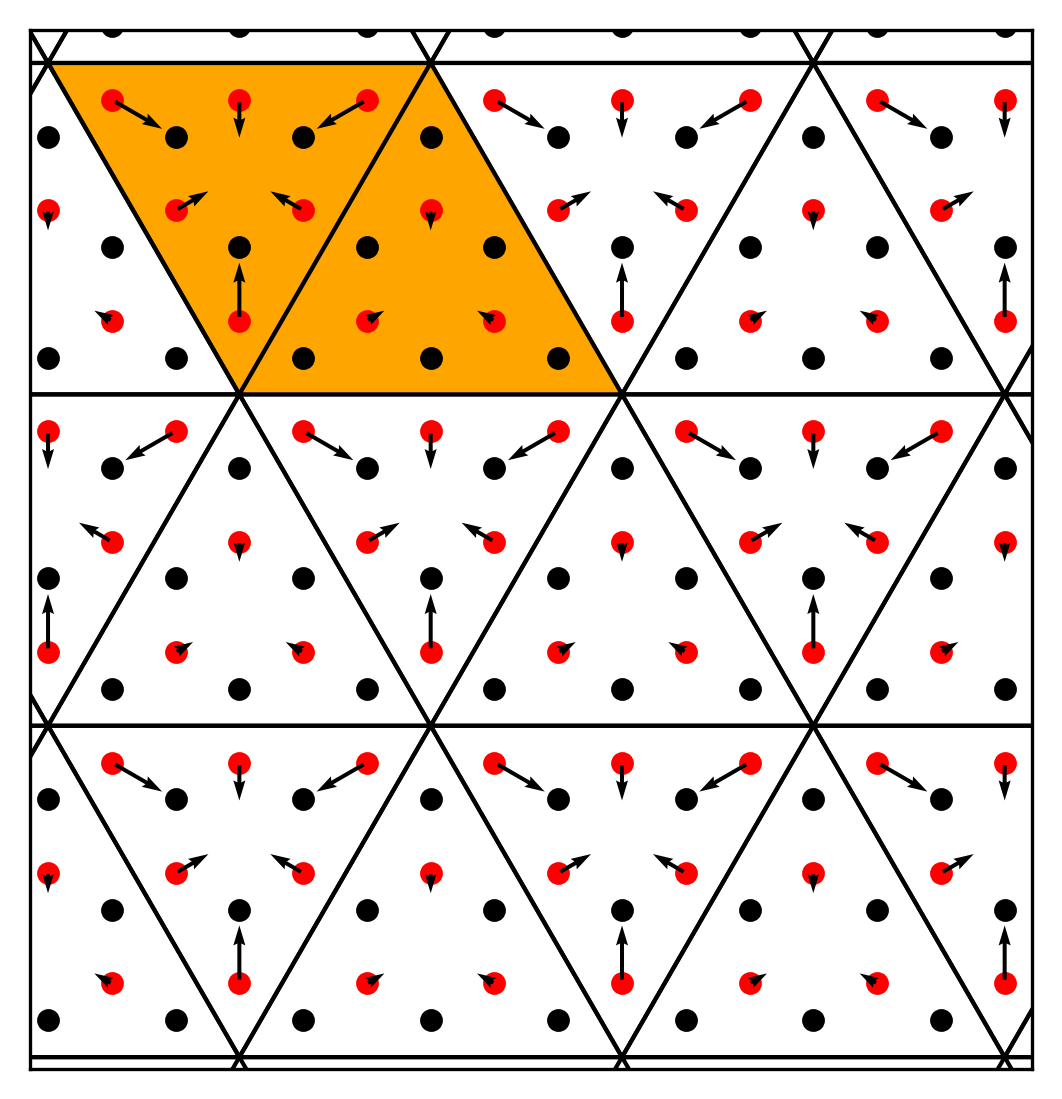}
    \end{subfigure}
    ~
    \begin{subfigure}[t]{0.235\textwidth}
        \centering
        \caption{\hspace{15pt}Hexagonal CDW}
        \includegraphics[width=\textwidth]{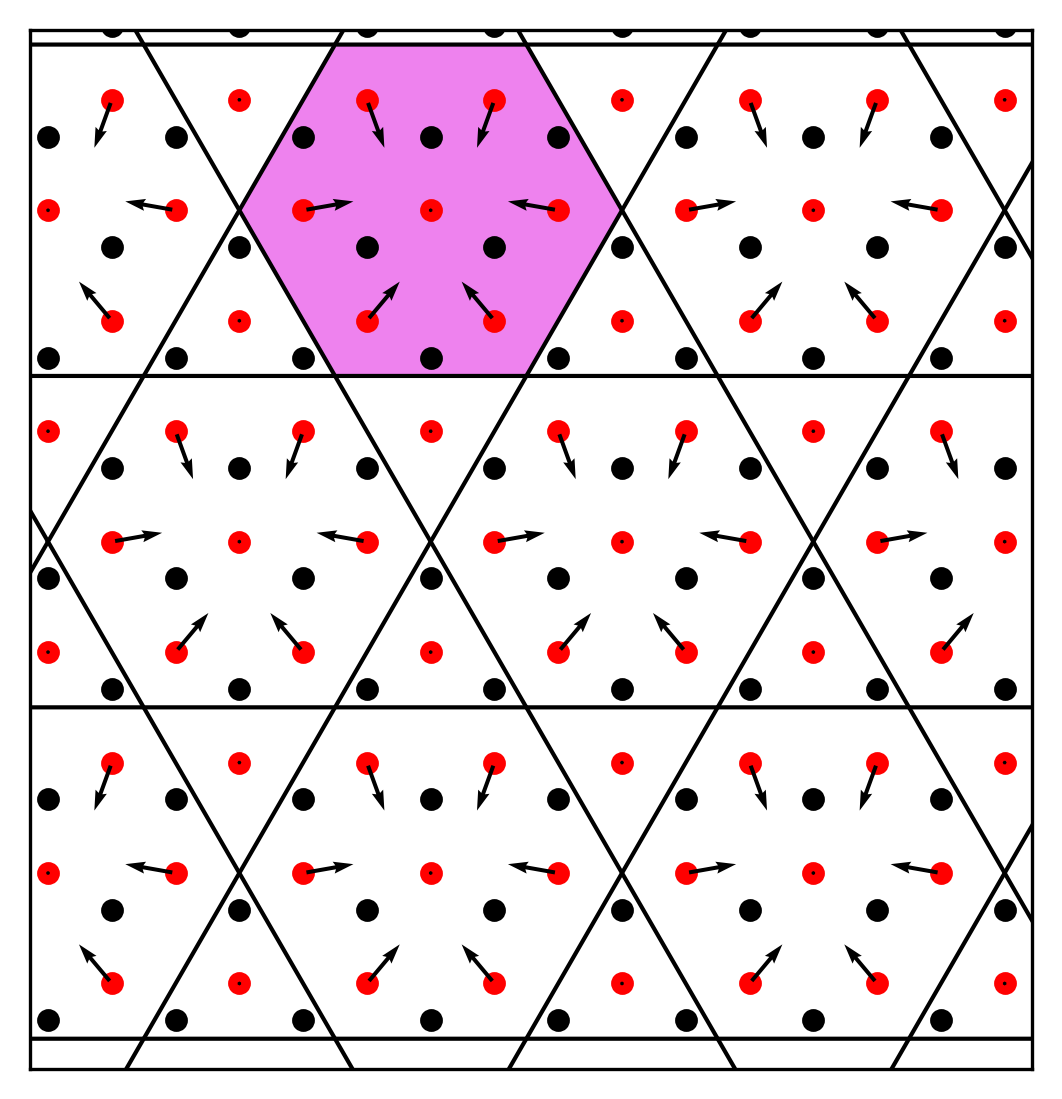}
    \end{subfigure}
    ~
    \begin{subfigure}[t]{0.235\textwidth}
        \centering
        \caption{\hspace{19pt}Stripe CDW}
        \includegraphics[width=\textwidth]{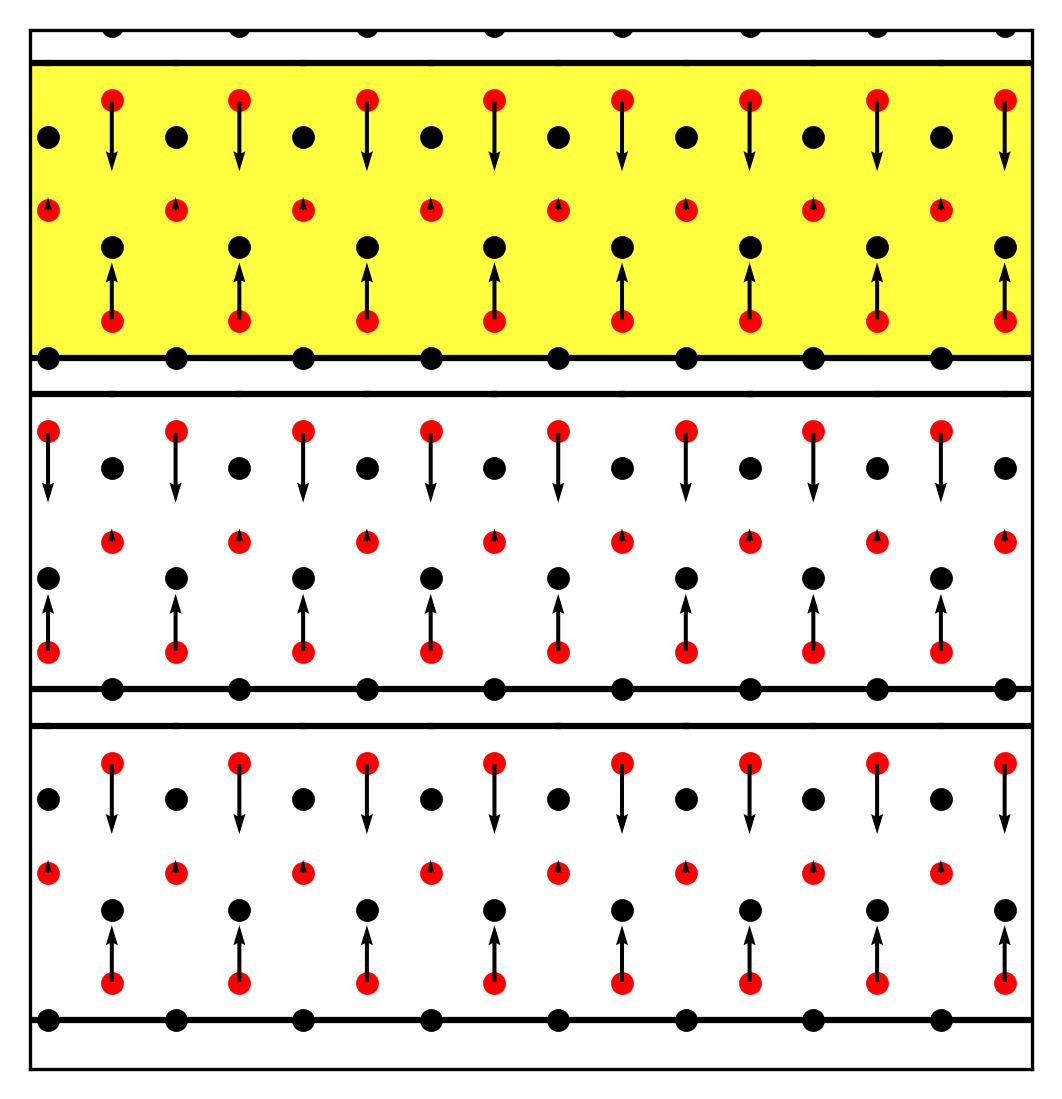}
    \end{subfigure}
    ~
    \begin{subfigure}[t]{0.235\textwidth}
        \centering
        \caption{}
        \includegraphics[width=\textwidth]{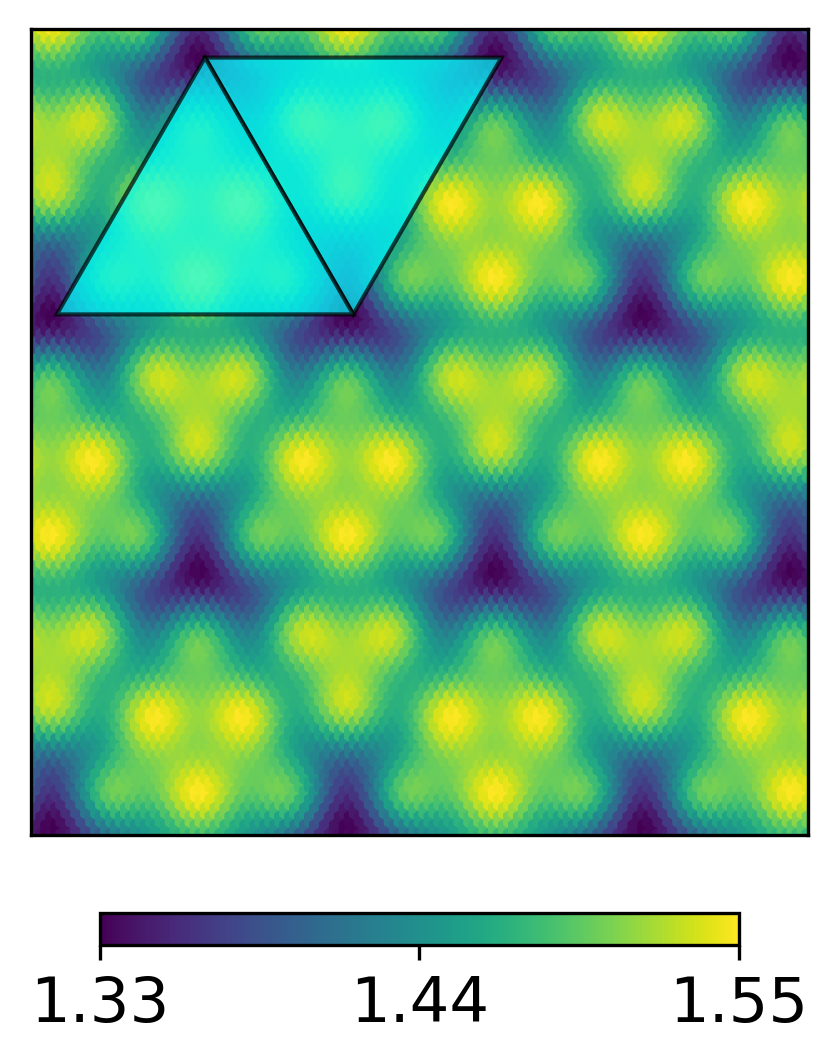}
    \end{subfigure}
    ~
    \begin{subfigure}[t]{0.235\textwidth}
        \centering
        \caption{}
        \includegraphics[width=\textwidth]{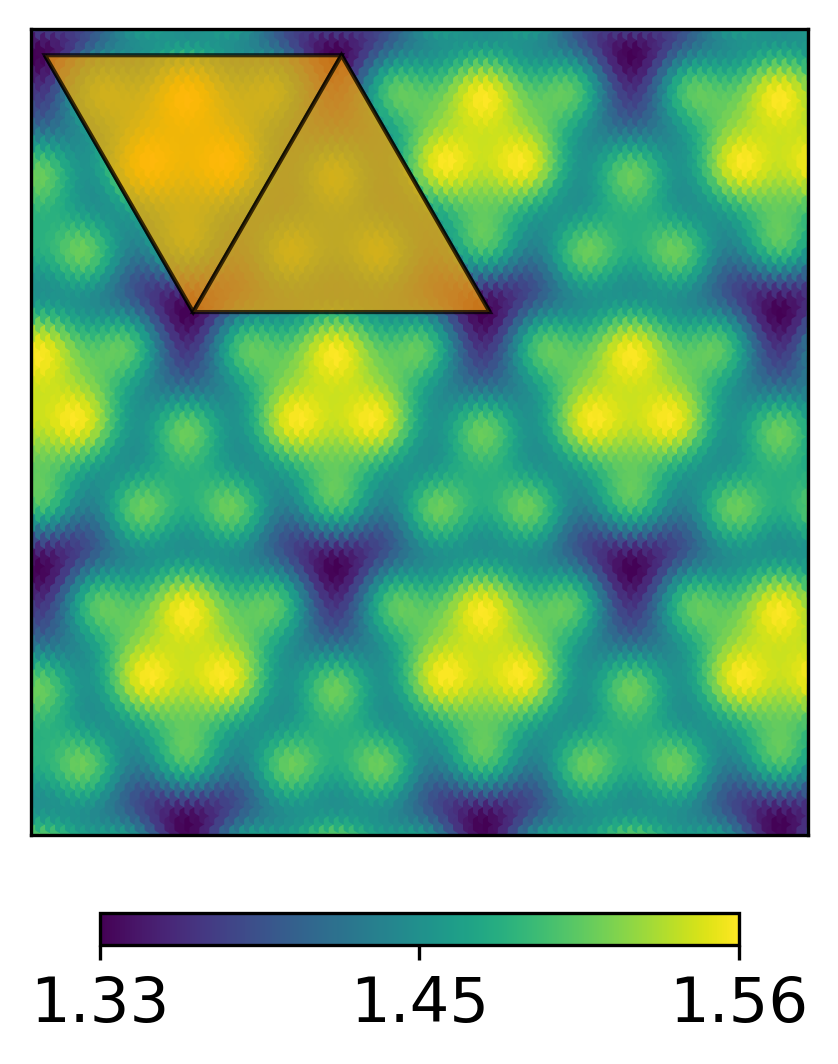}
    \end{subfigure}
    ~
    \begin{subfigure}[t]{0.235\textwidth}
        \centering
        \caption{}
        \includegraphics[width=\textwidth]{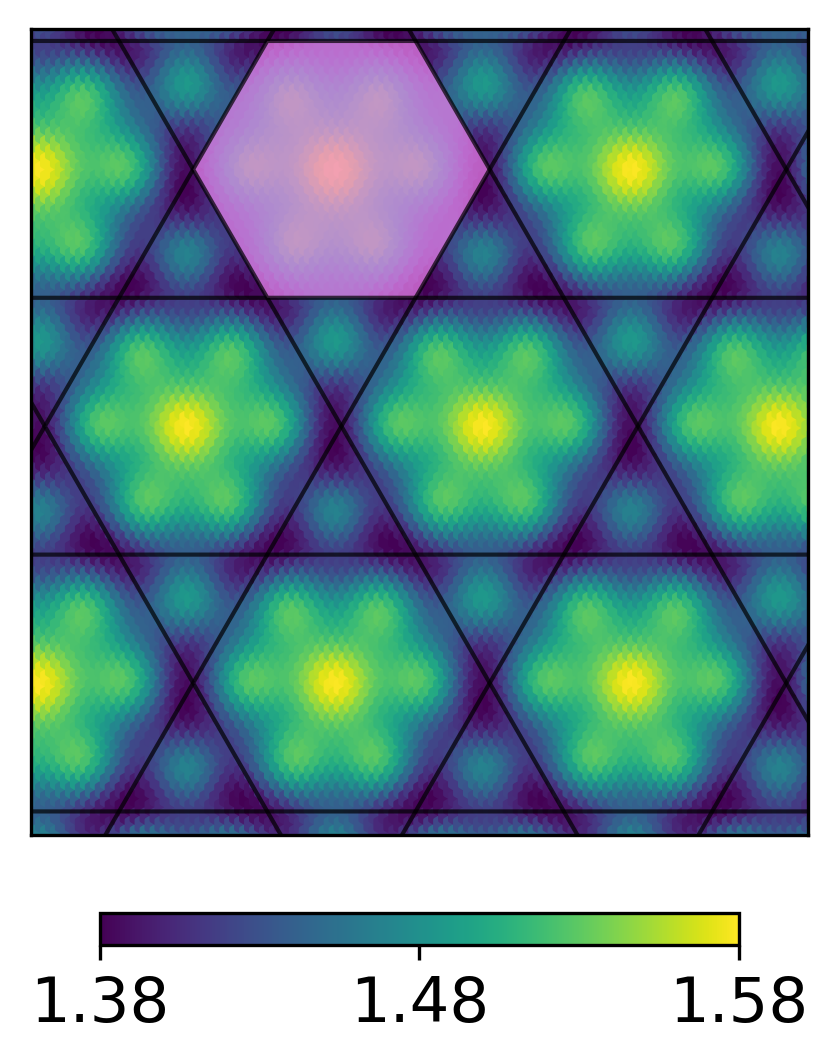}
    \end{subfigure}
    ~
    \begin{subfigure}[t]{0.235\textwidth}
~        \centering
        \caption{}
        \includegraphics[width=\textwidth]{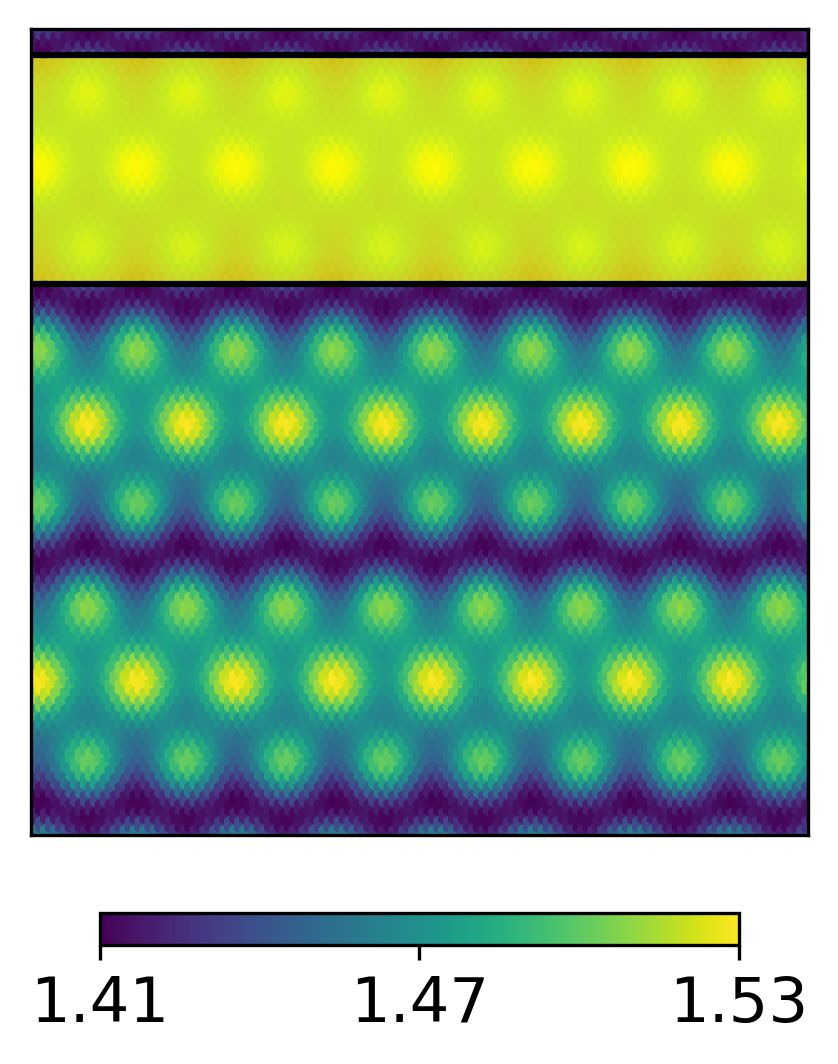}
    \end{subfigure}
    \caption{Atomic displacements and smeared Nb atomic densities of different charge density wave states. Atomic displacements of Nb atoms (red circles) due to a (a) filled-center CDW, (b) Hollow-center CDW, (c) hexagonal CDW and (d) stripe CDW in monolayer NbSe$_2$. The Se atoms (black circle) are also shown. Smeared Nb atomic densities of (e) filled-center CDW, (f) Hollow-center CDW, (g) hexagonal CDW and (h) stripe CDW in monolayer NbSe$_2$. The colored motifs used for highlighting the CDW motifs in Figures~\ref{fig:smear_tnbse2_n10}(b) and (d) are demonstrates in these figures. We use (a) and (e): cyan triangles for three- or six-atom filled-center CDW, (b) and (f): orange triangles for three- or six-atom hollow-center CDW, (c) and (g): purple hexagon for hexagonal CDW, and (d) and (h): yellow strips for stripe CDW.}
    \label{fig:cdw_motif_dem}
\end{figure*}

In contrast to the monolayer, it is difficult in a twisted bilayer to ascertain the presence of a CDW from the atomic displacement vectors relative to the initial structure without CDW: In a twisted bilayer system large in-plane atomic relaxations occur to reduce the area of high-energy stacking regions and increase the area of low-energy stacking regions. These displacements are often much larger than the displacements which give rise to the formation of CDWs and this makes the identification of a CDW challenging.

As an alternative way to visualise the presence of CDWs in twisted bilayer \ch{NbSe2}, we use the 
smeared atomic density of Nb atoms obtained by associating a Gaussian function with each Nb site, see Methods for details. For the monolayer, the formation of a $3 \times 3$ triangular CDW can easily be ascertained from an inspection of the smeared Nb density, see Fig.~\ref{fig:cdw_motif_dem}(e). The $3 \times 3$ triangular CDW is characterized by two sets of triangles: larger six-atom triangles and smaller three-atom triangles. If the triangles (highlighted in cyan color) are centered on interstitial sites, the CDW is referred to as a hollow-center $3 \times 3$ CDW, see Fig.~\ref{fig:cdw_motif_dem}(a). In contrast, if the triangles (highlighted in orange color) are centered on the Se atoms, it is called a filled-center $3 \times 3$ CDW~\cite{unveil_cdw_sc}, see Fig.~\ref{fig:cdw_motif_dem}(b). The bright triangles in the smeared atomic density are separated by dark blue lines  running along three directions with angles of $120\degree$ between them. These three directions correspond to the three wave vectors $\mathbf{q}_n$ (with $n=1,2,3$) at which the phonon spectrum of the unrelaxed monolayer exhibits instabilities~\cite{unveil_cdw_sc,weak_dim_dep_cdw}, see discussion below. Another CDW which has been predicted to be metastable in the monolayer~\cite{elas_cdw_guster} is the hexagonal CDW, where the nearest six nearest-neighbor Nb atoms move towards the central Nb atom as shown in  Fig.~\ref{fig:cdw_motif_dem}(c) giving rise to hexagons in the smeared atomic density (highlighted by purple color). Finally, one-dimensional stripe CDWs (highlighted in yellow color), see Fig.~\ref{fig:cdw_motif_dem}(d), have been predicted to occur in the monolayer under applied strain~\cite{Soumyanarayanan_Yee_He_van_wezel_Rahn_Rossnagel_Hudson_Norman_Hoffman_2013, unveil_cdw_imp}.

\begin{figure*}[htb!]
    \begin{subfigure}[t]{0.49\textwidth}
        \centering
        \caption{\hspace{75pt}Bottom layer}
        \includegraphics[width=\textwidth]{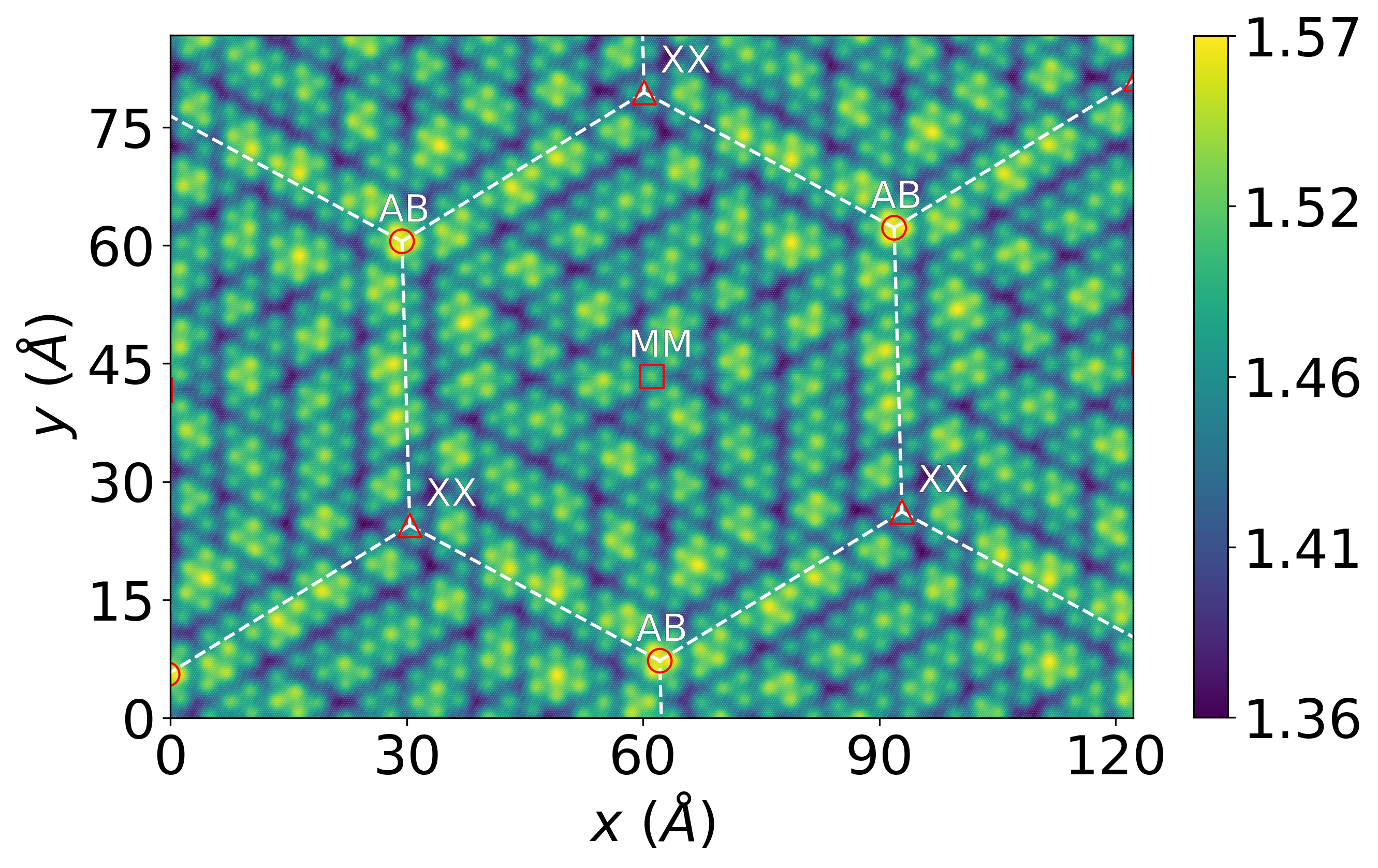}
        \label{fig:smear_tnbse2_bot}
    \end{subfigure}
    ~ 
    \begin{subfigure}[t]{0.432\textwidth}
        \vspace{1pt}
        \centering
        \caption{\hspace{75pt}Bottom layer}
        \includegraphics[width=\textwidth]{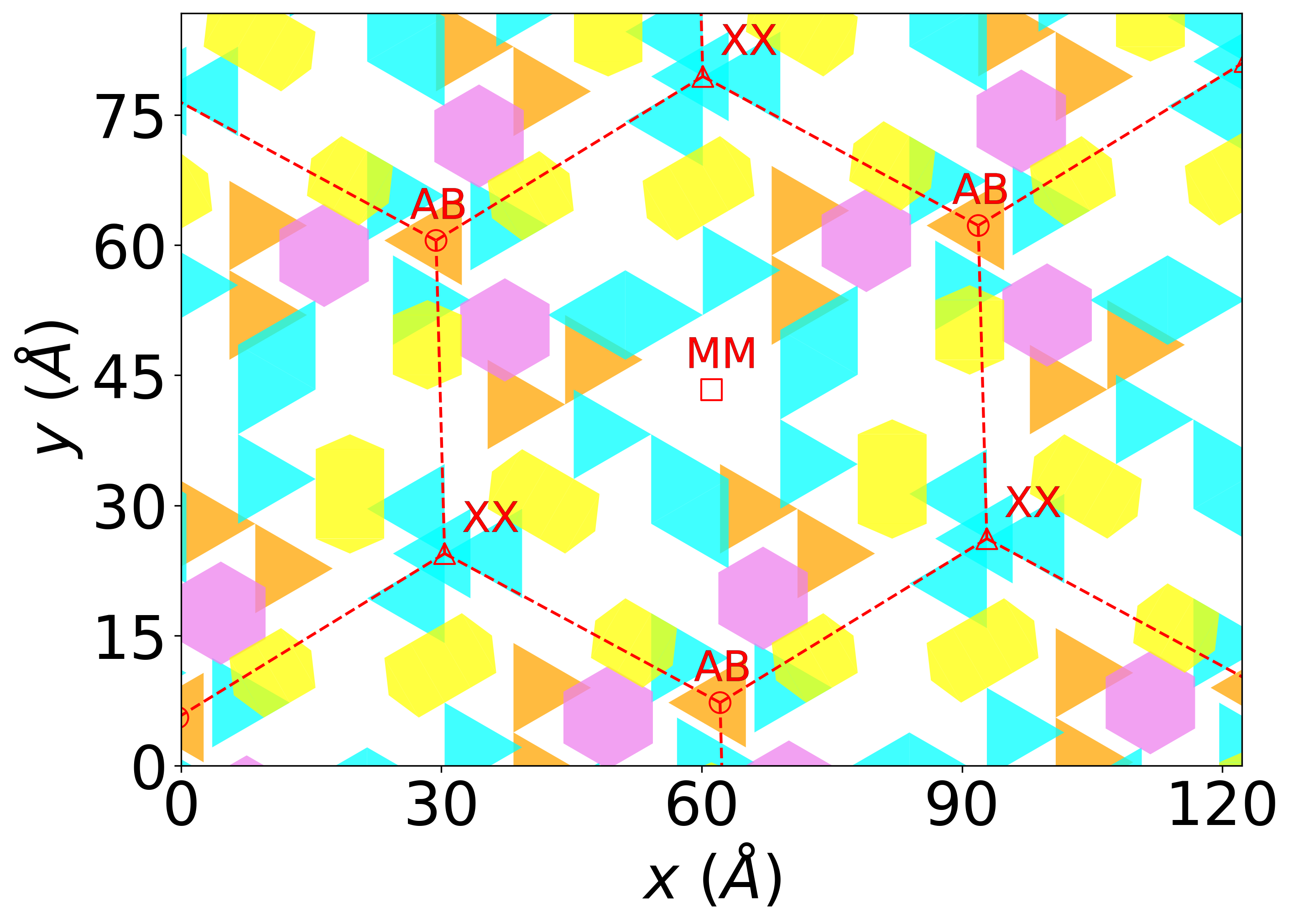}
        \label{fig:smear_tnbse2_label_bot}
    \end{subfigure}
    ~
    \begin{subfigure}[t]{0.49\textwidth}
        \centering
        \caption{\hspace{82pt}Top layer}
        \includegraphics[width=\textwidth]{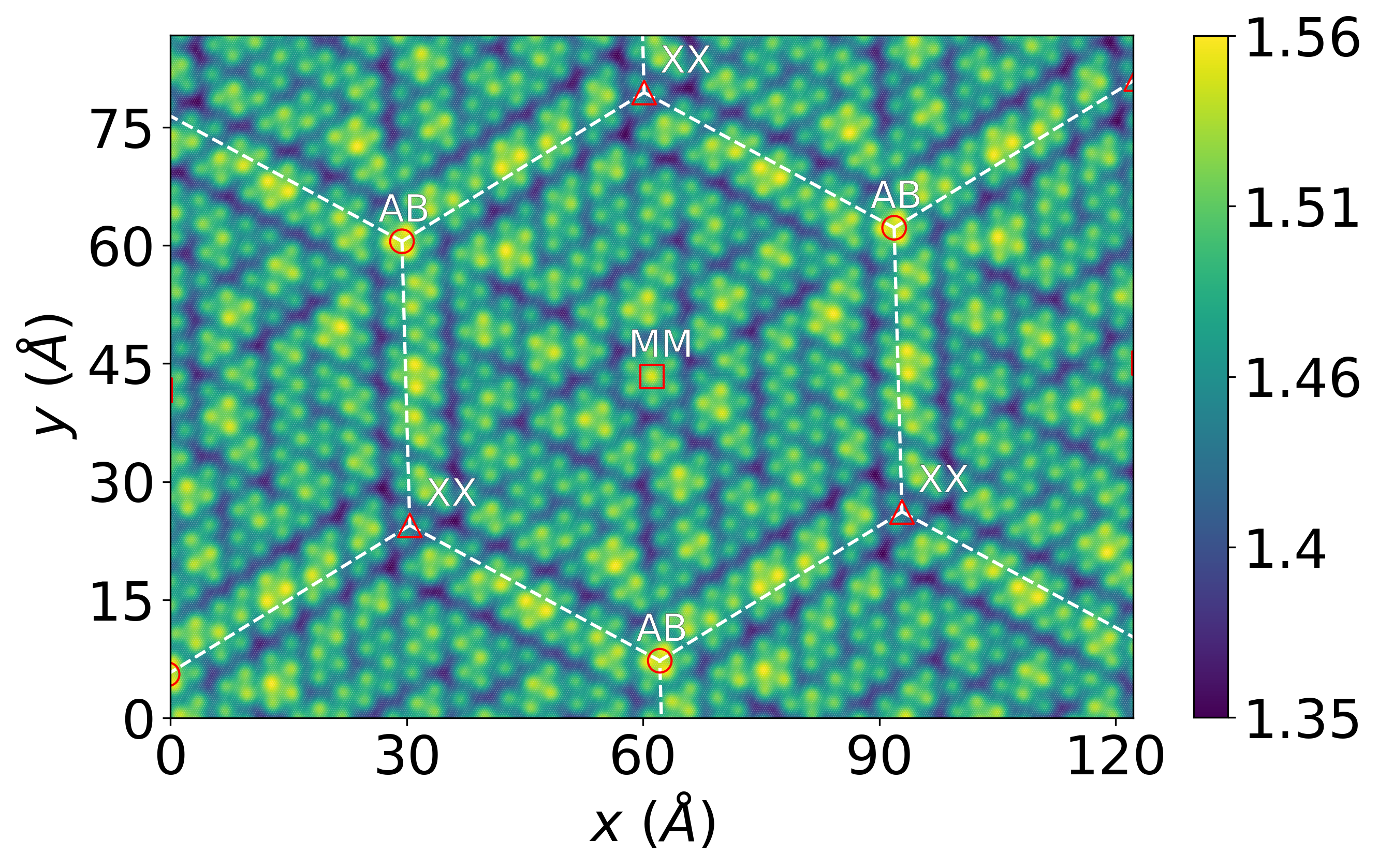}
        \label{fig:smear_tnbse2_top}
    \end{subfigure}
    ~
    \begin{subfigure}[t]{0.432\textwidth}
        \vspace{1pt}
        \centering
        \caption{\hspace{82pt}Top layer}
        \includegraphics[width=\textwidth]{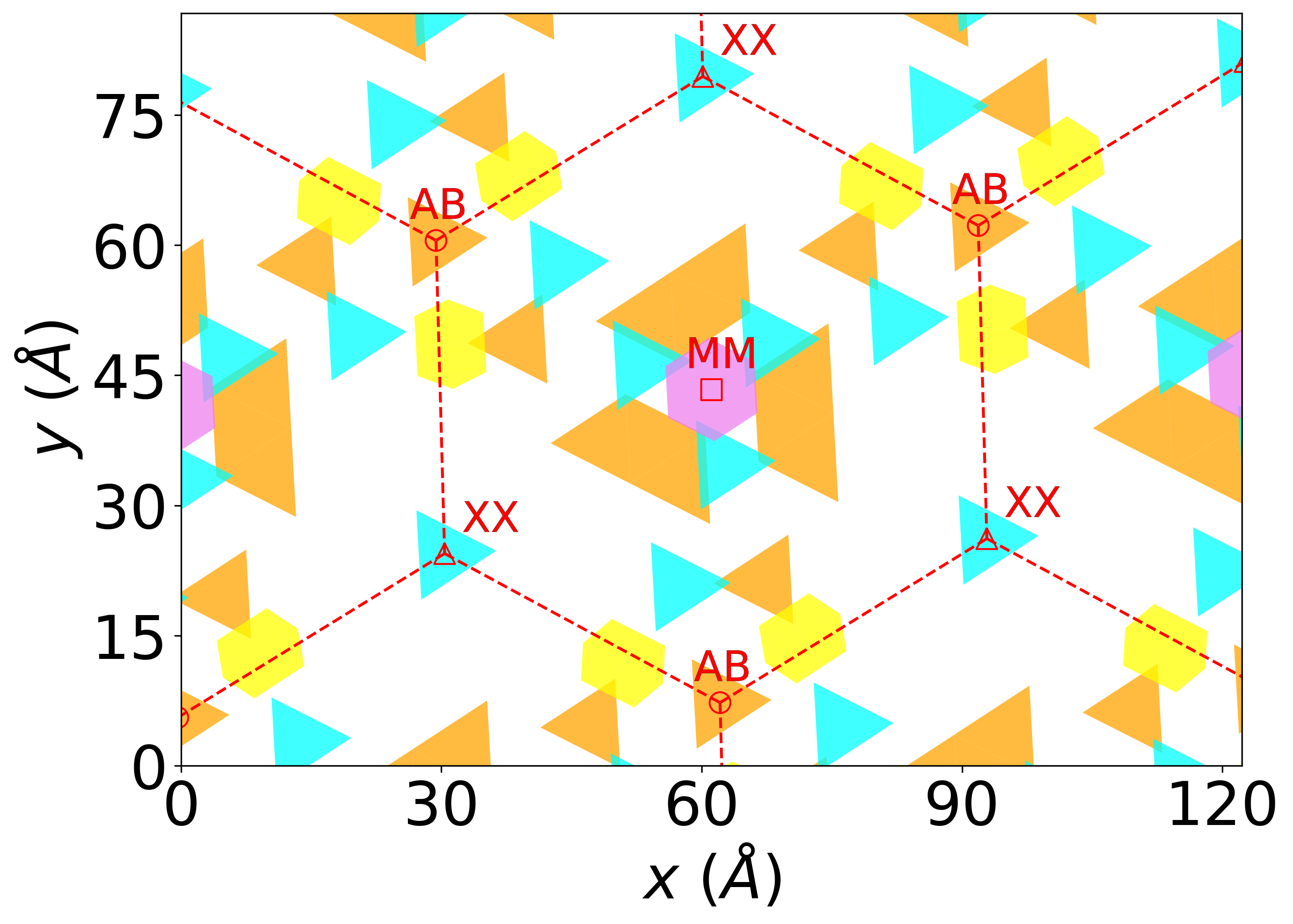}
        \label{fig:smear_tnbse2_label_top}
    \end{subfigure}
    \caption{Coexisting charge density waves in twisted bilayer \ch{NbSe2}. Smeared Nb atomic density of (a) the bottom layer and (c) the top layer of twisted bilayer \ch{NbSe2} at a twist angle of 3.14$^\circ$.  The hollow-center CDWs (orange triangles), filled-center CDWs (cyan triangles), hexagonal CDWs (violet hexagons) and stripe CDWs (yellow symbols) are indicated in (b) for the bottom layer and in (d) for the top layer. The red circles denote the centers of the AB stacking regions (metal on top of chalcogen and vice versa), the red squares denote the centers of the MM stacking region (metal on top of metal), and the red triangles denote the centers of the XX stacking regions (chalcogen on top of chalcogen). The white dashed lines join the AB and XX centers for visual aid in (a) and (c). In (b) and (d), red dashed lines are used.}
    \label{fig:smear_tnbse2_n10}
\end{figure*}


Figures~\ref{fig:smear_tnbse2_n10}(a) and (c) show the smeared Nb atomic densities of the bottom and top layers of twisted bilayer \ch{NbSe2}, respectively. In contrast to the monolayer which exhibits a uniform triangular $3\times 3$ CDW phase in its ground state, a variety of different CDW-like motifs can be observed in the twisted bilayer. To characterize these motifs, we have locally projected the displacements patterns obtained from our DFT relaxations onto reference displacement patterns characterizing the various CDWs shown in Fig.~\ref{fig:cdw_motif_dem}, see Methods for details. In the MM stacking region (which has the lowest energy and therefore increases in size upon relaxation), we find that hollow-center (orange triangles), filled-center (cyan triangles) and hexagonal (purple hexagons) CDWs coexist, as can be seen in Figs.~\ref{fig:smear_tnbse2_n10}(b) and (d). In the bottom layer (Fig.~\ref{fig:smear_tnbse2_n10}(b)), the MM center is surrounded by filled-center CDWs which are interspersed by a few hollow-center and hexagonal CDW motifs. At the center of the AB region, a hollow-center CDW motif is found which is surrounded by three filled-center CDW motifs; and a filled-center center motif is found as the center of the XX region. In the top layer (Fig.~\ref{fig:smear_tnbse2_n10}(d)), a hexagonal CDW motif is found at the MM center which is surrounded by hollow-center CDWs and a few filled-center CDW motifs. As in the bottom layer, a hollow-center (filled-center) CDW motif is located at the AB (XX) center. For comparison, Supplementary Fig.~\ref{fig:smear_tmose2} shows the smeared Mo atomic density of relaxed twisted bilayer \ch{MoSe2}. In this system, no CDW motifs can be observed.

\begin{figure}
    \centering
    \includegraphics[width=0.48\textwidth]{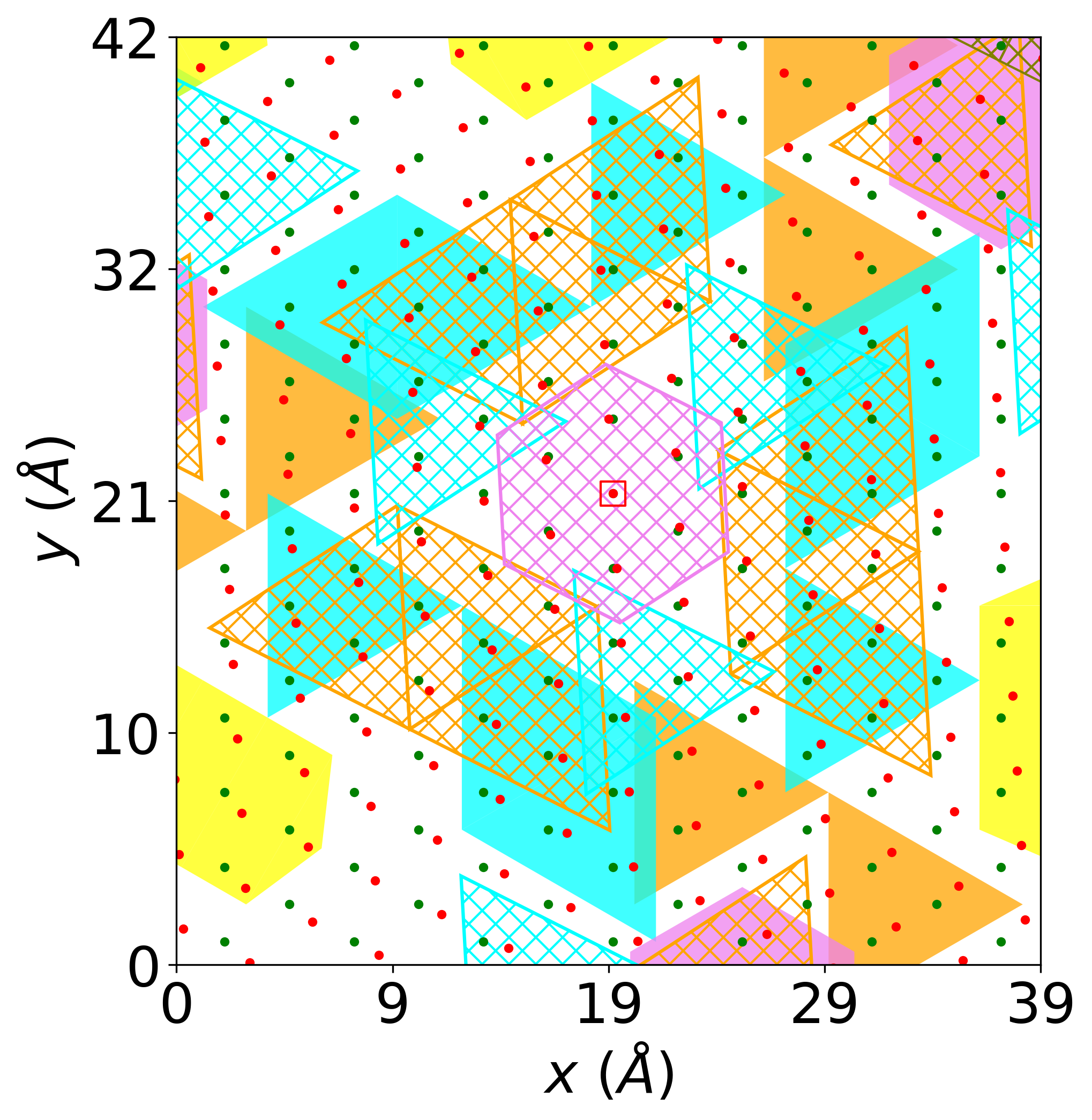}
    \caption{Interlayer stacking of CDW motifs in twisted bilayer \ch{NbSe2} in the MM region. The green circles represent the Nb atoms in the bottom layer, and the red circles represent the Nb atoms in the top layer. The cyan (orange) solid triangles represents the filled-center (hollow-center) CDW motifs in the bottom layer. Purple solid hexagons represent the hexagonal CDW motifs in the bottom layer. The cyan (orange) hatched triangles and purple hatched hexagons represent the filled-center (hollow-center) and hexagonal CDW motifs in the top layer. The yellow solid strips show the stripe CDW in the bottom layer.}
    \label{fig:at_over_MM}
\end{figure}

In Fig.~\ref{fig:at_over_MM}, we superimpose the CDW motifs of the top and bottom layers in and around the MM region. We observe that a hollow-center CDW motif (orange triangle) is found in the top layer whenever a filled-center CDW motif (cyan triangle) is located directly underneath in the bottom layer and vice versa. This means that Nb atoms in the two layers move in opposite directions, i.e., if a Nb atom in the bottom layer moves towards an interstitial site, its nearest neighbor in the top layer moves towards an Se site and vice versa. This coordinated motion of Nb atoms in both layers reduces the steric repulsion between Se atoms: the in-plane movement of the Nb atoms forming CDWs induces a corresponding out-of-plane displacement of the neighboring Se atoms. If Nb atoms in both layers move in the same direction, the neighboring Se atoms would move towards each other increase the total energy due to steric repulsion.

\begin{figure*}[htb!]
    \centering
    \begin{subfigure}[t]{0.32\textwidth}
        \centering
        \caption{}
        \includegraphics[width=\textwidth]{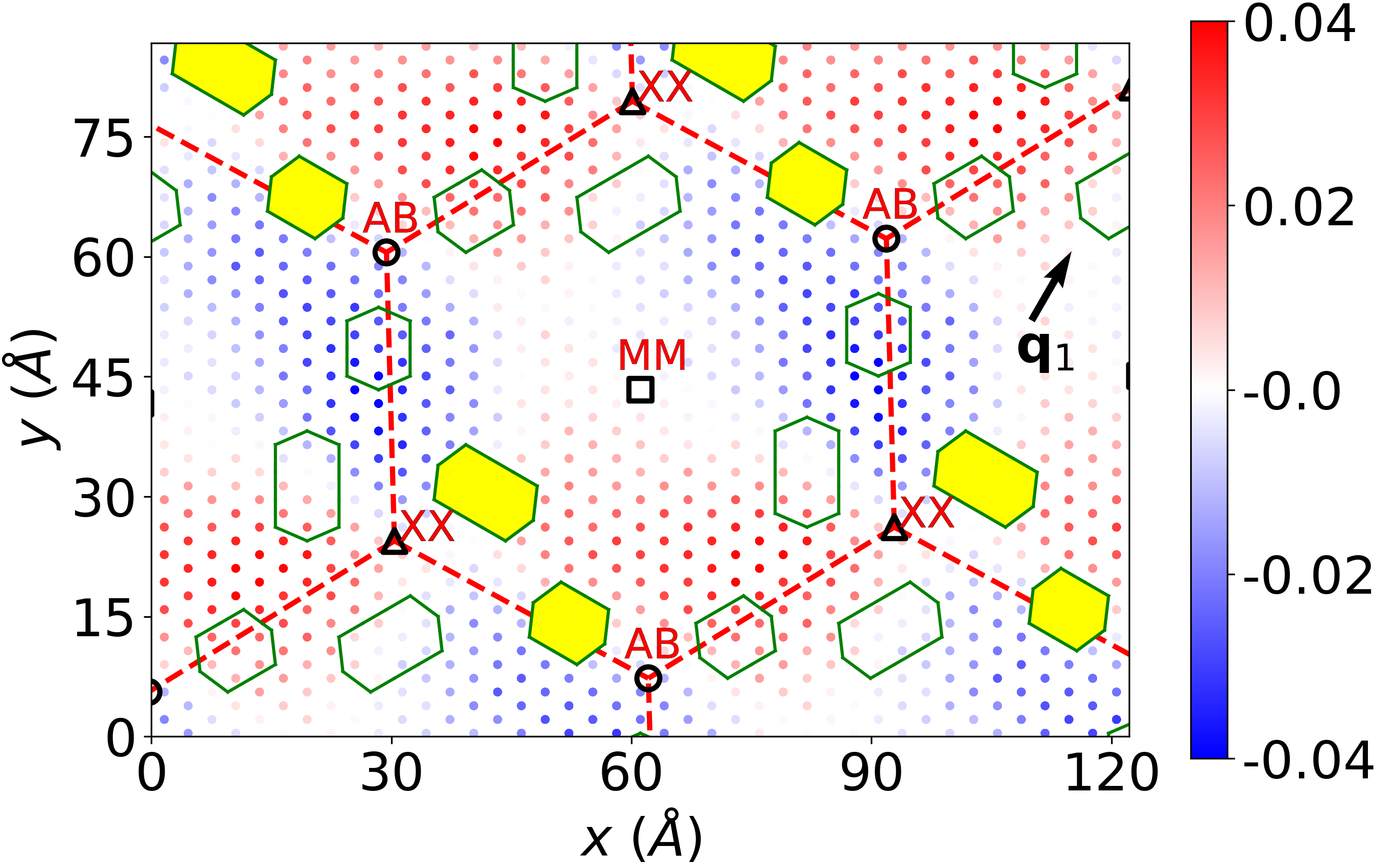}
    \end{subfigure}
    ~
    \begin{subfigure}[t]{0.32\textwidth}
        \centering
        \caption{}
        \includegraphics[width=\textwidth]{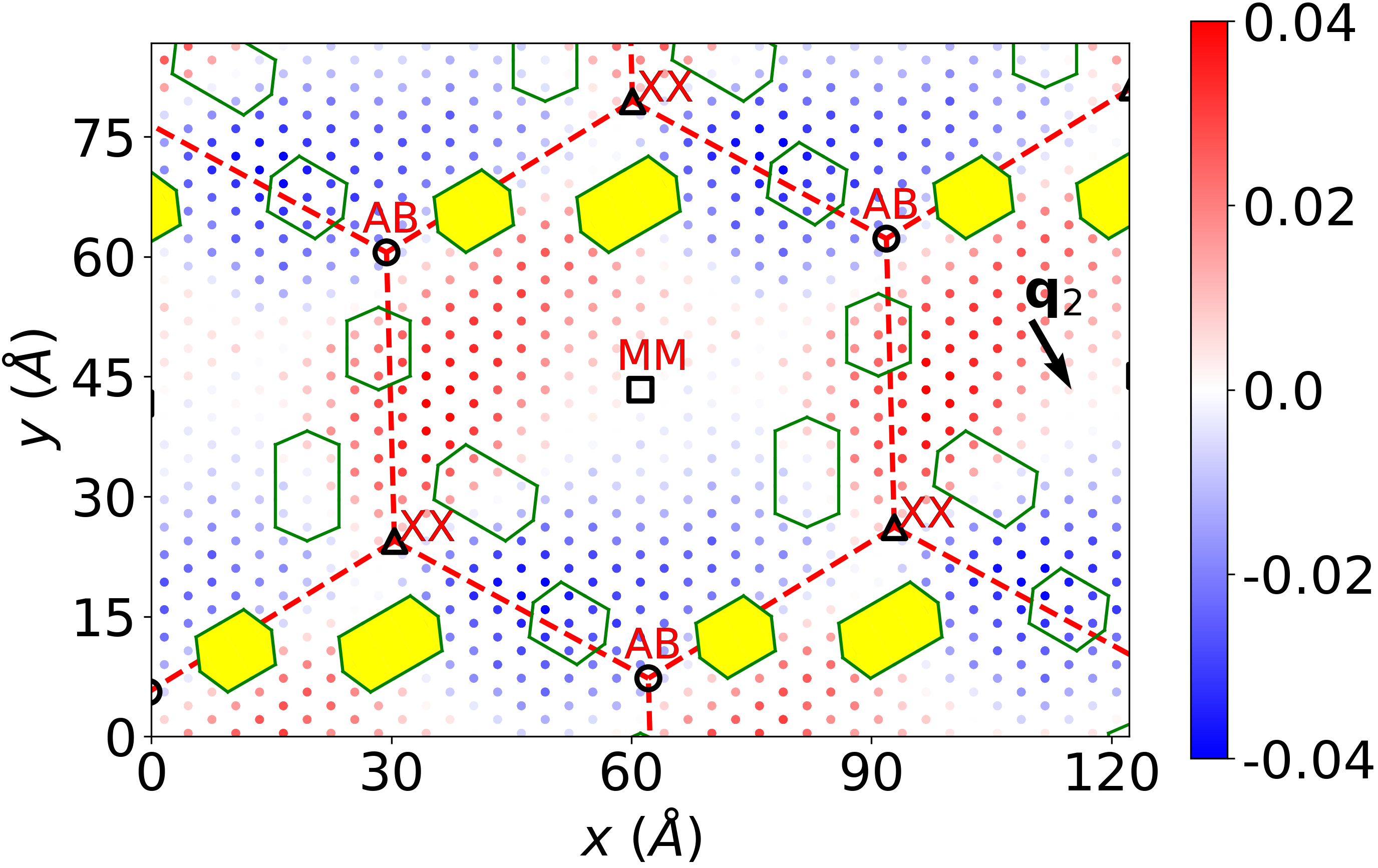}
    \end{subfigure}
    ~
    \begin{subfigure}[t]{0.32\textwidth}
        \centering
        \caption{}
        \includegraphics[width=\textwidth]{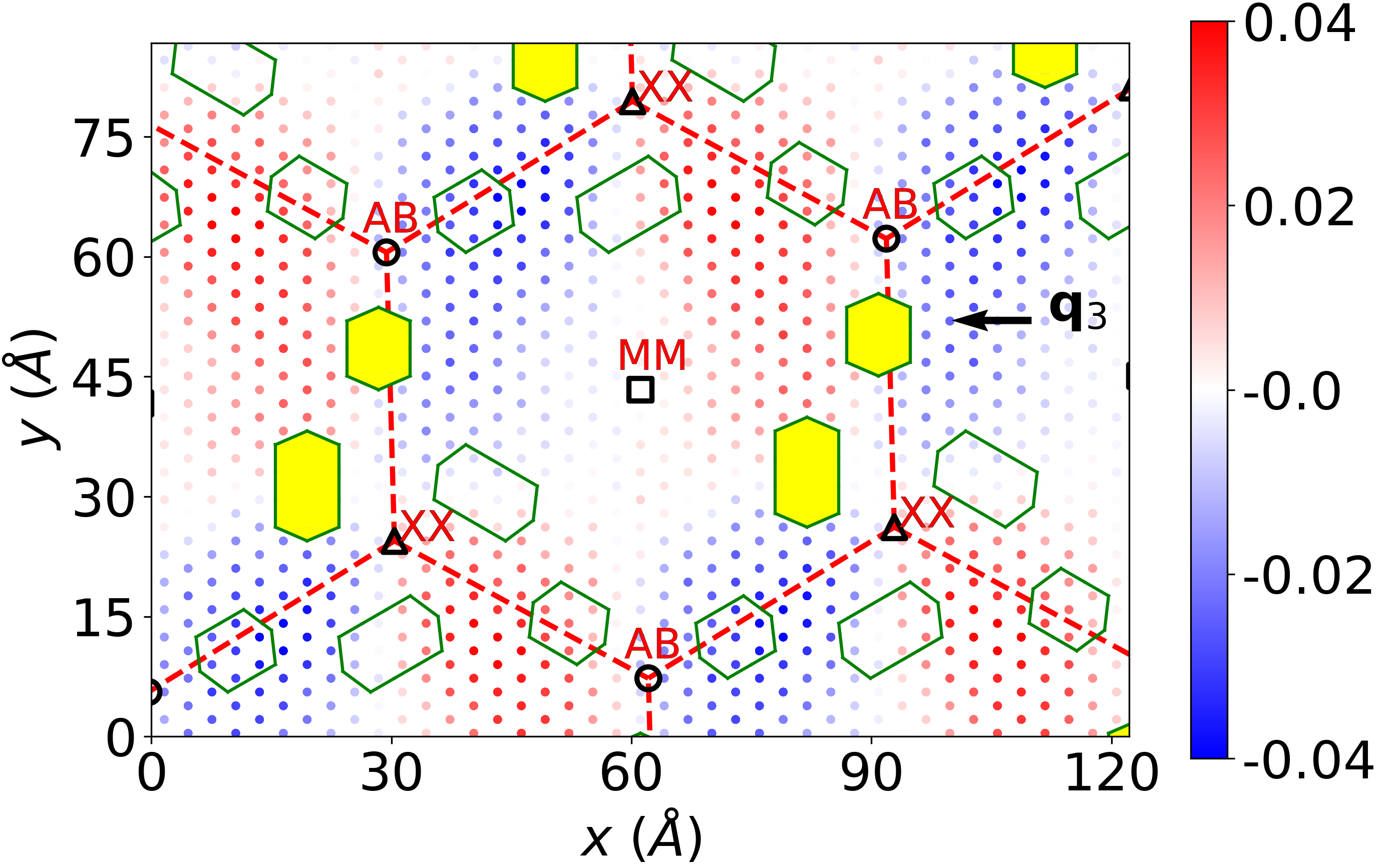}
    \end{subfigure}
    \caption{Local strain and stripe charge density waves in twisted bilayer NbSe$_2$. Distribution of strain in the bottom layer along each of the three charge density wave vectors: (a) $\mathbf{q}_1$, (b) $\mathbf{q}_2$, and (c) $\mathbf{q}_3$. Green strips indicate the stripe CDWs. In each panel, the hexagons describing stripe CDWs characterized by the wave vector $\mathbf{q}_1$ (a), $\mathbf{q}_2$ (b) and $\mathbf{q}_3$ (c) are colored in yellow. The color of the solid circles represents the magnitude of the strain. The black circles denote the centers of the AB stacking regions (metal on top of chalcogen and vice versa), the black squares denote the centers of the MM stacking regions (metal on top of metal), and the black triangles denote the centers of the XX stacking regions (chalcogen on top of chalcogen). The red dashed lines join the AB and XX centers for visual aid.}
    \label{fig:strain_cdw_tnbse2}
\end{figure*}

Surrounding the regions of MM stacking are domains walls which connect the XX points to the AB points. In these domain wall regions, we observe stripe CDWs (yellow strips) which exhibit an extended character along the direction of the domain wall. The formation of one-dimensional CDWs along the domain walls is a consequence of the local strain. In particular, we find that if the atoms experience a local strain along one of the three CDW wavevectors $\{\mathbf{q}_n\}$, the contribution to the CDW from this wavevector is suppressed. Fig.~\ref{fig:strain_cdw_tnbse2} shows the strain along each of the three CDW wavevectors (see Methods for details). This demonstrates that a one-dimensional CDW with a specific wavevector, say $\mathbf{q}_1$, occurs in regions where there is no strain along $\mathbf{q}_1$, but significant strain along $\mathbf{q}_2$ and $\mathbf{q}_3$. This is consistent with previous findings that non-uniform strain induces one-dimensional stripe charge density waves \cite{uni_strain_cdw_flicker,strain_stripe_cdw}.

Our calculations suggest the following picture for the fate of the CDW in twisted bilayer \ch{NbSe2}: atomic relaxations driven by the energetics of different stacking arrangements give rise to a strain pattern that acts as a template for the formation of CDWs. Since the strain pattern is commensurate with the moir\'e unit cell, the resulting CDW pattern will also have this property (even if the the periodicity of the moir\'e unit cell is not commensurate with $3 \times 3$ CDW which has the lowest energy in the monolayer). To further support this conclusion, we have carried out a calculation for a $3 \times 3$ supercell (which is commensurate with the $3 \times 3$ CDW) of a twisted bilayer \ch{NbSe2} at a twist angle of $\theta=9.43\degree$. As expected, we find that each of the nine moir\'e unit cells contained in the supercell have the same atomic structure, see Supplementary Fig.~\ref{fig:smear_den_3x3_n3}. These findings highlights the importance of strain on the CDW in a twisted bilayer. We note that a previous study by Goodwin and Falko neglected this effect~\cite{goodwin_falko_cdw_moire}.

\section{Conclusions}
In this study, we have investigated the relaxed atomic structure of twisted bilayer \ch{NbSe2} using large-scale ab initio density-functional theory calculations. We observe an interesting interplay of in-plane and out-of-plane relaxations as well as the formation of coexisting charge density waves. In particular, low-energy stacking regions exhibit both triangular and hexagonal charge density waves with a $3\times 3$ periodicity similar to the monolayer. Steric repulsion between Se atoms results in the stacking of hollow-center triangular CDWs on top of filled-center triangular CDWs and vice versa. In the domain walls surrounding the low-energy stacking domains stripe CDWs are formed as a consequence of anisotropic localized strain which suppresses contributions to the CDW from all but one wavevector. Our work sheds light on the complex interplay of atomic displacements driven by stacking energetics and those driven by the formation of charge density waves and also challenges the fundamental assumptions of previous investigations in which the interaction of strain with CDWs was neglected. Our predictions can be tested by scanning tunneling microscopy experiments. Besides CDWs, NbSe$_2$ also exhibits superconductivity at low temperatures. Both the transition temperature and the symmetry of the superconducting state depend on the number of layers and are influenced by the properties of the CDW state~\cite{hamill2021two,yokoya2001fermi,dreher2021proximity,unveil_cdw_sc}. Therefore, the insights from our study provide a starting point for future investigations of superconductivity in twisted bilayers with CDW order. 

\section{Acknowledgements}
We thank Adolfo Fumega for useful discussions. C. T. S. C. acknowledges funding from the Croucher Foundation and an Imperial College President's Scholarship. This work used the ARCHER2 UK National Supercomputing Service
(https://www.archer2.ac.uk) and the Imperial College Research Computing Service facility CX1 (DOI: 10.14469/hpc/2232). We acknowledge the Thomas Young Centre under Grant No. TYC-101.

\section{Methods}

\subsection{Construction of the moir\'e unit cell}
To construct the starting configuration for the atomic relaxations, we stack and twist two flat anti-parallel \ch{NbSe2} monolayers that do not feature a charge density wave. The resulting moir\'e unit cell is spanned by the lattice vectors $\vc{t}_1$ and $\vc{t}_2$ given by 
\begin{align}
    \notag
    \vc{t}_1 &= n\vc{a}_1 + m\vc{a}_2, \\
    \vc{t}_2 &= -m\vc{a}_1 + (n+m)\vc{a}_2,
    \label{eq:moiret}
\end{align}
where $\vc{a}_1=a_0/2(\sqrt{3},1,0)$ and $\vc{a}_2=a_0/2(\sqrt{3},-1,0)$, and $a_0$ is the lattice constant of monolayer \ch{NbSe2} without any charge density wave. The value of the lattice constant is found by relaxing the \ch{NbSe2} unit cell using DFT as described in the next section. We found the equilibrium lattice constant to be 3.44~$\ang$, in good agreement with other values reported in literature \cite{falko_tNbSe2,weak_dim_dep_cdw,elas_cdw_guster}.
The integers $n$, $m$ are the number of unit cells along each moir\'e lattice vector. The twist angle is given by $\cos{\theta}=(n^2+4nm+m^2)/[2(n^2+nm+m^2)]$ \cite{moire_def_dft}. Thus, choosing the pair of integers $(n,m)$ as $(10,11)$ gives the twist angle of $3.14\degree$. The number of atoms in the moiré unit cell is $N=6(n^2+nm+m^2)$, which is 1,986 at the chosen twist angle. 

\subsection{Computational parameters}
To determine the relaxed atomic structure of twisted bilayer \ch{NbSe2}, we use ab initio density-functional theory as implemented in the SIESTA code \cite{Soler_Artacho_Gale_García_Junquera_Ordejón_Sánchez-Portal_2002}. To capture van der Waals interactions, the exchange-correlation functional developed by Cooper is used \cite{cooper_xcf}. We use Troullier-Martins pseudopotentials \cite{TMPP}, a double-$\zeta$ polarized basis and $\Gamma$-point sampling of the first Brillouin zone. In the self-consistent cycle, the Hamiltonian and the density matrix are converged to 1~$\mu$eV and $1\times 10^{-7}$ per atom, respectively. The force tolerance is set to $5\times 10^{-2}~\mathrm{eV}/\ang$. 

\subsection{Smeared Nb atomic density}
The smeared Nb atomic density for a given layer is the sum of Gaussian densities associated with each Nb atomic site, and is given by
\begin{equation}
    \rho(\vc{r}) = \mathlarger{{\sum_{i=1}^{N_\textrm{Nb}}}}e^{-|\vc{r}-
\vc{R}_i|^2/(2\sigma^2)},
\label{eq:smearden}
\end{equation}
where $N_\textrm{Nb}$ is the number of Nb atoms in the layer, $\vc{R}_i$ is the relaxed atomic position of the $i$-th Nb atom, and $\sigma$ is a parameter that controls the width of the Gaussian. Supplementary Fig.~\ref{fig:dem_cdw_smear} shows the smeared Nb density for different values of $\sigma$ for a \ch{NbSe2} monolayer. For very small values of $\sigma$ such as 0.2~$a_0$ (see Supplementary Fig.~\ref{fig:dem_cdw_smear}(a)), it is difficult to see whether a CDW is present or not. However, if $\sigma$ is increased to 0.45~$a_0$ (see Supplementary Fig.~\ref{fig:dem_cdw_smear}(b)), the $3 \times 3$ modulation becomes evident. If $\sigma$ is further increased to 0.9~$a_0$ (see Supplementary Fig.~\ref{fig:dem_cdw_smear}(c)), the atomic resolution is lost, but the $3\times 3$ modulation is still visible. We use $\sigma=0.45 a_0$ in this paper as this choice both provides atomic resolution and enables easy identification of charge density waves. 

\subsection{Order parameter for charge density wave motifs}
As described in the main text, the identification of CDW motifs based on the Nb displacement vectors relative to the initial structure is challenging in a twisted bilayer NbSe$_2$. In contrast to the monolayer where the Nb displacement vectors relative to the high-symmetry structure without CDW can be used to assess the presence of a CDW, the displacement vectors in  twisted bilayer are dominated by the movement of the atoms to avoid high-energy stacking arrangements. 

To determine the smaller local displacements of the Nb atoms due the formation of a CDW, we note that in the high-symmetry phase without CDW the position of a Nb atom coincides with the centroid of its six neighboring Nb atoms, i.e. the position of a Nb atom is equal to the average of the positions of its nearest Nb neighbors. When a CDW is formed, the Nb atom position no longer coincides with the centroid of the nearest Nb neighbors. 

To assess the presence of a CDW in a twisted bilayer, we therefore define the local displacement  $\mathbf{u}_i$ of the $i$-th Nb atom according
\begin{equation}
    \mathbf{u}_i=\mathbf{R}_i-\frac{1}{6}\sum_{l=1}^{6}\mathbf{R}_l ,
    \label{eq:cdwdispl}
\end{equation}
where $\mathbf{R}_i$ is the relaxed position of the $i$-th Nb atom and $\mathbf{R}_l$ are the relaxed positions of its six nearest Nb neighbors. The second term on the right hand side is the centroid of the nearest Nb neighbors. The local displacements of the Nb atoms are shown in Supplementary Figure~\ref{fig:dv_cdw_moire}.

To assess the presence of a specific CDW motif (labeled by the index $\alpha$) in the twisted bilayer, we search through all possible motif centers (labeled by the index $i$) and project the local displacements of the neighboring Nb atoms (labeled by the index $l$) onto a reference displacement pattern $\mathbf{u}_l^{\text{ref}(\alpha)}$, shown in Fig.~\ref{fig:cdw_motif_dem}(a)-(d). For the filled-center CDWs, the motif centers are the Se atoms. For the hollow-center CDWs, the motif centers are the interstitial sites. For the hexagonal and stripe CDWs, the motif centers are the Nb atoms. The local order parameter for $i$-th motif center is then obtained as 
\begin{equation}
    p^\alpha_i = \frac{1}{N_{\mathrm{at}}}\sum_{l=1}^{N_{\mathrm{at}}}p_{il}^\alpha,
    \label{eq:cdw_op}
\end{equation}
where $N_{\mathrm{at}}$ denotes the number of Nb atoms near the motif center whose displacements are taken into account, and $p_{il}^\alpha$ is the contribution to the total order parameter from $l$-th Nb atom given by
\begin{align}
    p_{il}^\alpha &= \exp{\left(-\frac{\arccos\left({\hat{\bm{u}}_l^{\text{ref}(\alpha)}\cdot \hat{\bm{u}}_l}\right)^2}{2\gamma^2}\right)},
    \label{eq:cdw_op_at_proj}
\end{align}
where $\gamma$ is an smearing parameter. Here, we use a value of $\pi/8$ for this parameter. Note that $1 \leq p_i^\alpha \leq 0$. We have verified that $p_i^\alpha$ is close to unity if the order parameter is calculated for a monolayer with a uniform CDW.



In a twisted bilayer, we consider a CDW motif of type $\alpha$ to be present at the $i$-th motif center, if $p_i^\alpha$ exceeds a critical value $p^\alpha_{\mathrm{c}}$.
For the filled-center (hollow-center) CDW motifs, for each Se atom (interstitial site), separate order parameters are calculated for the smaller three-atom triangles ($N_{\mathrm{at}}=3$) and the larger six-atom triangles ($N_{\mathrm{at}}=6$). If $p_i^\alpha$ for the three-atom triangle exceeds $p^\alpha_\mathrm{c}=2/3$ (i.e. more than two of the three atoms in the triangle follow the reference displacement pattern) and $p_i^\alpha$ for the six-atom triangle is smaller than $p^\alpha_\mathrm{c}=4/6$ (i.e. fewer than four of the six atoms in the triangle follow the reference displacement pattern), the three-atom motif is assumed to be present. If $p_i^\alpha$ for both the three- and the six-atom triangle exceed $p^\alpha_\mathrm{c}=2/3$, a six-atom triangle is assumed to be present. 

For hexagonal CDW motifs, the order parameter is calculated for each Nb atom so the contributions from all neighboring Nb atoms are taken into account ($N_{\mathrm{at}}=6$). We use $p^\alpha_\mathrm{c}=1/2$ (i.e. more than three of the six Nb atoms follow the reference displacement pattern). 

For stripe CDW motifs along a wave vector $\bm{q}_n$, we consider pairs of Nb atoms ($N_{\mathrm{at}}=2$) whose distance vector is parallel to $\bm{q}_n$. If at least three neighboring pairs of Nb atoms have an order parameter exceeding $p^\alpha_\mathrm{c}=1/2$ (i.e. more than one out of the two atoms follow the reference displacement pattern), we consider a stripe CDW to be present. 

This procedure results in a scalar field of order parameters associated with the Se atoms for the filled-center CDW, interstitial sites for the hollow-center CDW, Nb atoms for the hexagonal, and mid-points between the next-nearest Nb atoms stripe CDWs. The scalar fields obtained from calculating the order parameter of a chosen type of CDW on a monolayer with the same type of CDW are shown in Supplementary Fig.~\ref{fig:cdw_on_same_cdw}. We have also checked that applying the order parameter calculation of a chosen type of CDW (e.g. hexagonal CDW) on a monolayer with a different type of CDW (e.g. hollow-center CDW) does not lead to any spurious identification of the chosen type of CDW. In other words, the order parameter calculation identifies a CDW motif if and only if that same type of CDW is present.  


\subsection{Determining the strain}
To estimate the local strain at the $i$-th Nb atom due to moiré relaxations, we partition the total displacement of the Nb atoms from their high-symmetry positions to their relaxed positions into contributions from moiré displacements and local displacements. The moiré displacements are then given by 

\begin{equation}
    \mathbf{u}_i^\text{moir}= \Delta\mathbf{R}_i-  \mathbf{u}_i,
    \label{eq:mdispl}
\end{equation}
with $\Delta \mathbf{R}_i$ denoting the total displacement from the high-symmetry position. The moir\'e displacements are shown in Supplementary Figure~\ref{fig:cdw_on_same_cdw}. 

The local strain along the direction $\hat{\mathbf{q}}_n$ at the $i$-th Nb atom to moiré displacement is then estimated as
\begin{equation}
    \bm{\varepsilon}_{i}^{(n)}= \left(\frac{ \mathbf{u}^\text{moir}_{i-1,n}+\mathbf{u}^\text{moir}_{i+1,n}}{\sqrt{3}a_0}\right)\cdot\hat{\bm{q}}_n, 
    \label{eq:strain}
\end{equation}
where $\sqrt{3}a_0$ denotes the distance between the two Nb atoms in the initial configuration and $\mathbf{u}^\text{moir}_{i \pm 1,n}$ denote the moir\'e displacements of the two Nb nearest neighbors of atom $i$ along the direction of $\hat{\mathbf{q}}_n$. 

\clearpage
\onecolumngrid
\section{Supplementary Information}
\setcounter{figure}{0}
\setcounter{page}{1}
\renewcommand{\figurename}{SUPPLEMENTARY FIG.}
\renewcommand{\thefigure}{S\arabic{figure}}

\begin{figure}[htb!]
    \centering
    \begin{subfigure}[t]{0.49\textwidth}
        \centering
        \caption{}
        \includegraphics[width=\textwidth]{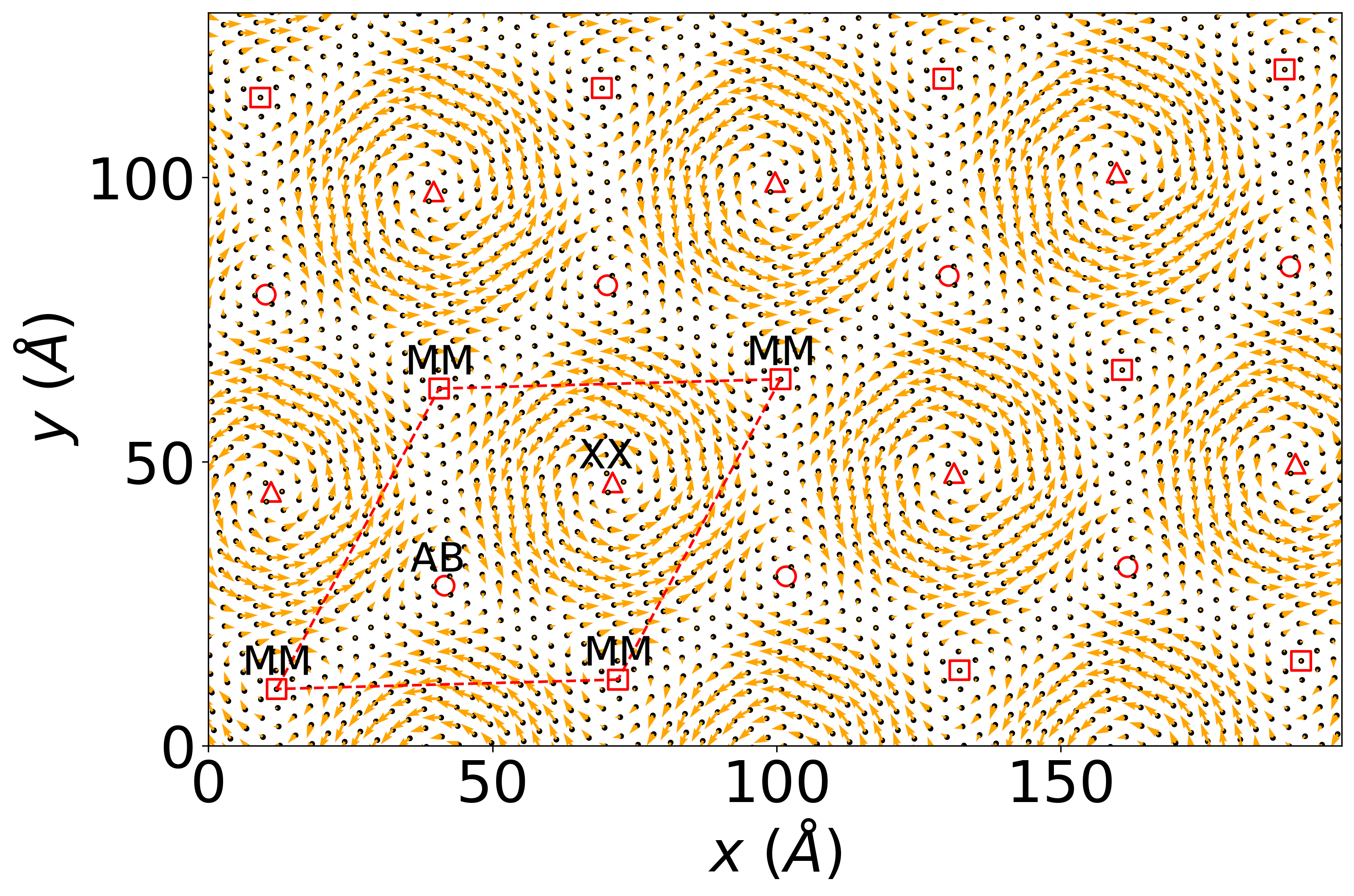}
    \end{subfigure}
    ~
    \begin{subfigure}[t]{0.49\textwidth}
        \centering
        \caption{}
        \includegraphics[width=\textwidth]{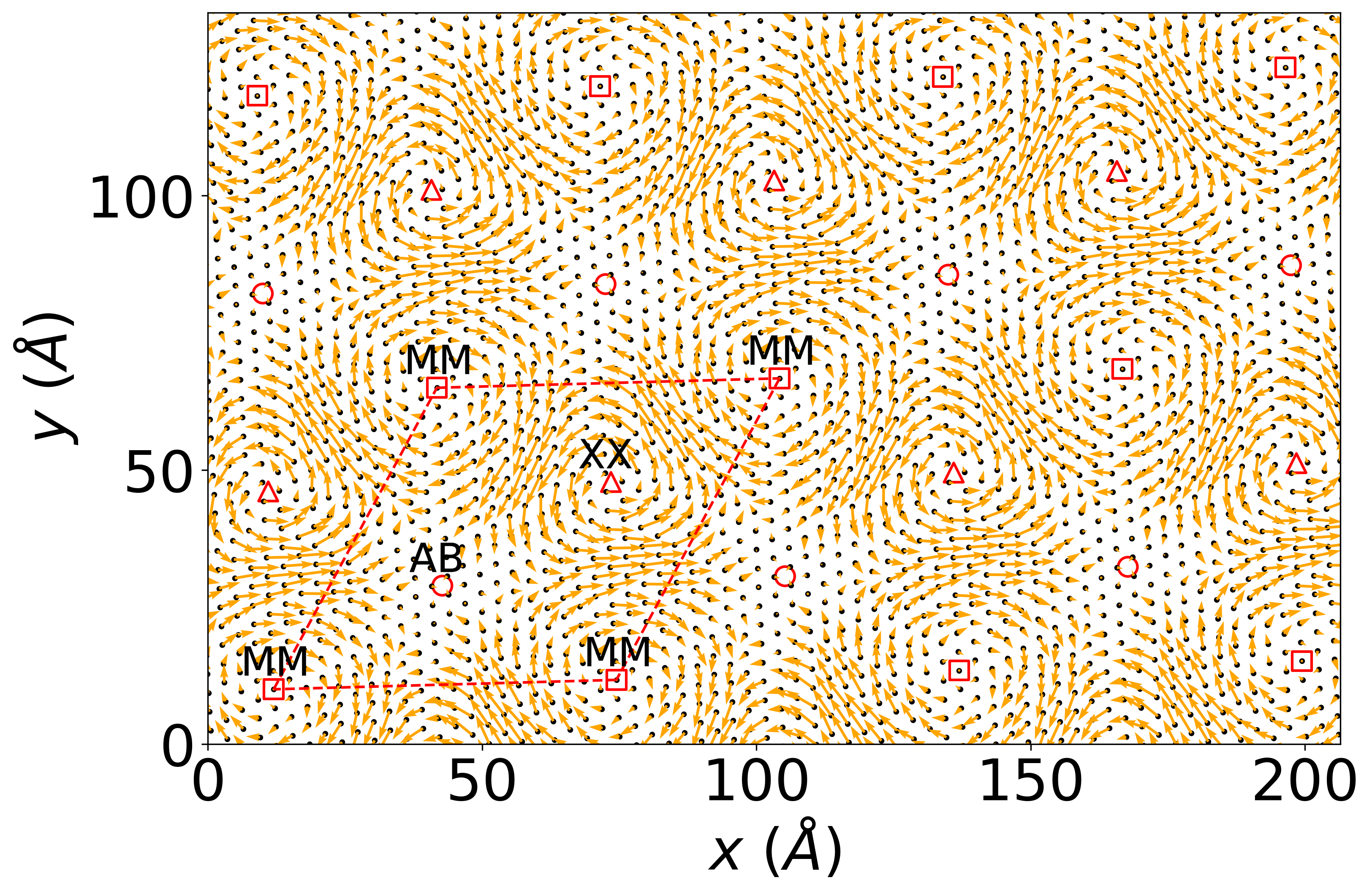}
    \end{subfigure}
    \caption[justification=raggedright]{In-plane atomic relaxations of twisted bilayers \ch{MoSe2} and \ch{NbSe2}. In-plane displacements of metal atoms relative to unrelaxed initial structure in the top layer for (c) twisted bilayer \ch{MoSe2} and (d) twisted bilayer \ch{NbSe2}. The red circles denote the centers of the AB stacking regions (metal on top of chalcogen and vice versa), the red squares denote the centers of the MM stacking region (metal on top of metal), and the red triangles denote the centers of the XX stacking region (chalcogen on top of chalcogen).}
    \label{fig:displ_top}
\end{figure}

\begin{figure}[htb!]
    \centering
    \includegraphics[width=0.5\textwidth]{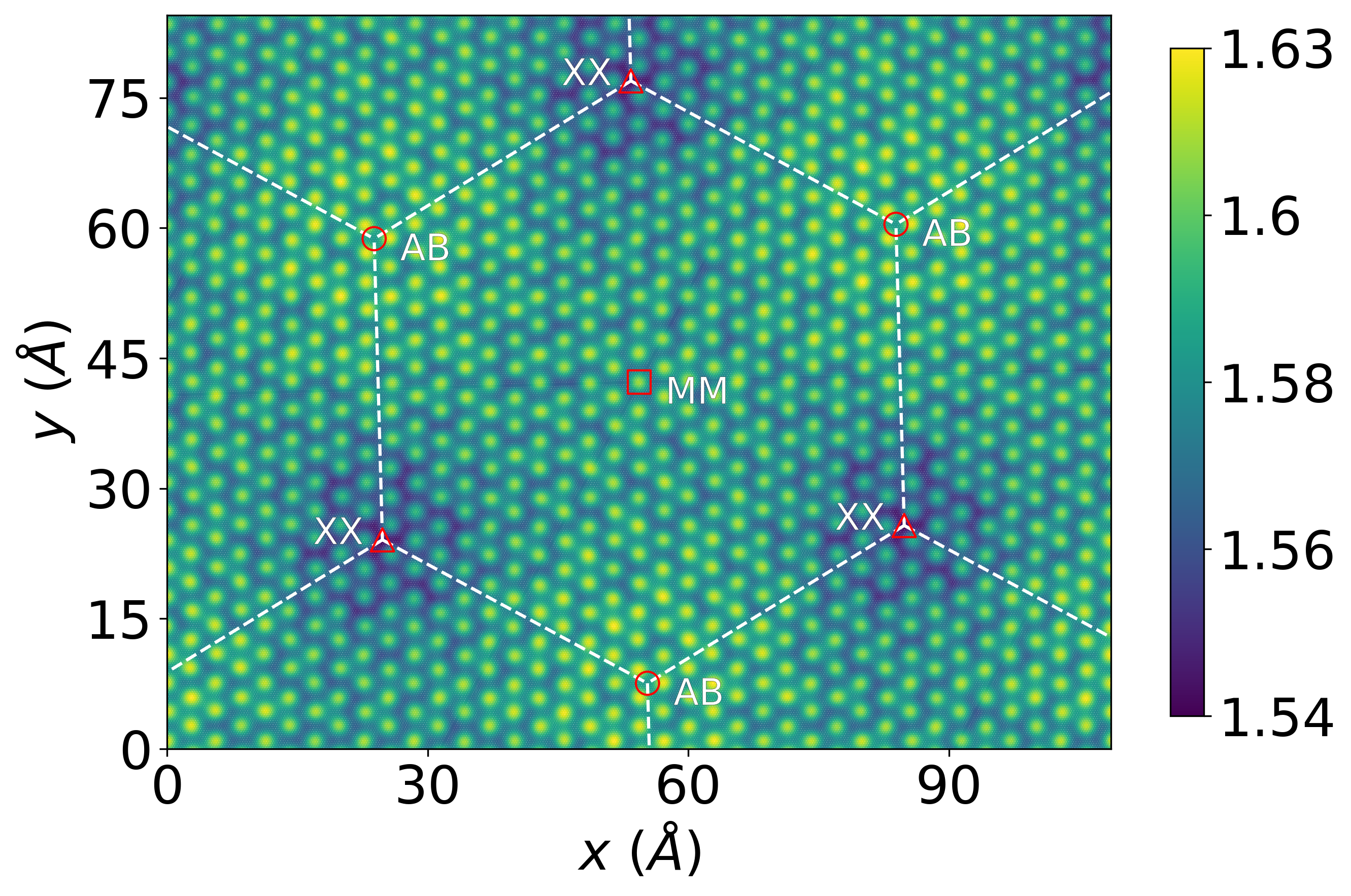}
    \caption{The smeared Mo atomic density in the bottom layer of relaxed twisted \ch{MoSe2/MoSe2}. The red cirles, triangles, and squares mark the centers of the XX, AB and MM stacking regions respectively. White dashed lines join the AB and XX centers for visual aid.}
    \label{fig:smear_tmose2}
\end{figure}

\begin{figure*}[htb!]
    \begin{subfigure}[t]{0.51\textwidth}
        \centering
        \caption{}
        \includegraphics[width=\textwidth]{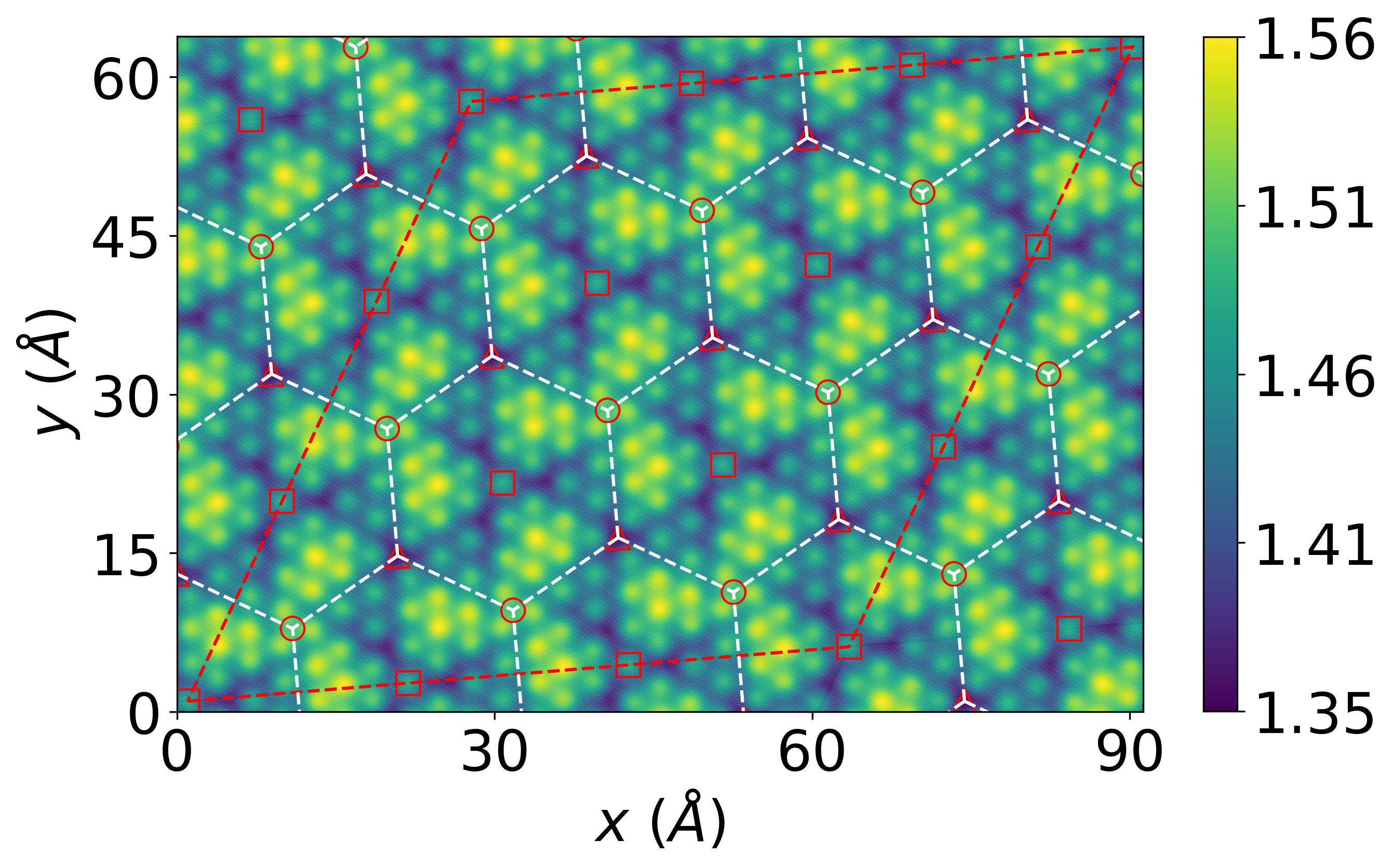}
        \label{fig:smear_tnbse2_bot_3x3}
    \end{subfigure}
    ~
    \begin{subfigure}[t]{0.432\textwidth}
        \vspace{1pt}
        \centering
        \caption{}
        \includegraphics[width=\textwidth]{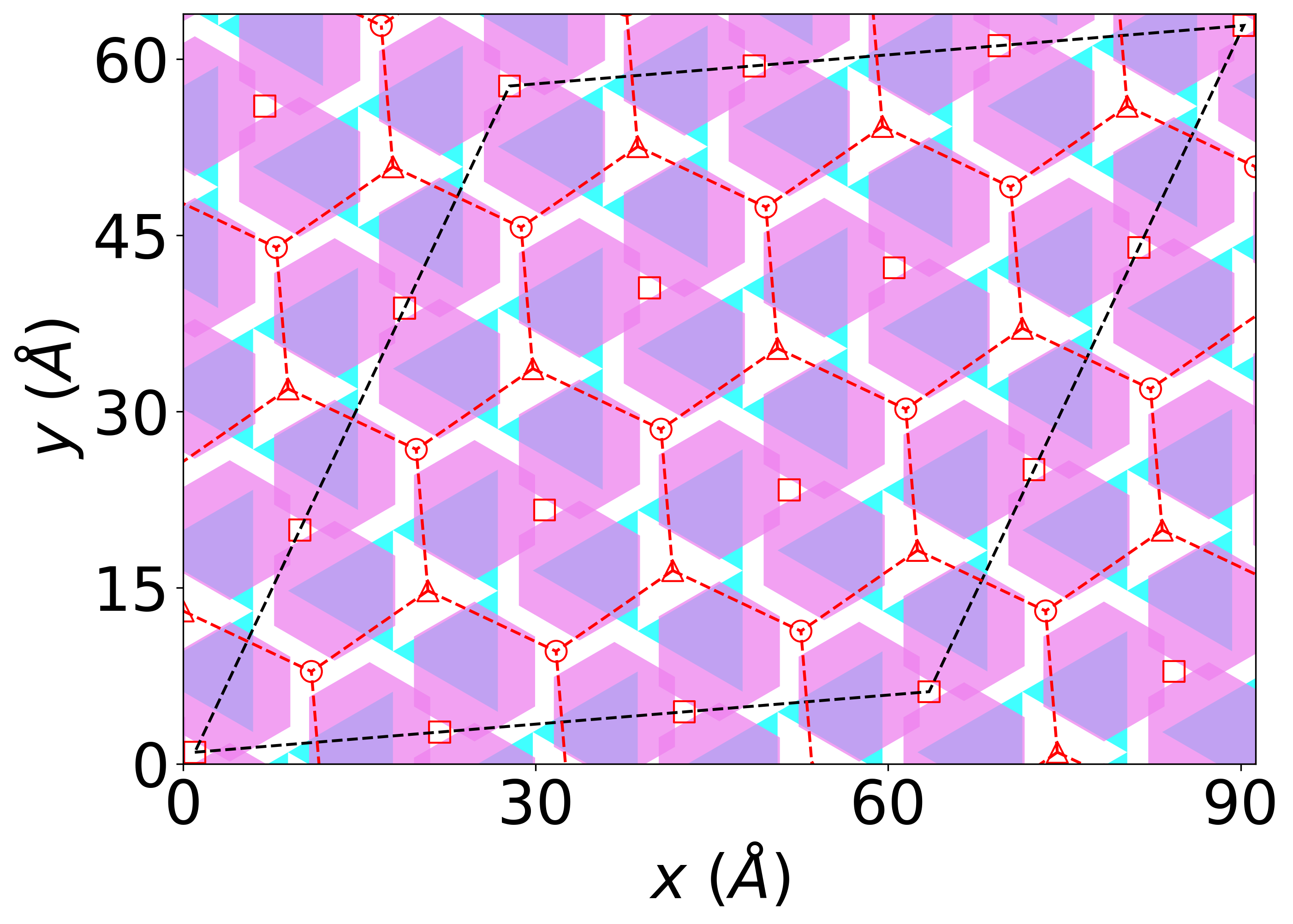}
        \label{fig:smear_tnbse2_label_bot_3x3}
    \end{subfigure}
    ~
    \begin{subfigure}[t]{0.51\textwidth}
        \centering
        \caption{}
        \includegraphics[width=\textwidth]{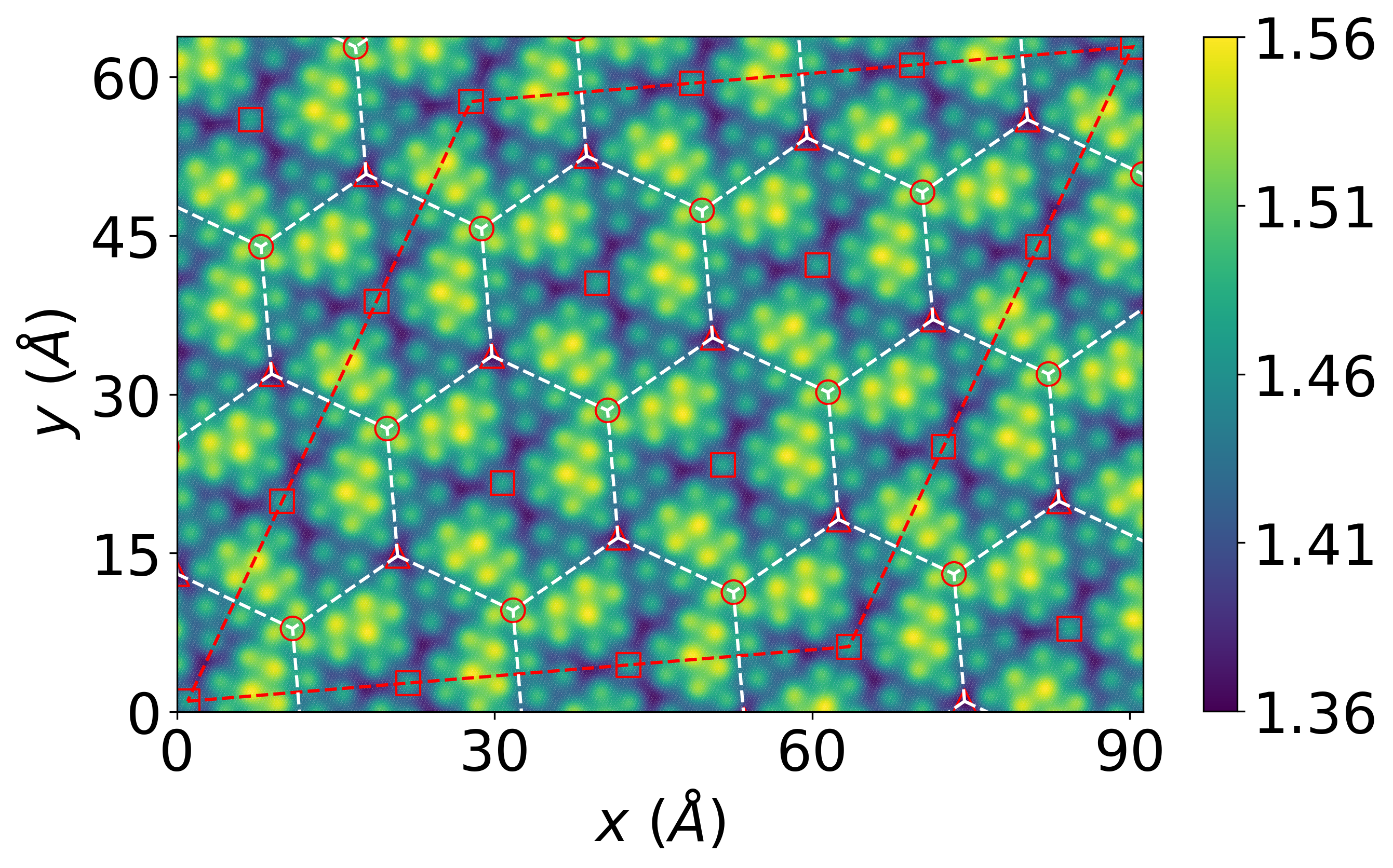}
        \label{fig:smear_tnbse2_top3x3}
    \end{subfigure}
    ~
    \begin{subfigure}[t]{0.432\textwidth}
        \vspace{1pt}
        \centering
        \caption{}
        \includegraphics[width=\textwidth]{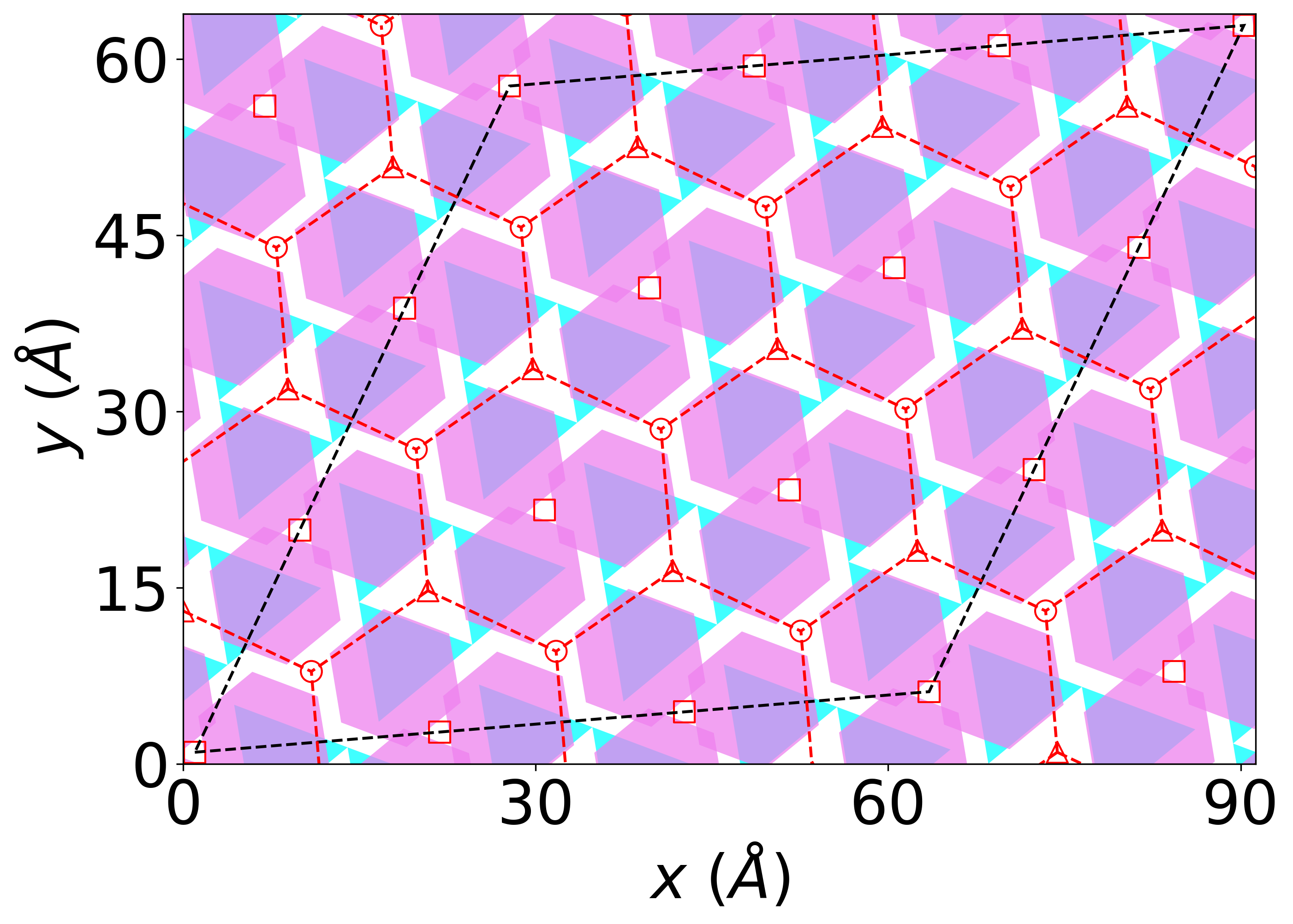}
        \label{fig:smear_tnbse2_label_top_3x3}
    \end{subfigure}
    \caption{Charge density wave in twisted bilayer \ch{NbSe2} at a twist angle of 9.43$^\circ$ from a supercell calculation. Smeared Nb atomic density of (a) the bottom layer and (c) the top layer. The red parallelogram represents the $3 \times 3$ supercell of moir\'e unit cells used in the calculation. The white dashed lines join the AB and XX centers for visual aid. Motifs corresponding to filled-center triangular (hexagonal) CDWs are shown as cyan (purple) symbols for (b) the bottom layer and (d) the top layer. The black parallelogram represents the $3 \times 3$ supercell of moir\'e unit cells used in the calculation. The red circles denote the centers of the AB stacking regions (metal on top of chalcogen and vice versa), the red squares denote the centers of the MM stacking regions (metal on top of metal), and the red triangles denote the centers of the XX stacking regions (chalcogen on top of chalcogen). The red dashed lines join the AB and XX centers for visual aid.}
    \label{fig:smear_den_3x3_n3}
\end{figure*}

\begin{figure}[htb!]
    \centering
    \begin{subfigure}[t]{0.5\textwidth}
        \centering
        \caption{}
        \includegraphics[width=\textwidth]{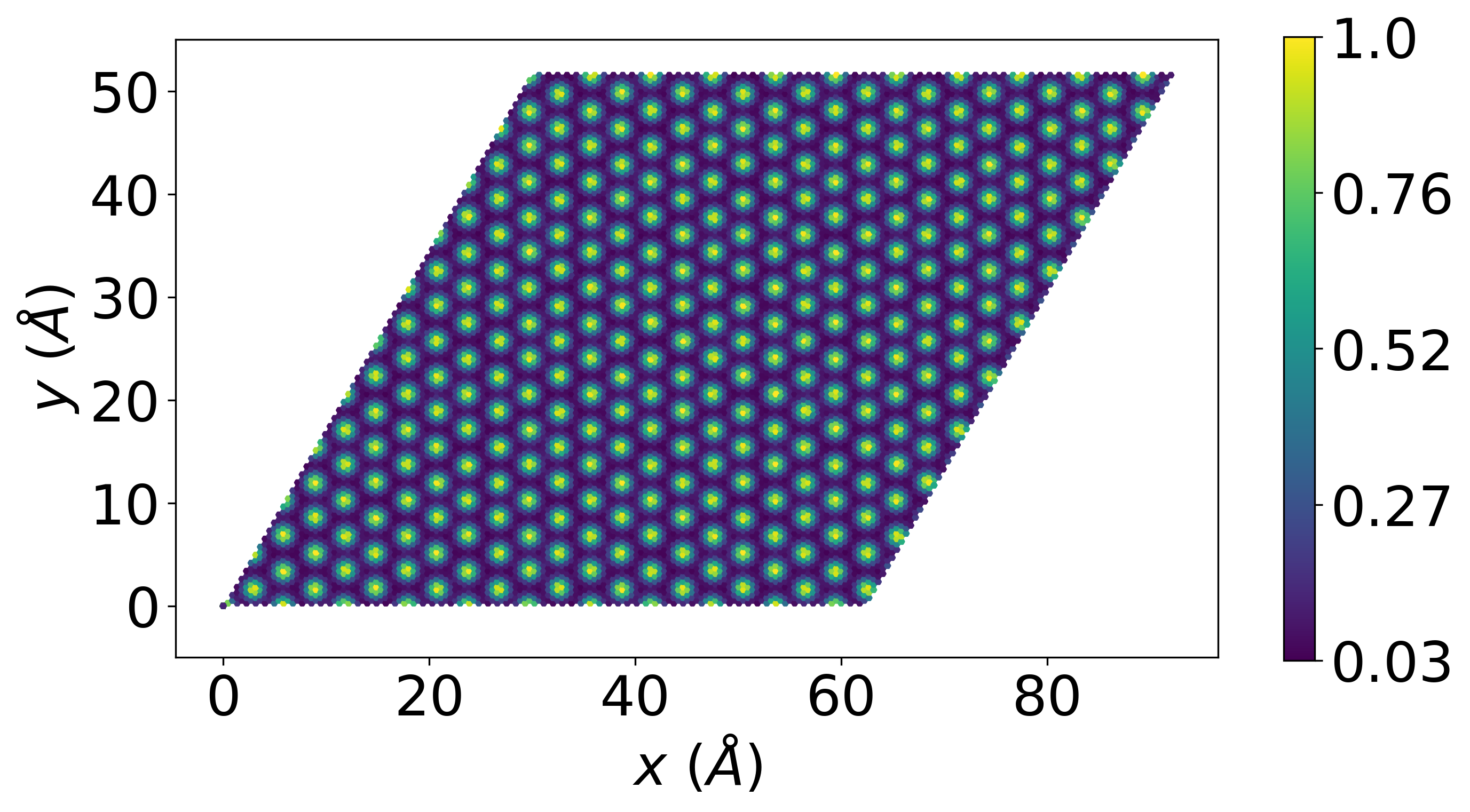}
    \end{subfigure}
    ~
    \begin{subfigure}[t]{0.5\textwidth}
        \centering
        \caption{}
        \includegraphics[width=\textwidth]{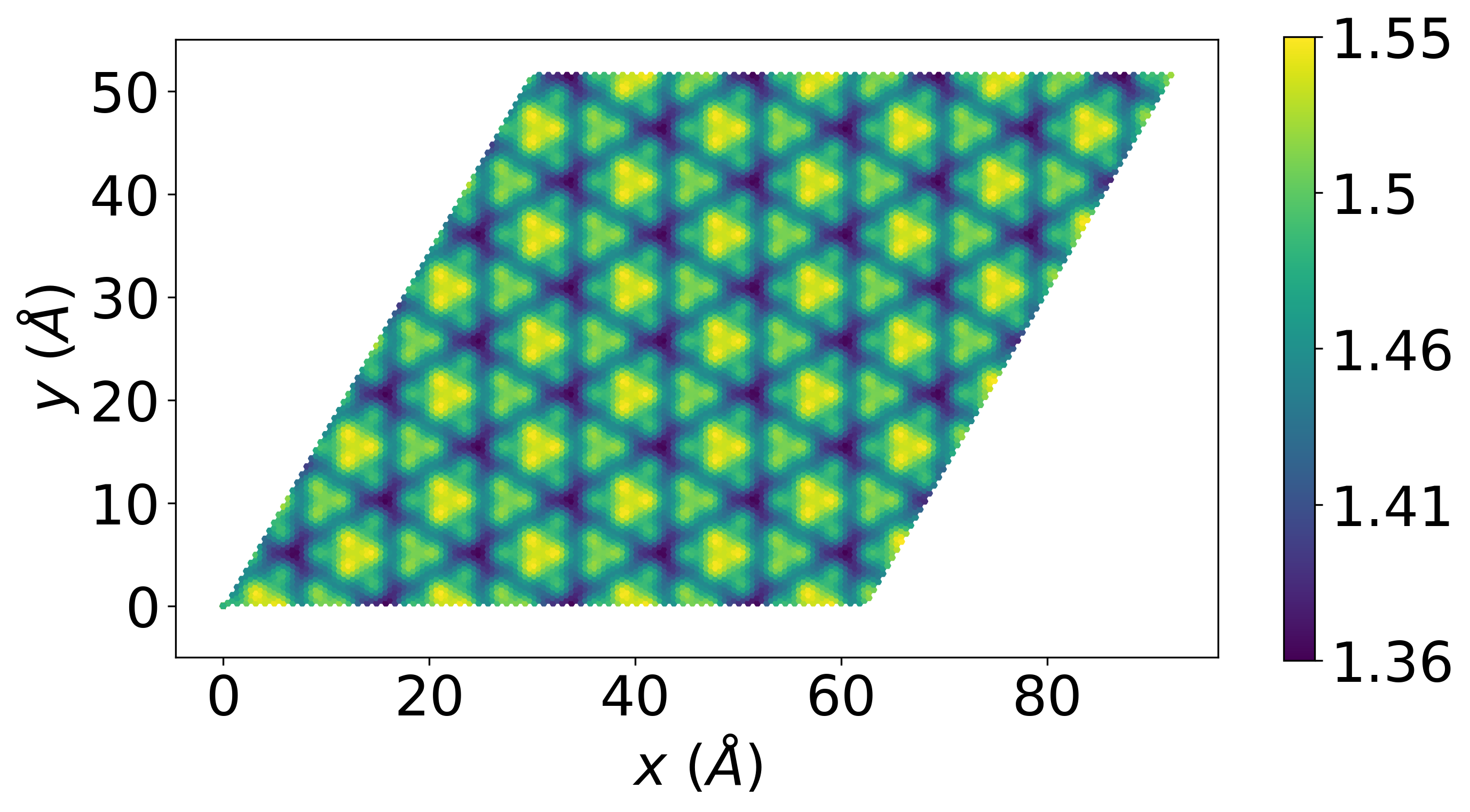}
    \end{subfigure}
    ~
    \begin{subfigure}[t]{0.5\textwidth}
        \centering
        \caption{}
        \includegraphics[width=\textwidth]{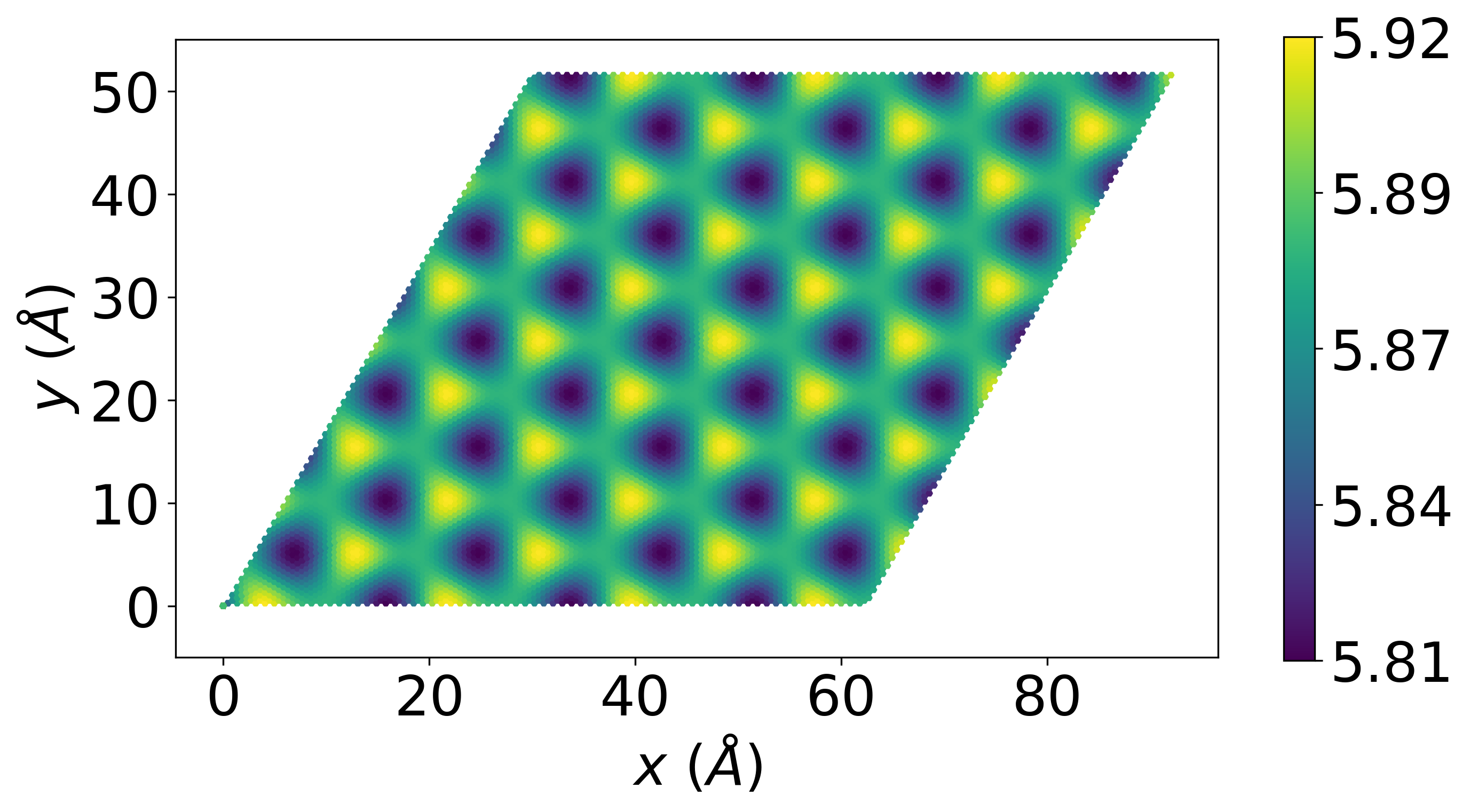}
    \end{subfigure}
    \caption[justification=raggedright]{The smeared Nb atomic densities for a monolayer with a $3 \times 3$ hollow-center charge density wave of amplitude 0.07 $\mathrm{\AA}$ for different values of the smearing parameter $\sigma$: (a): 0.2 $a_0$, (b): 0.45 $a_0$, and (c) 0.9 $a_0$ with $a_0 = 3.44 \ang$.}
    \label{fig:dem_cdw_smear}
\end{figure}

\begin{figure*}[htb!]
    \centering
    \begin{subfigure}[t]{0.49\textwidth}
        \centering
        \caption{}
        \includegraphics[width=\textwidth]{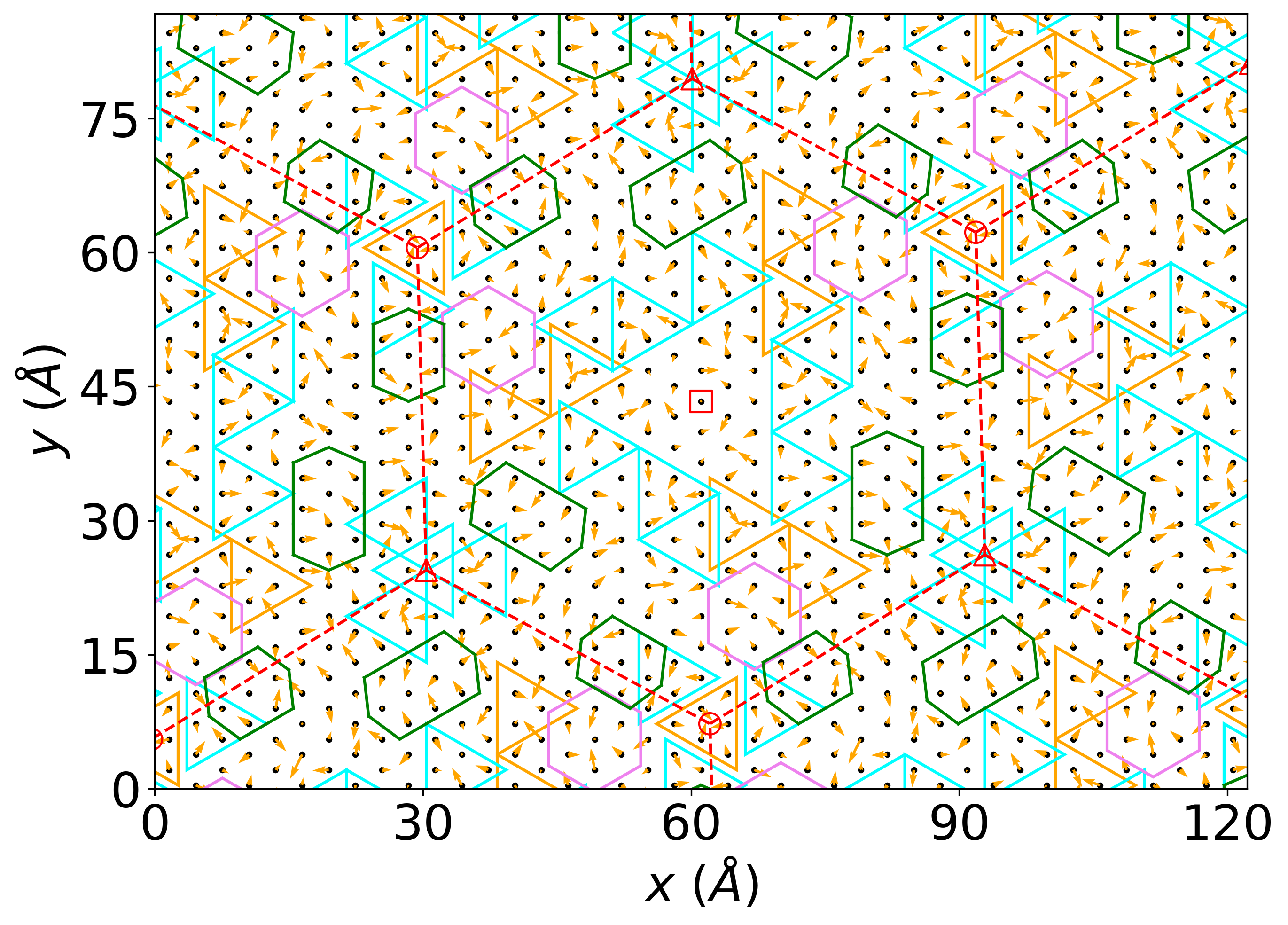}
        \label{fig:cdwvbot}
    \end{subfigure}
    ~
    \begin{subfigure}[t]{0.49\textwidth}
        \centering
        \caption{}
        \includegraphics[width=\textwidth]{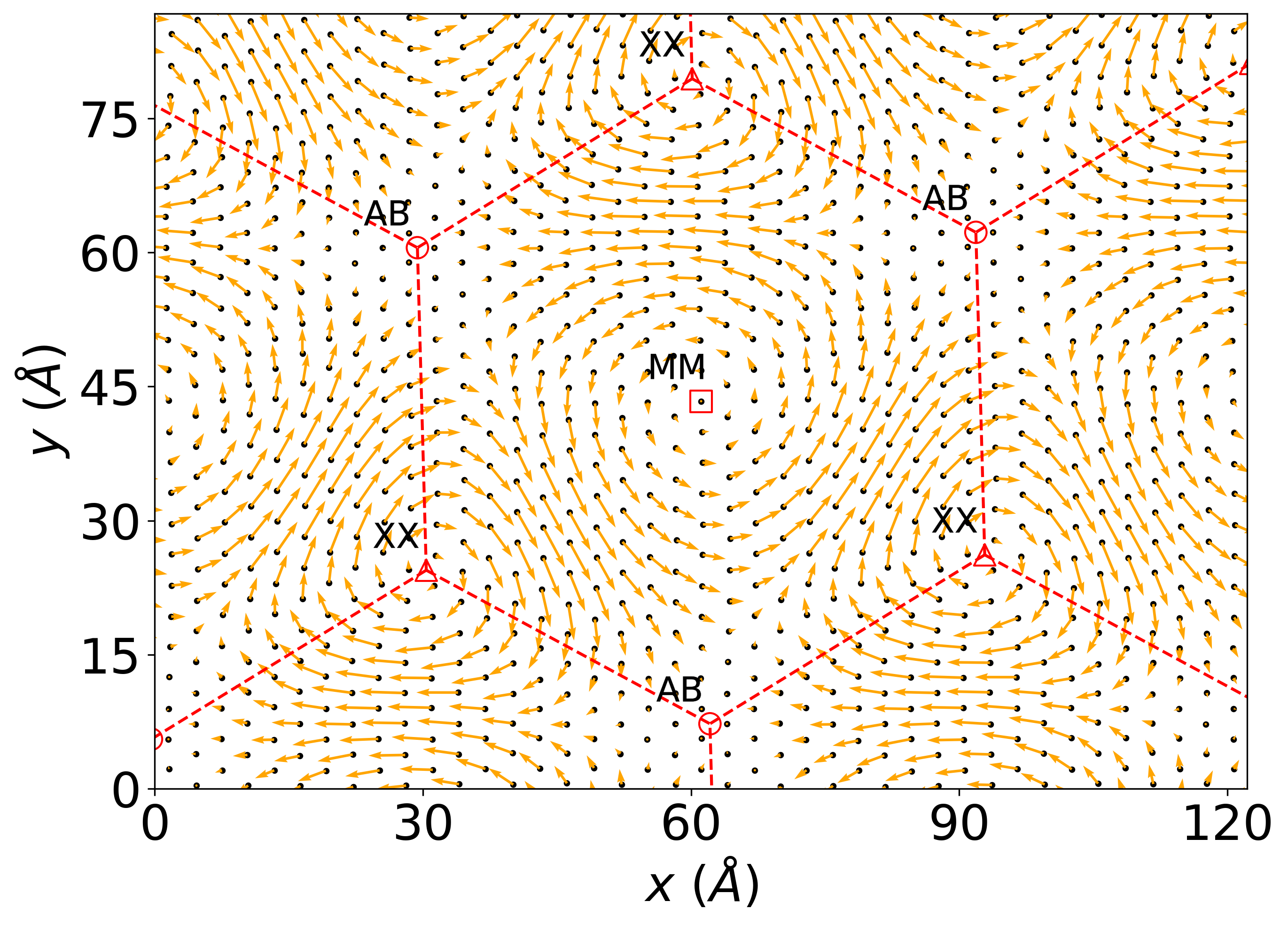}
        \label{fig:moirevbot}
    \end{subfigure}
    ~
    \begin{subfigure}[t]{0.49\textwidth}
        \centering
        \caption{}
        \includegraphics[width=\textwidth]{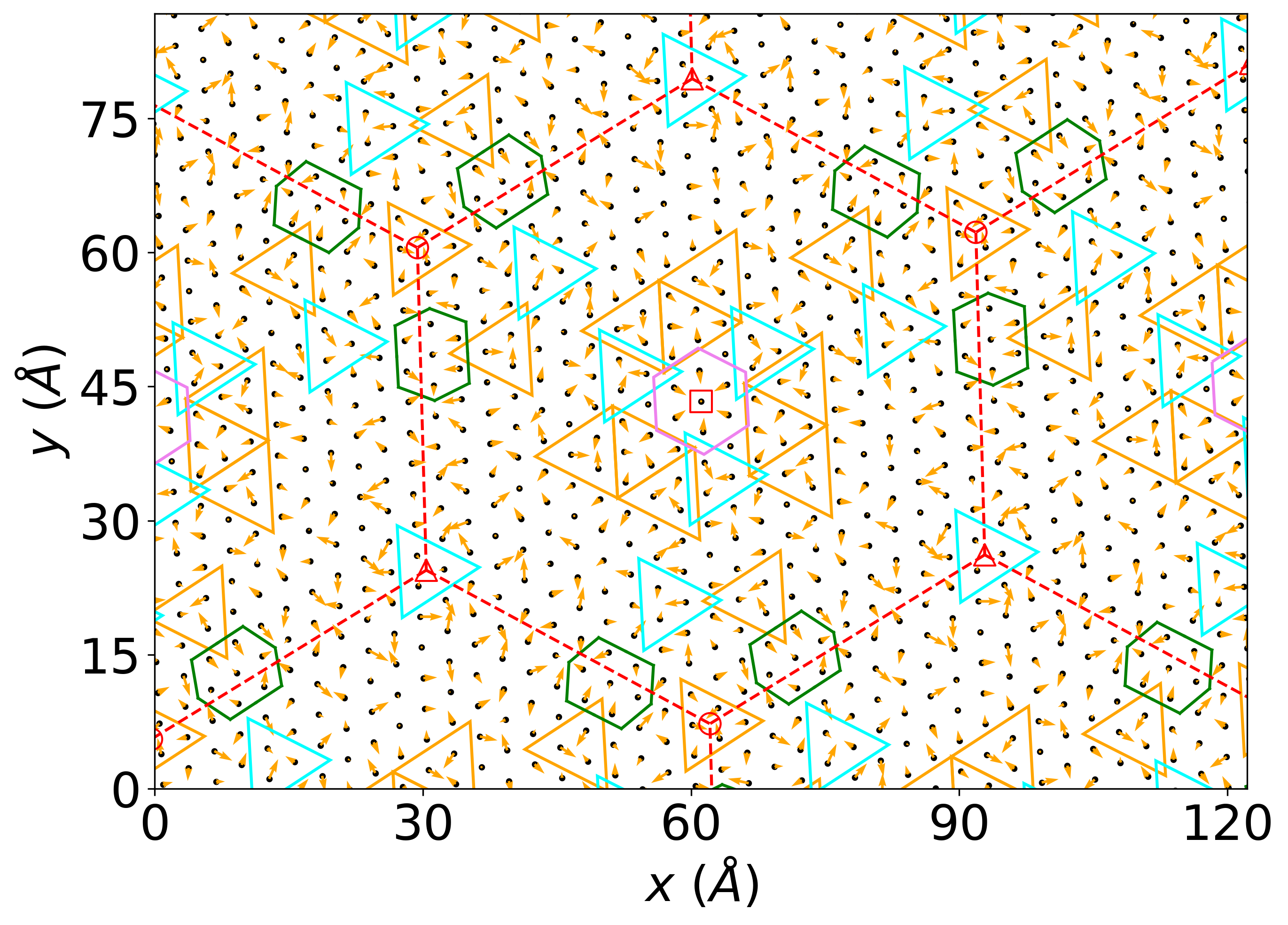}
        \label{fig:cdwvtop}
    \end{subfigure}
    ~
    \begin{subfigure}[t]{0.49\textwidth}
        \centering
        \caption{}
        \includegraphics[width=\textwidth]{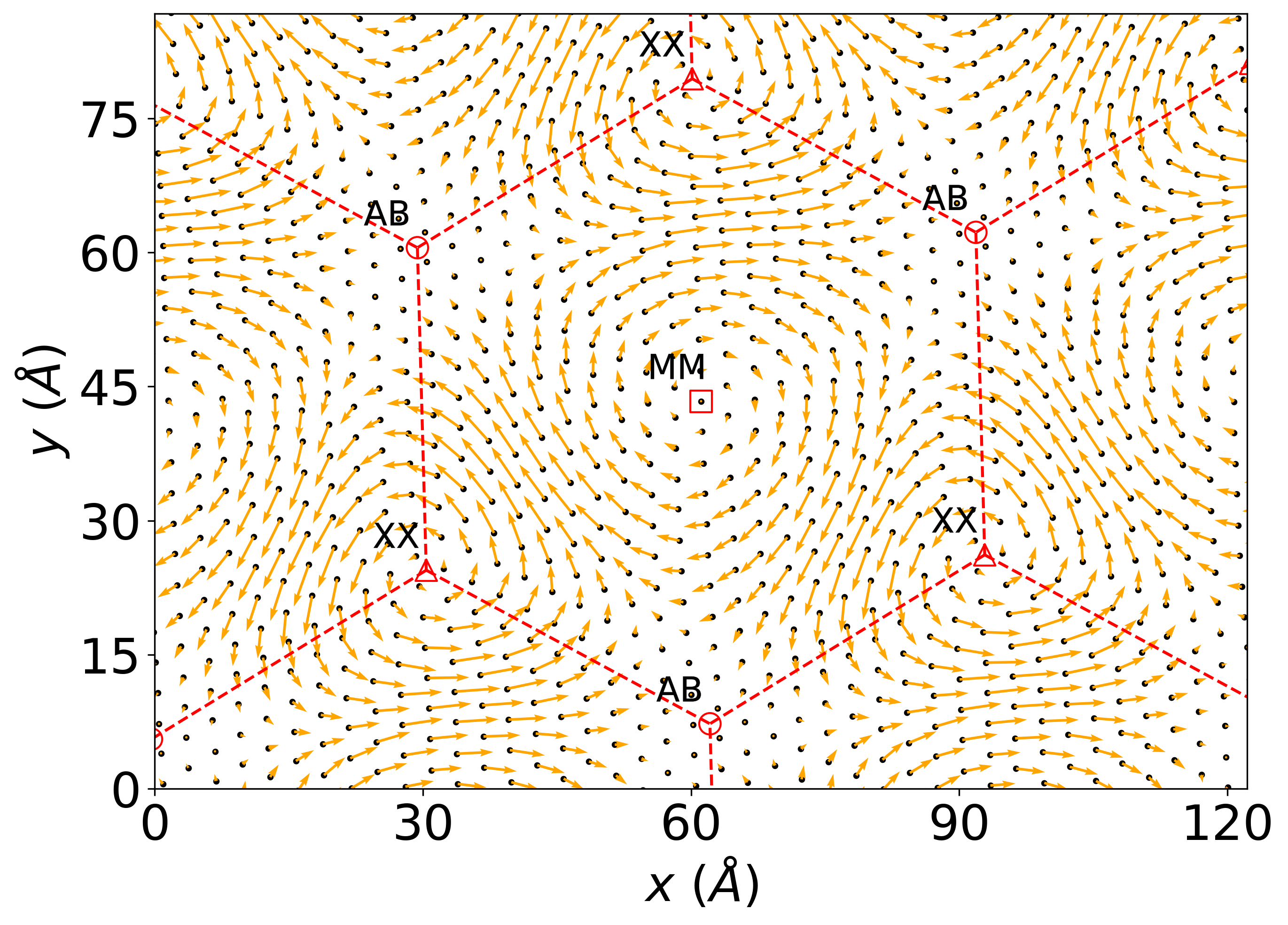}
        \label{fig:moirevtop}
    \end{subfigure}
    \caption{The charge density wave displacements for the bottom layer (a) and the top layer (c). The moir\'e displacements obtained using for the bottom (b) and the top layer (d). Purple hexagons, cyan triangles, orange trianges, green strips incides hexagonal CDW motifs, filled-center CDW motifs, hollow-center CDW motifs and stripe CDW motifs, respectively. The red circles denote the centers of the AB stacking regions (metal on top of chalcogen and vice versa), the red squares denote the centers of the MM stacking regions (metal on top of metal), and the red triangles denote the centers of the XX stacking regions (chalcogen on top of chalcogen). The red dashed lines join the AB and XX centers for visual aid.}
    \label{fig:dv_cdw_moire}
\end{figure*}

\begin{figure*}[htb!]
    \centering
    \begin{subfigure}[t]{0.35\textwidth}
        \centering
        \caption{\hspace{3pt}3-atom filled-center CDW }
        \includegraphics[width=\textwidth]{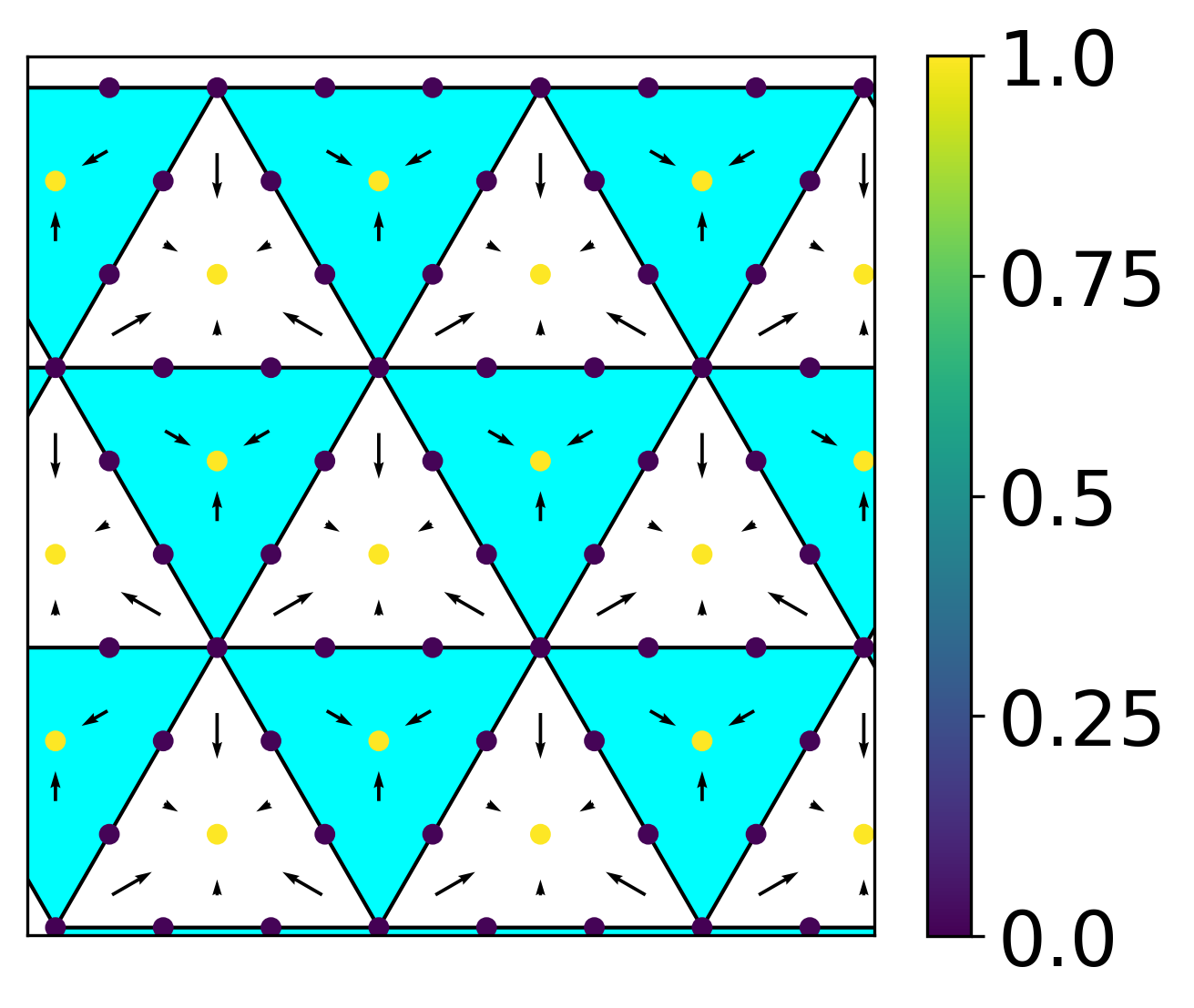}
    \end{subfigure}
    ~
    \begin{subfigure}[t]{0.35\textwidth}
        \centering
        \caption{\hspace{3pt}6-atom filled-center CDW }
        \includegraphics[width=\textwidth]{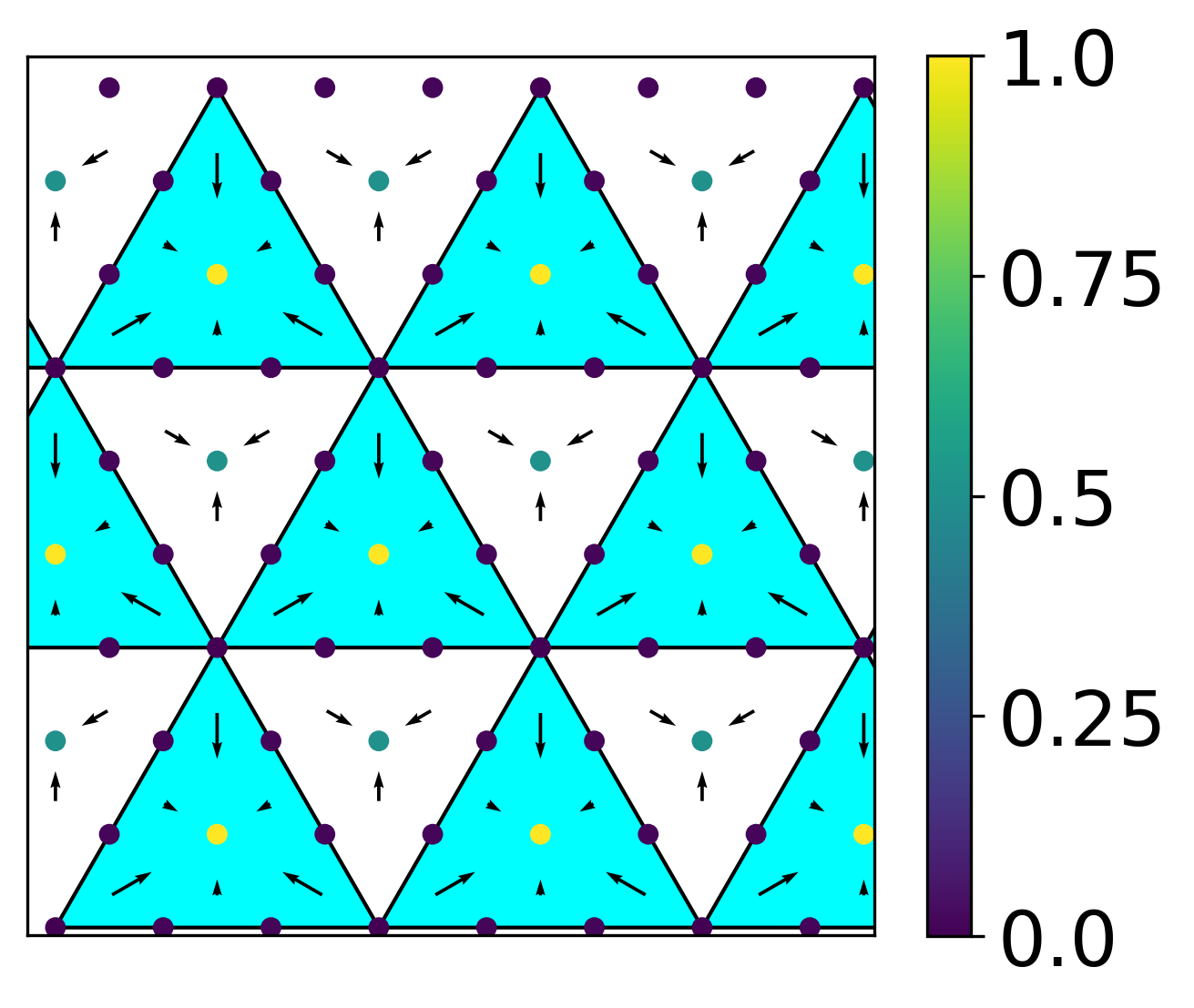}
    \end{subfigure}
    ~
    \begin{subfigure}[t]{0.35\textwidth}
        \centering
        \caption{\hspace{3pt}3-atom hollow-center CDW }
        \includegraphics[width=\textwidth]{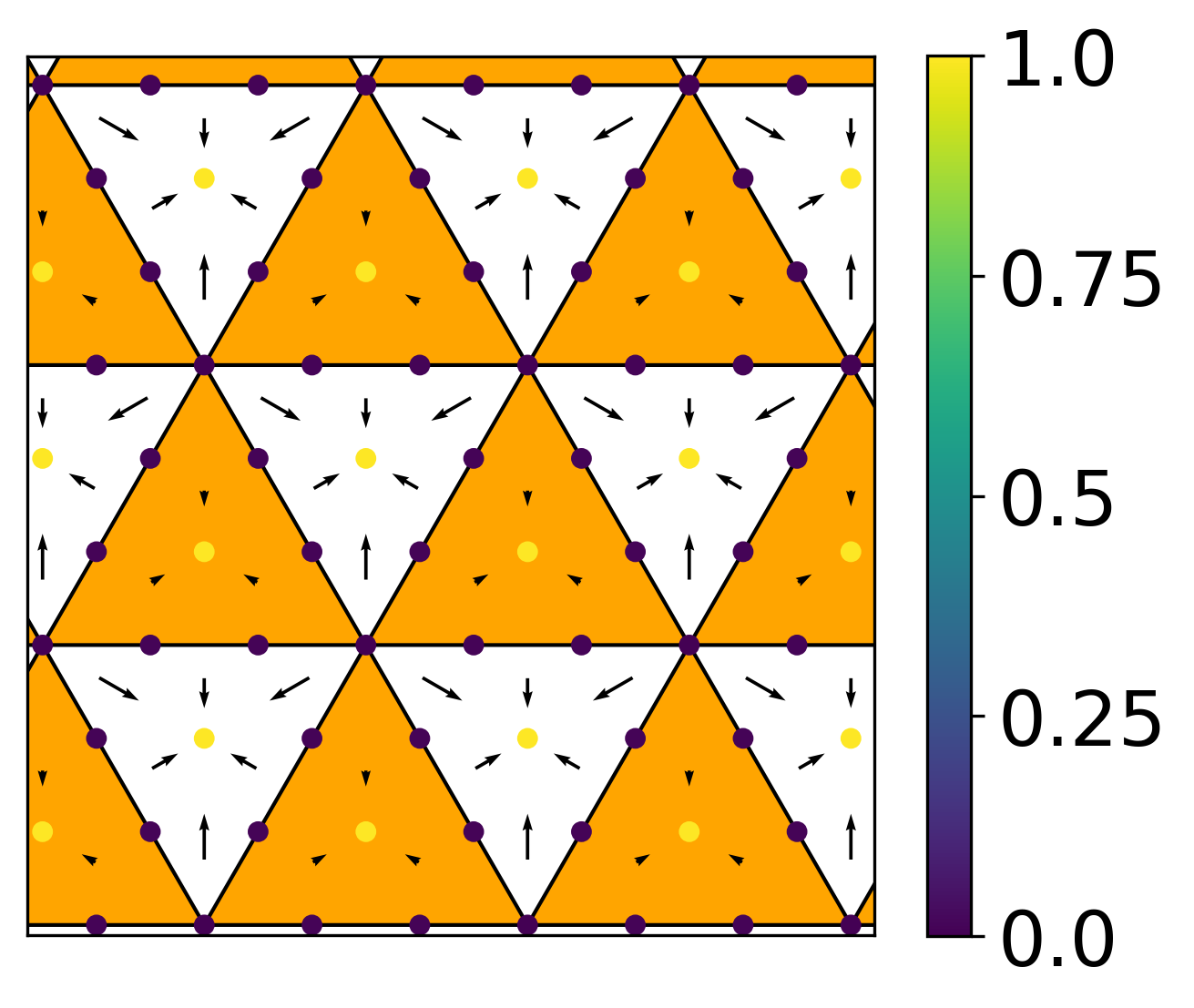}
    \end{subfigure}
    ~
    \begin{subfigure}[t]{0.35\textwidth}
        \centering
        \caption{\hspace{3pt}6-atom hollow-center CDW }
        \includegraphics[width=\textwidth]{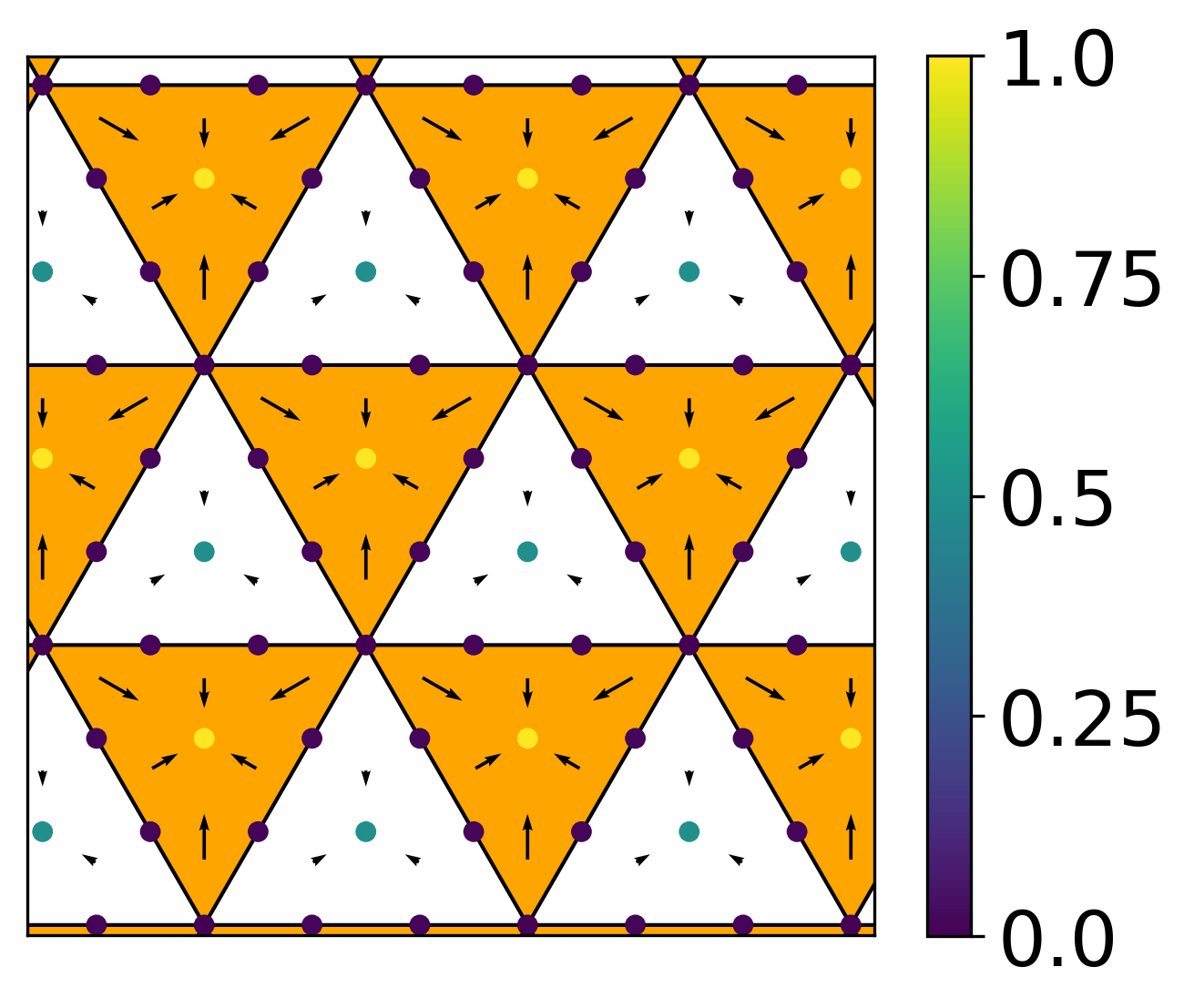}
    \end{subfigure}
    ~
    \begin{subfigure}[t]{0.35\textwidth}
        \centering
        \caption{\hspace{23pt}Hexagonal CDW}
        \includegraphics[width=\textwidth]{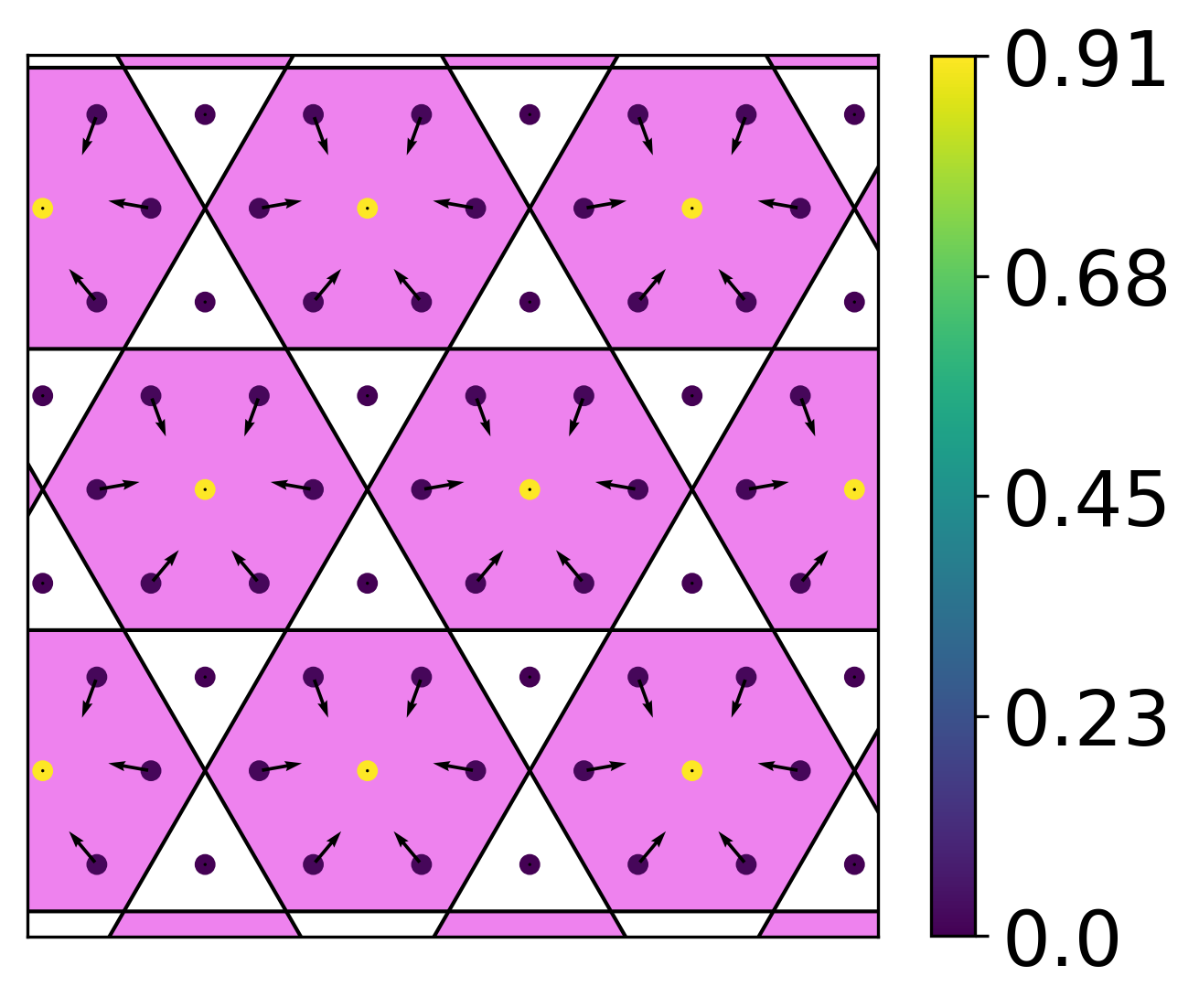}
    \end{subfigure}
    ~
    \begin{subfigure}[t]{0.35\textwidth}
        \centering
        \caption{\hspace{27pt}Stripe CDW}
        \includegraphics[width=\textwidth]{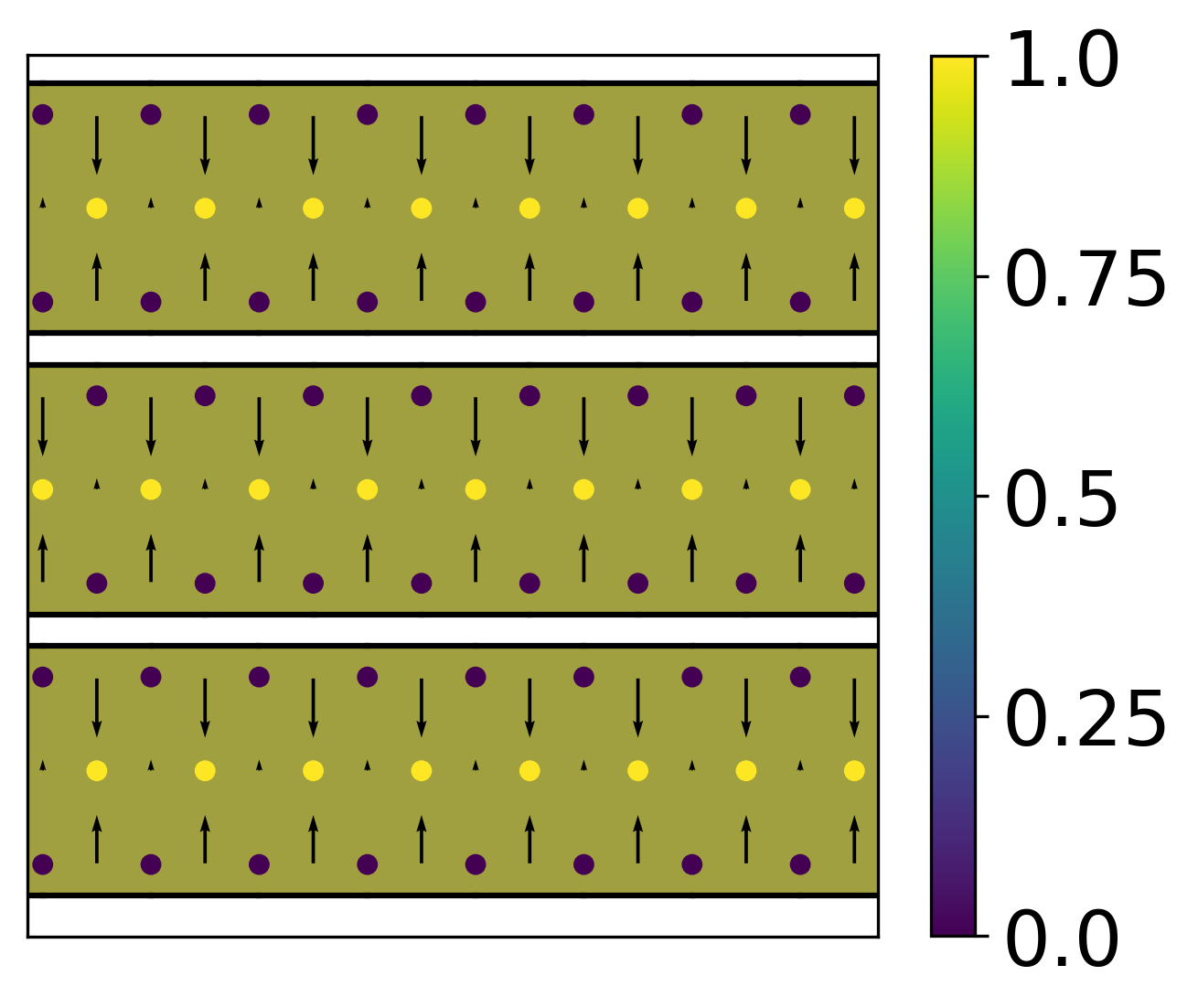}
    \end{subfigure}
    \caption{Order parameters for the different charge density wave motifs and corresponding reference displacement patterns. (a): Order parameter for three-atom triangle of the filled-center CDW evaluated in a monolayer with a filled-center CDW. (b): Order parameter for six-atom triangle of the filled-center CDW evaluated in a monolayer with a filled-center CDW; (c): Order parameter for three-atom triangle of the hollow-center CDW evaluated in a monolayer with a hollow-center CDW. (b): Order parameter for six-atom triangle of the hollow-center CDW evaluated in a monolayer with a hollow-center CDW; (e) Order parameter for hexagonal CDW evaluated for a monolayer with a hexagonal CDW; and (e): Order parameter for a stripe CDW evaluated for a monolayer with a stripe CDW. Colored solid circles denote the positions of the motif centers; i.e. Se atoms in (a) and (b), interstitial sites in (c) and (d); Nb atoms in (e) and the mid-points between neighboring Nb atom pair in (f). The color of the solid circles represents the value of the order parameters. The black arrows represent the displacements of the Nb atoms. The corresponding symbols (cyan triangles for filled-center CDW; orange triangles for hollow-center CDW; purple hexagons for hexagonal CDW and green stripe for stripe CDW) are shown in each of the panels.}
    \label{fig:cdw_on_same_cdw}
\end{figure*}

\clearpage
\twocolumngrid
\bibliography{cdw,dft,general_moire,optical_moire,vib_moire}

\begin{thebibliography}{51}%
\makeatletter
\providecommand \@ifxundefined [1]{%
 \@ifx{#1\undefined}
}%
\providecommand \@ifnum [1]{%
 \ifnum #1\expandafter \@firstoftwo
 \else \expandafter \@secondoftwo
 \fi
}%
\providecommand \@ifx [1]{%
 \ifx #1\expandafter \@firstoftwo
 \else \expandafter \@secondoftwo
 \fi
}%
\providecommand \natexlab [1]{#1}%
\providecommand \enquote  [1]{``#1''}%
\providecommand \bibnamefont  [1]{#1}%
\providecommand \bibfnamefont [1]{#1}%
\providecommand \citenamefont [1]{#1}%
\providecommand \href@noop [0]{\@secondoftwo}%
\providecommand \href [0]{\begingroup \@sanitize@url \@href}%
\providecommand \@href[1]{\@@startlink{#1}\@@href}%
\providecommand \@@href[1]{\endgroup#1\@@endlink}%
\providecommand \@sanitize@url [0]{\catcode `\\12\catcode `\$12\catcode `\&12\catcode `\#12\catcode `\^12\catcode `\_12\catcode `\%12\relax}%
\providecommand \@@startlink[1]{}%
\providecommand \@@endlink[0]{}%
\providecommand \url  [0]{\begingroup\@sanitize@url \@url }%
\providecommand \@url [1]{\endgroup\@href {#1}{\urlprefix }}%
\providecommand \urlprefix  [0]{URL }%
\providecommand \Eprint [0]{\href }%
\providecommand \doibase [0]{https://doi.org/}%
\providecommand \selectlanguage [0]{\@gobble}%
\providecommand \bibinfo  [0]{\@secondoftwo}%
\providecommand \bibfield  [0]{\@secondoftwo}%
\providecommand \translation [1]{[#1]}%
\providecommand \BibitemOpen [0]{}%
\providecommand \bibitemStop [0]{}%
\providecommand \bibitemNoStop [0]{.\EOS\space}%
\providecommand \EOS [0]{\spacefactor3000\relax}%
\providecommand \BibitemShut  [1]{\csname bibitem#1\endcsname}%
\let\auto@bib@innerbib\@empty
\bibitem [{\citenamefont {Bistritzer}\ and\ \citenamefont {MacDonald}(2011)}]{Bistritzer_MacDonald_2011}%
  \BibitemOpen
  \bibfield  {author} {\bibinfo {author} {\bibfnamefont {R.}~\bibnamefont {Bistritzer}}\ and\ \bibinfo {author} {\bibfnamefont {A.~H.}\ \bibnamefont {MacDonald}},\ }\href {https://doi.org/10.1073/pnas.1108174108} {\bibfield  {journal} {\bibinfo  {journal} {Proceedings of the {N}ational {A}cademy of {S}ciences}\ }\textbf {\bibinfo {volume} {108}},\ \bibinfo {pages} {12233–12237} (\bibinfo {year} {2011})}\BibitemShut {NoStop}%
\bibitem [{\citenamefont {Cao}\ \emph {et~al.}(2018{\natexlab{a}})\citenamefont {Cao}, \citenamefont {Fatemi}, \citenamefont {Demir}, \citenamefont {Fang}, \citenamefont {Tomarken}, \citenamefont {Luo}, \citenamefont {Sanchez-Yamagishi}, \citenamefont {Watanabe}, \citenamefont {Taniguchi}, \citenamefont {Kaxiras},\ and\ \citenamefont {et~al.}}]{Cao_Fatemi_Demir_Fang_Tomarken_Luo_Sanchez-Yamagishi_Watanabe_Taniguchi_Kaxiras_2018}%
  \BibitemOpen
  \bibfield  {author} {\bibinfo {author} {\bibfnamefont {Y.}~\bibnamefont {Cao}}, \bibinfo {author} {\bibfnamefont {V.}~\bibnamefont {Fatemi}}, \bibinfo {author} {\bibfnamefont {A.}~\bibnamefont {Demir}}, \bibinfo {author} {\bibfnamefont {S.}~\bibnamefont {Fang}}, \bibinfo {author} {\bibfnamefont {S.~L.}\ \bibnamefont {Tomarken}}, \bibinfo {author} {\bibfnamefont {J.~Y.}\ \bibnamefont {Luo}}, \bibinfo {author} {\bibfnamefont {J.~D.}\ \bibnamefont {Sanchez-Yamagishi}}, \bibinfo {author} {\bibfnamefont {K.}~\bibnamefont {Watanabe}}, \bibinfo {author} {\bibfnamefont {T.}~\bibnamefont {Taniguchi}}, \bibinfo {author} {\bibfnamefont {E.}~\bibnamefont {Kaxiras}},\ and\ \bibinfo {author} {\bibnamefont {et~al.}},\ }\href {https://doi.org/10.1038/nature26154} {\bibfield  {journal} {\bibinfo  {journal} {Nature}\ }\textbf {\bibinfo {volume} {556}},\ \bibinfo {pages} {80–84} (\bibinfo {year} {2018}{\natexlab{a}})}\BibitemShut {NoStop}%
\bibitem [{\citenamefont {Cao}\ \emph {et~al.}(2018{\natexlab{b}})\citenamefont {Cao}, \citenamefont {Fatemi}, \citenamefont {Fang}, \citenamefont {Watanabe}, \citenamefont {Taniguchi}, \citenamefont {Kaxiras},\ and\ \citenamefont {Jarillo-Herrero}}]{Cao_Fatemi_Fang_Watanabe_Taniguchi_Kaxiras_Jarillo-Herrero_2018}%
  \BibitemOpen
  \bibfield  {author} {\bibinfo {author} {\bibfnamefont {Y.}~\bibnamefont {Cao}}, \bibinfo {author} {\bibfnamefont {V.}~\bibnamefont {Fatemi}}, \bibinfo {author} {\bibfnamefont {S.}~\bibnamefont {Fang}}, \bibinfo {author} {\bibfnamefont {K.}~\bibnamefont {Watanabe}}, \bibinfo {author} {\bibfnamefont {T.}~\bibnamefont {Taniguchi}}, \bibinfo {author} {\bibfnamefont {E.}~\bibnamefont {Kaxiras}},\ and\ \bibinfo {author} {\bibfnamefont {P.}~\bibnamefont {Jarillo-Herrero}},\ }\href {https://doi.org/10.1038/nature26160} {\bibfield  {journal} {\bibinfo  {journal} {Nature}\ }\textbf {\bibinfo {volume} {556}},\ \bibinfo {pages} {43–50} (\bibinfo {year} {2018}{\natexlab{b}})}\BibitemShut {NoStop}%
\bibitem [{\citenamefont {Goodwin}\ \emph {et~al.}(2019)\citenamefont {Goodwin}, \citenamefont {Corsetti}, \citenamefont {Mostofi},\ and\ \citenamefont {Lischner}}]{attrac_elel_goodwin}%
  \BibitemOpen
  \bibfield  {author} {\bibinfo {author} {\bibfnamefont {Z.~A.~H.}\ \bibnamefont {Goodwin}}, \bibinfo {author} {\bibfnamefont {F.}~\bibnamefont {Corsetti}}, \bibinfo {author} {\bibfnamefont {A.~A.}\ \bibnamefont {Mostofi}},\ and\ \bibinfo {author} {\bibfnamefont {J.}~\bibnamefont {Lischner}},\ }\href {https://doi.org/10.1103/PhysRevB.100.235424} {\bibfield  {journal} {\bibinfo  {journal} {Phys. {R}ev. {B}}\ }\textbf {\bibinfo {volume} {100}},\ \bibinfo {pages} {235424} (\bibinfo {year} {2019})}\BibitemShut {NoStop}%
\bibitem [{\citenamefont {Goodwin}\ \emph {et~al.}(2021)\citenamefont {Goodwin}, \citenamefont {Klebl}, \citenamefont {Vitale}, \citenamefont {Liang}, \citenamefont {Gogtay}, \citenamefont {van Gorp}, \citenamefont {Kennes}, \citenamefont {Mostofi},\ and\ \citenamefont {Lischner}}]{flat_mag_goodwin}%
  \BibitemOpen
  \bibfield  {author} {\bibinfo {author} {\bibfnamefont {Z.~A.~H.}\ \bibnamefont {Goodwin}}, \bibinfo {author} {\bibfnamefont {L.}~\bibnamefont {Klebl}}, \bibinfo {author} {\bibfnamefont {V.}~\bibnamefont {Vitale}}, \bibinfo {author} {\bibfnamefont {X.}~\bibnamefont {Liang}}, \bibinfo {author} {\bibfnamefont {V.}~\bibnamefont {Gogtay}}, \bibinfo {author} {\bibfnamefont {X.}~\bibnamefont {van Gorp}}, \bibinfo {author} {\bibfnamefont {D.~M.}\ \bibnamefont {Kennes}}, \bibinfo {author} {\bibfnamefont {A.~A.}\ \bibnamefont {Mostofi}},\ and\ \bibinfo {author} {\bibfnamefont {J.}~\bibnamefont {Lischner}},\ }\href {https://doi.org/10.1103/PhysRevMaterials.5.084008} {\bibfield  {journal} {\bibinfo  {journal} {Phys. {R}ev. {M}ater.}\ }\textbf {\bibinfo {volume} {5}},\ \bibinfo {pages} {084008} (\bibinfo {year} {2021})}\BibitemShut {NoStop}%
\bibitem [{\citenamefont {Farrar}\ \emph {et~al.}(2021)\citenamefont {Farrar}, \citenamefont {Nevill}, \citenamefont {Lim}, \citenamefont {Balakrishnan}, \citenamefont {Dale},\ and\ \citenamefont {Bending}}]{sqi_tmd_vdw}%
  \BibitemOpen
  \bibfield  {author} {\bibinfo {author} {\bibfnamefont {L.~S.}\ \bibnamefont {Farrar}}, \bibinfo {author} {\bibfnamefont {A.}~\bibnamefont {Nevill}}, \bibinfo {author} {\bibfnamefont {Z.~J.}\ \bibnamefont {Lim}}, \bibinfo {author} {\bibfnamefont {G.}~\bibnamefont {Balakrishnan}}, \bibinfo {author} {\bibfnamefont {S.}~\bibnamefont {Dale}},\ and\ \bibinfo {author} {\bibfnamefont {S.~J.}\ \bibnamefont {Bending}},\ }\href {https://doi.org/10.1021/acs.nanolett.1c00152} {\bibfield  {journal} {\bibinfo  {journal} {Nano {L}etters}\ }\textbf {\bibinfo {volume} {21}},\ \bibinfo {pages} {6725} (\bibinfo {year} {2021})},\ \bibinfo {note} {pMID: 34428907},\ \Eprint {https://arxiv.org/abs/https://doi.org/10.1021/acs.nanolett.1c00152} {https://doi.org/10.1021/acs.nanolett.1c00152} \BibitemShut {NoStop}%
\bibitem [{\citenamefont {Kennes}\ \emph {et~al.}(2021)\citenamefont {Kennes}, \citenamefont {Claassen}, \citenamefont {Xian}, \citenamefont {Georges}, \citenamefont {Millis}, \citenamefont {Hone}, \citenamefont {Dean}, \citenamefont {Basov}, \citenamefont {Pasupathy},\ and\ \citenamefont {Rubio}}]{Kennes_Claassen_Xian_Georges_Millis_Hone_Dean_Basov_Pasupathy_Rubio_2021}%
  \BibitemOpen
  \bibfield  {author} {\bibinfo {author} {\bibfnamefont {D.~M.}\ \bibnamefont {Kennes}}, \bibinfo {author} {\bibfnamefont {M.}~\bibnamefont {Claassen}}, \bibinfo {author} {\bibfnamefont {L.}~\bibnamefont {Xian}}, \bibinfo {author} {\bibfnamefont {A.}~\bibnamefont {Georges}}, \bibinfo {author} {\bibfnamefont {A.~J.}\ \bibnamefont {Millis}}, \bibinfo {author} {\bibfnamefont {J.}~\bibnamefont {Hone}}, \bibinfo {author} {\bibfnamefont {C.~R.}\ \bibnamefont {Dean}}, \bibinfo {author} {\bibfnamefont {D.~N.}\ \bibnamefont {Basov}}, \bibinfo {author} {\bibfnamefont {A.~N.}\ \bibnamefont {Pasupathy}},\ and\ \bibinfo {author} {\bibfnamefont {A.}~\bibnamefont {Rubio}},\ }\href {https://doi.org/10.1038/s41567-020-01154-3} {\bibfield  {journal} {\bibinfo  {journal} {Nature {P}hysics}\ }\textbf {\bibinfo {volume} {17}},\ \bibinfo {pages} {155–163} (\bibinfo {year} {2021})}\BibitemShut {NoStop}%
\bibitem [{\citenamefont {Zheng}\ \emph {et~al.}(2023)\citenamefont {Zheng}, \citenamefont {Wu}, \citenamefont {Li}, \citenamefont {Ding}, \citenamefont {He}, \citenamefont {Liu}, \citenamefont {Wang}, \citenamefont {Wang}, \citenamefont {Pan},\ and\ \citenamefont {Liu}}]{Zheng_Wu_Li_Ding_He_Liu_Wang_Wang_Pan_Liu_2023}%
  \BibitemOpen
  \bibfield  {author} {\bibinfo {author} {\bibfnamefont {H.}~\bibnamefont {Zheng}}, \bibinfo {author} {\bibfnamefont {B.}~\bibnamefont {Wu}}, \bibinfo {author} {\bibfnamefont {S.}~\bibnamefont {Li}}, \bibinfo {author} {\bibfnamefont {J.}~\bibnamefont {Ding}}, \bibinfo {author} {\bibfnamefont {J.}~\bibnamefont {He}}, \bibinfo {author} {\bibfnamefont {Z.}~\bibnamefont {Liu}}, \bibinfo {author} {\bibfnamefont {C.-T.}\ \bibnamefont {Wang}}, \bibinfo {author} {\bibfnamefont {J.-T.}\ \bibnamefont {Wang}}, \bibinfo {author} {\bibfnamefont {A.}~\bibnamefont {Pan}},\ and\ \bibinfo {author} {\bibfnamefont {Y.}~\bibnamefont {Liu}},\ }\bibfield  {journal} {\bibinfo  {journal} {Light: {S}cience \& {A}pplications}\ }\textbf {\bibinfo {volume} {12}},\ \href {https://doi.org/10.1038/s41377-023-01171-w} {10.1038/s41377-023-01171-w} (\bibinfo {year} {2023})\BibitemShut {NoStop}%
\bibitem [{\citenamefont {Lian}\ \emph {et~al.}(2023)\citenamefont {Lian}, \citenamefont {Meng}, \citenamefont {Ma}, \citenamefont {Maity}, \citenamefont {Yan}, \citenamefont {Wu}, \citenamefont {Huang}, \citenamefont {Chen}, \citenamefont {Chen}, \citenamefont {Chen},\ and\ \citenamefont {et~al.}}]{Lian_Meng_Ma_Maity_Yan_Wu_Huang_Chen_Chen_Chen_2023}%
  \BibitemOpen
  \bibfield  {author} {\bibinfo {author} {\bibfnamefont {Z.}~\bibnamefont {Lian}}, \bibinfo {author} {\bibfnamefont {Y.}~\bibnamefont {Meng}}, \bibinfo {author} {\bibfnamefont {L.}~\bibnamefont {Ma}}, \bibinfo {author} {\bibfnamefont {I.}~\bibnamefont {Maity}}, \bibinfo {author} {\bibfnamefont {L.}~\bibnamefont {Yan}}, \bibinfo {author} {\bibfnamefont {Q.}~\bibnamefont {Wu}}, \bibinfo {author} {\bibfnamefont {X.}~\bibnamefont {Huang}}, \bibinfo {author} {\bibfnamefont {D.}~\bibnamefont {Chen}}, \bibinfo {author} {\bibfnamefont {X.}~\bibnamefont {Chen}}, \bibinfo {author} {\bibfnamefont {X.}~\bibnamefont {Chen}},\ and\ \bibinfo {author} {\bibnamefont {et~al.}},\ }\href {https://doi.org/10.1038/s41567-023-02266-2} {\bibfield  {journal} {\bibinfo  {journal} {Nature {P}hysics}\ }\textbf {\bibinfo {volume} {20}},\ \bibinfo {pages} {34–39} (\bibinfo {year} {2023})}\BibitemShut {NoStop}%
\bibitem [{\citenamefont {Chen}\ \emph {et~al.}(2022)\citenamefont {Chen}, \citenamefont {Lian}, \citenamefont {Huang}, \citenamefont {Su}, \citenamefont {Rashetnia}, \citenamefont {Ma}, \citenamefont {Yan}, \citenamefont {Blei}, \citenamefont {Xiang}, \citenamefont {Taniguchi},\ and\ \citenamefont {et~al.}}]{Chen_Lian_Huang_Su_Rashetnia_Ma_Yan_Blei_Xiang_Taniguchi_et_2022}%
  \BibitemOpen
  \bibfield  {author} {\bibinfo {author} {\bibfnamefont {D.}~\bibnamefont {Chen}}, \bibinfo {author} {\bibfnamefont {Z.}~\bibnamefont {Lian}}, \bibinfo {author} {\bibfnamefont {X.}~\bibnamefont {Huang}}, \bibinfo {author} {\bibfnamefont {Y.}~\bibnamefont {Su}}, \bibinfo {author} {\bibfnamefont {M.}~\bibnamefont {Rashetnia}}, \bibinfo {author} {\bibfnamefont {L.}~\bibnamefont {Ma}}, \bibinfo {author} {\bibfnamefont {L.}~\bibnamefont {Yan}}, \bibinfo {author} {\bibfnamefont {M.}~\bibnamefont {Blei}}, \bibinfo {author} {\bibfnamefont {L.}~\bibnamefont {Xiang}}, \bibinfo {author} {\bibfnamefont {T.}~\bibnamefont {Taniguchi}},\ and\ \bibinfo {author} {\bibnamefont {et~al.}},\ }\href {https://doi.org/10.1038/s41567-022-01703-y} {\bibfield  {journal} {\bibinfo  {journal} {Nature {P}hysics}\ }\textbf {\bibinfo {volume} {18}},\ \bibinfo {pages} {1171–1176} (\bibinfo {year} {2022})}\BibitemShut {NoStop}%
\bibitem [{\citenamefont {Brotons-Gisbert}\ \emph {et~al.}(2021)\citenamefont {Brotons-Gisbert}, \citenamefont {Baek}, \citenamefont {Campbell}, \citenamefont {Watanabe}, \citenamefont {Taniguchi},\ and\ \citenamefont {Gerardot}}]{Brotons-Gisbert_Baek_Campbell_Watanabe_Taniguchi_Gerardot_2021}%
  \BibitemOpen
  \bibfield  {author} {\bibinfo {author} {\bibfnamefont {M.}~\bibnamefont {Brotons-Gisbert}}, \bibinfo {author} {\bibfnamefont {H.}~\bibnamefont {Baek}}, \bibinfo {author} {\bibfnamefont {A.}~\bibnamefont {Campbell}}, \bibinfo {author} {\bibfnamefont {K.}~\bibnamefont {Watanabe}}, \bibinfo {author} {\bibfnamefont {T.}~\bibnamefont {Taniguchi}},\ and\ \bibinfo {author} {\bibfnamefont {B.~D.}\ \bibnamefont {Gerardot}},\ }\bibfield  {journal} {\bibinfo  {journal} {Physical {R}eview {X}}\ }\textbf {\bibinfo {volume} {11}},\ \href {https://doi.org/10.1103/physrevx.11.031033} {10.1103/physrevx.11.031033} (\bibinfo {year} {2021})\BibitemShut {NoStop}%
\bibitem [{\citenamefont {Li}\ \emph {et~al.}(2021)\citenamefont {Li}, \citenamefont {Lu}, \citenamefont {Cordovilla~Leon}, \citenamefont {Lyu}, \citenamefont {Xie}, \citenamefont {Hou}, \citenamefont {Lu}, \citenamefont {Guo}, \citenamefont {Kaczmarek}, \citenamefont {Taniguchi}, \citenamefont {Watanabe}, \citenamefont {Zhao}, \citenamefont {Yang},\ and\ \citenamefont {Deotare}}]{interlayer_x_transport}%
  \BibitemOpen
  \bibfield  {author} {\bibinfo {author} {\bibfnamefont {Z.}~\bibnamefont {Li}}, \bibinfo {author} {\bibfnamefont {X.}~\bibnamefont {Lu}}, \bibinfo {author} {\bibfnamefont {D.~F.}\ \bibnamefont {Cordovilla~Leon}}, \bibinfo {author} {\bibfnamefont {Z.}~\bibnamefont {Lyu}}, \bibinfo {author} {\bibfnamefont {H.}~\bibnamefont {Xie}}, \bibinfo {author} {\bibfnamefont {J.}~\bibnamefont {Hou}}, \bibinfo {author} {\bibfnamefont {Y.}~\bibnamefont {Lu}}, \bibinfo {author} {\bibfnamefont {X.}~\bibnamefont {Guo}}, \bibinfo {author} {\bibfnamefont {A.}~\bibnamefont {Kaczmarek}}, \bibinfo {author} {\bibfnamefont {T.}~\bibnamefont {Taniguchi}}, \bibinfo {author} {\bibfnamefont {K.}~\bibnamefont {Watanabe}}, \bibinfo {author} {\bibfnamefont {L.}~\bibnamefont {Zhao}}, \bibinfo {author} {\bibfnamefont {L.}~\bibnamefont {Yang}},\ and\ \bibinfo {author} {\bibfnamefont {P.~B.}\ \bibnamefont {Deotare}},\ }\href {https://doi.org/10.1021/acsnano.0c08981} {\bibfield  {journal} {\bibinfo  {journal} {{ACS} {N}ano}\ }\textbf {\bibinfo
  {volume} {15}},\ \bibinfo {pages} {1539} (\bibinfo {year} {2021})},\ \bibinfo {note} {pMID: 33417424},\ \Eprint {https://arxiv.org/abs/https://doi.org/10.1021/acsnano.0c08981} {https://doi.org/10.1021/acsnano.0c08981} \BibitemShut {NoStop}%
\bibitem [{\citenamefont {Maity}\ \emph {et~al.}(2023)\citenamefont {Maity}, \citenamefont {Mostofi},\ and\ \citenamefont {Lischner}}]{phason_maity}%
  \BibitemOpen
  \bibfield  {author} {\bibinfo {author} {\bibfnamefont {I.}~\bibnamefont {Maity}}, \bibinfo {author} {\bibfnamefont {A.~A.}\ \bibnamefont {Mostofi}},\ and\ \bibinfo {author} {\bibfnamefont {J.}~\bibnamefont {Lischner}},\ }\href {https://doi.org/10.1021/acs.nanolett.3c00490} {\bibfield  {journal} {\bibinfo  {journal} {Nano {L}etters}\ }\textbf {\bibinfo {volume} {23}},\ \bibinfo {pages} {4870} (\bibinfo {year} {2023})},\ \bibinfo {note} {pMID: 37235740},\ \Eprint {https://arxiv.org/abs/https://doi.org/10.1021/acs.nanolett.3c00490} {https://doi.org/10.1021/acs.nanolett.3c00490} \BibitemShut {NoStop}%
\bibitem [{\citenamefont {Maity}\ \emph {et~al.}(2022)\citenamefont {Maity}, \citenamefont {Mostofi},\ and\ \citenamefont {Lischner}}]{chiral_phon_maity}%
  \BibitemOpen
  \bibfield  {author} {\bibinfo {author} {\bibfnamefont {I.}~\bibnamefont {Maity}}, \bibinfo {author} {\bibfnamefont {A.~A.}\ \bibnamefont {Mostofi}},\ and\ \bibinfo {author} {\bibfnamefont {J.}~\bibnamefont {Lischner}},\ }\href {https://doi.org/10.1103/PhysRevB.105.L041408} {\bibfield  {journal} {\bibinfo  {journal} {Phys. {R}ev. {B}}\ }\textbf {\bibinfo {volume} {105}},\ \bibinfo {pages} {L041408} (\bibinfo {year} {2022})}\BibitemShut {NoStop}%
\bibitem [{\citenamefont {Trambly~de Laissardière}\ \emph {et~al.}(2010)\citenamefont {Trambly~de Laissardière}, \citenamefont {Mayou},\ and\ \citenamefont {Magaud}}]{local_dirac_el}%
  \BibitemOpen
  \bibfield  {author} {\bibinfo {author} {\bibfnamefont {G.}~\bibnamefont {Trambly~de Laissardière}}, \bibinfo {author} {\bibfnamefont {D.}~\bibnamefont {Mayou}},\ and\ \bibinfo {author} {\bibfnamefont {L.}~\bibnamefont {Magaud}},\ }\href {https://doi.org/10.1021/nl902948m} {\bibfield  {journal} {\bibinfo  {journal} {Nano {L}etters}\ }\textbf {\bibinfo {volume} {10}},\ \bibinfo {pages} {804} (\bibinfo {year} {2010})},\ \bibinfo {note} {pMID: 20121163},\ \Eprint {https://arxiv.org/abs/https://doi.org/10.1021/nl902948m} {https://doi.org/10.1021/nl902948m} \BibitemShut {NoStop}%
\bibitem [{\citenamefont {Mele}(2010)}]{commensuration_mele}%
  \BibitemOpen
  \bibfield  {author} {\bibinfo {author} {\bibfnamefont {E.~J.}\ \bibnamefont {Mele}},\ }\href {https://doi.org/10.1103/PhysRevB.81.161405} {\bibfield  {journal} {\bibinfo  {journal} {Phys. {R}ev. {B}}\ }\textbf {\bibinfo {volume} {81}},\ \bibinfo {pages} {161405} (\bibinfo {year} {2010})}\BibitemShut {NoStop}%
\bibitem [{\citenamefont {Xu}\ \emph {et~al.}(2022)\citenamefont {Xu}, \citenamefont {Ray}, \citenamefont {Shao}, \citenamefont {Jiang}, \citenamefont {Lee}, \citenamefont {Weber}, \citenamefont {Goldberger}, \citenamefont {Watanabe}, \citenamefont {Taniguchi}, \citenamefont {Muller} \emph {et~al.}}]{xu2022coexisting}%
  \BibitemOpen
  \bibfield  {author} {\bibinfo {author} {\bibfnamefont {Y.}~\bibnamefont {Xu}}, \bibinfo {author} {\bibfnamefont {A.}~\bibnamefont {Ray}}, \bibinfo {author} {\bibfnamefont {Y.-T.}\ \bibnamefont {Shao}}, \bibinfo {author} {\bibfnamefont {S.}~\bibnamefont {Jiang}}, \bibinfo {author} {\bibfnamefont {K.}~\bibnamefont {Lee}}, \bibinfo {author} {\bibfnamefont {D.}~\bibnamefont {Weber}}, \bibinfo {author} {\bibfnamefont {J.~E.}\ \bibnamefont {Goldberger}}, \bibinfo {author} {\bibfnamefont {K.}~\bibnamefont {Watanabe}}, \bibinfo {author} {\bibfnamefont {T.}~\bibnamefont {Taniguchi}}, \bibinfo {author} {\bibfnamefont {D.~A.}\ \bibnamefont {Muller}}, \emph {et~al.},\ }\href@noop {} {\bibfield  {journal} {\bibinfo  {journal} {Nature {N}anotechnology}\ }\textbf {\bibinfo {volume} {17}},\ \bibinfo {pages} {143} (\bibinfo {year} {2022})}\BibitemShut {NoStop}%
\bibitem [{\citenamefont {Hejazi}\ \emph {et~al.}(2020)\citenamefont {Hejazi}, \citenamefont {Luo},\ and\ \citenamefont {Balents}}]{hejazi2020noncollinear}%
  \BibitemOpen
  \bibfield  {author} {\bibinfo {author} {\bibfnamefont {K.}~\bibnamefont {Hejazi}}, \bibinfo {author} {\bibfnamefont {Z.-X.}\ \bibnamefont {Luo}},\ and\ \bibinfo {author} {\bibfnamefont {L.}~\bibnamefont {Balents}},\ }\href@noop {} {\bibfield  {journal} {\bibinfo  {journal} {Proceedings of the {N}ational {A}cademy of {S}ciences}\ }\textbf {\bibinfo {volume} {117}},\ \bibinfo {pages} {10721} (\bibinfo {year} {2020})}\BibitemShut {NoStop}%
\bibitem [{\citenamefont {Tong}\ \emph {et~al.}(2018)\citenamefont {Tong}, \citenamefont {Liu}, \citenamefont {Xiao},\ and\ \citenamefont {Yao}}]{tong2018skyrmions}%
  \BibitemOpen
  \bibfield  {author} {\bibinfo {author} {\bibfnamefont {Q.}~\bibnamefont {Tong}}, \bibinfo {author} {\bibfnamefont {F.}~\bibnamefont {Liu}}, \bibinfo {author} {\bibfnamefont {J.}~\bibnamefont {Xiao}},\ and\ \bibinfo {author} {\bibfnamefont {W.}~\bibnamefont {Yao}},\ }\href@noop {} {\bibfield  {journal} {\bibinfo  {journal} {Nano {L}etters}\ }\textbf {\bibinfo {volume} {18}},\ \bibinfo {pages} {7194} (\bibinfo {year} {2018})}\BibitemShut {NoStop}%
\bibitem [{\citenamefont {Akram}\ and\ \citenamefont {Erten}(2021)}]{akram2021skyrmions}%
  \BibitemOpen
  \bibfield  {author} {\bibinfo {author} {\bibfnamefont {M.}~\bibnamefont {Akram}}\ and\ \bibinfo {author} {\bibfnamefont {O.}~\bibnamefont {Erten}},\ }\href@noop {} {\bibfield  {journal} {\bibinfo  {journal} {Physical {R}eview {B}}\ }\textbf {\bibinfo {volume} {103}},\ \bibinfo {pages} {L140406} (\bibinfo {year} {2021})}\BibitemShut {NoStop}%
\bibitem [{\citenamefont {Xiao}\ \emph {et~al.}(2021)\citenamefont {Xiao}, \citenamefont {Chen},\ and\ \citenamefont {Tong}}]{xiao2021magnetization}%
  \BibitemOpen
  \bibfield  {author} {\bibinfo {author} {\bibfnamefont {F.}~\bibnamefont {Xiao}}, \bibinfo {author} {\bibfnamefont {K.}~\bibnamefont {Chen}},\ and\ \bibinfo {author} {\bibfnamefont {Q.}~\bibnamefont {Tong}},\ }\href@noop {} {\bibfield  {journal} {\bibinfo  {journal} {Physical {R}eview {R}esearch}\ }\textbf {\bibinfo {volume} {3}},\ \bibinfo {pages} {013027} (\bibinfo {year} {2021})}\BibitemShut {NoStop}%
\bibitem [{\citenamefont {Lian}\ \emph {et~al.}(2018)\citenamefont {Lian}, \citenamefont {Si},\ and\ \citenamefont {Duan}}]{unveil_cdw_sc}%
  \BibitemOpen
  \bibfield  {author} {\bibinfo {author} {\bibfnamefont {C.-S.}\ \bibnamefont {Lian}}, \bibinfo {author} {\bibfnamefont {C.}~\bibnamefont {Si}},\ and\ \bibinfo {author} {\bibfnamefont {W.}~\bibnamefont {Duan}},\ }\href {https://doi.org/10.1021/acs.nanolett.8b00237} {\bibfield  {journal} {\bibinfo  {journal} {Nano {Le}tters}\ }\textbf {\bibinfo {volume} {18}},\ \bibinfo {pages} {2924} (\bibinfo {year} {2018})},\ \bibinfo {note} {pMID: 29652158},\ \Eprint {https://arxiv.org/abs/https://doi.org/10.1021/acs.nanolett.8b00237} {https://doi.org/10.1021/acs.nanolett.8b00237} \BibitemShut {NoStop}%
\bibitem [{\citenamefont {Cossu}\ \emph {et~al.}(2018)\citenamefont {Cossu}, \citenamefont {Moghaddam}, \citenamefont {Kim}, \citenamefont {Tahini}, \citenamefont {Di~Marco}, \citenamefont {Yeom},\ and\ \citenamefont {Akbari}}]{unveil_cdw_imp}%
  \BibitemOpen
  \bibfield  {author} {\bibinfo {author} {\bibfnamefont {F.}~\bibnamefont {Cossu}}, \bibinfo {author} {\bibfnamefont {A.~G.}\ \bibnamefont {Moghaddam}}, \bibinfo {author} {\bibfnamefont {K.}~\bibnamefont {Kim}}, \bibinfo {author} {\bibfnamefont {H.~A.}\ \bibnamefont {Tahini}}, \bibinfo {author} {\bibfnamefont {I.}~\bibnamefont {Di~Marco}}, \bibinfo {author} {\bibfnamefont {H.-W.}\ \bibnamefont {Yeom}},\ and\ \bibinfo {author} {\bibfnamefont {A.}~\bibnamefont {Akbari}},\ }\href {https://doi.org/10.1103/PhysRevB.98.195419} {\bibfield  {journal} {\bibinfo  {journal} {Phys. {R}ev. {B}}\ }\textbf {\bibinfo {volume} {98}},\ \bibinfo {pages} {195419} (\bibinfo {year} {2018})}\BibitemShut {NoStop}%
\bibitem [{\citenamefont {Guster}\ \emph {et~al.}(2019)\citenamefont {Guster}, \citenamefont {Rubio-Verdú}, \citenamefont {Robles}, \citenamefont {Zaldívar}, \citenamefont {Dreher}, \citenamefont {Pruneda}, \citenamefont {Silva-Guillén}, \citenamefont {Choi}, \citenamefont {Pascual}, \citenamefont {Ugeda}, \citenamefont {Ordejón},\ and\ \citenamefont {Canadell}}]{elas_cdw_guster}%
  \BibitemOpen
  \bibfield  {author} {\bibinfo {author} {\bibfnamefont {B.}~\bibnamefont {Guster}}, \bibinfo {author} {\bibfnamefont {C.}~\bibnamefont {Rubio-Verdú}}, \bibinfo {author} {\bibfnamefont {R.}~\bibnamefont {Robles}}, \bibinfo {author} {\bibfnamefont {J.}~\bibnamefont {Zaldívar}}, \bibinfo {author} {\bibfnamefont {P.}~\bibnamefont {Dreher}}, \bibinfo {author} {\bibfnamefont {M.}~\bibnamefont {Pruneda}}, \bibinfo {author} {\bibfnamefont {J.~Ã.}\ \bibnamefont {Silva-Guillén}}, \bibinfo {author} {\bibfnamefont {D.-J.}\ \bibnamefont {Choi}}, \bibinfo {author} {\bibfnamefont {J.~I.}\ \bibnamefont {Pascual}}, \bibinfo {author} {\bibfnamefont {M.~M.}\ \bibnamefont {Ugeda}}, \bibinfo {author} {\bibfnamefont {P.}~\bibnamefont {Ordejón}},\ and\ \bibinfo {author} {\bibfnamefont {E.}~\bibnamefont {Canadell}},\ }\href {https://doi.org/10.1021/acs.nanolett.9b00268} {\bibfield  {journal} {\bibinfo  {journal} {Nano {Le}tters}\ }\textbf {\bibinfo {volume} {19}},\ \bibinfo {pages} {3027} (\bibinfo {year} {2019})},\ \bibinfo
  {note} {pMID: 30998364},\ \Eprint {https://arxiv.org/abs/https://doi.org/10.1021/acs.nanolett.9b00268} {https://doi.org/10.1021/acs.nanolett.9b00268} \BibitemShut {NoStop}%
\bibitem [{\citenamefont {Ugeda}\ \emph {et~al.}(2015)\citenamefont {Ugeda}, \citenamefont {{B}radley}, \citenamefont {Zhang}, \citenamefont {Onishi}, \citenamefont {Chen}, \citenamefont {Ruan}, \citenamefont {Ojeda-Aristizabal}, \citenamefont {Ryu}, \citenamefont {Edmonds}, \citenamefont {Tsai},\ and\ \citenamefont {et~al.}}]{Ugeda_Bradley_Zhang_Onishi_Chen_Ruan_Ojeda-Aristizabal_Ryu_Edmonds_Tsai_etal._2015}%
  \BibitemOpen
  \bibfield  {author} {\bibinfo {author} {\bibfnamefont {M.~M.}\ \bibnamefont {Ugeda}}, \bibinfo {author} {\bibfnamefont {A.~J.}\ \bibnamefont {{B}radley}}, \bibinfo {author} {\bibfnamefont {Y.}~\bibnamefont {Zhang}}, \bibinfo {author} {\bibfnamefont {S.}~\bibnamefont {Onishi}}, \bibinfo {author} {\bibfnamefont {Y.}~\bibnamefont {Chen}}, \bibinfo {author} {\bibfnamefont {W.}~\bibnamefont {Ruan}}, \bibinfo {author} {\bibfnamefont {C.}~\bibnamefont {Ojeda-Aristizabal}}, \bibinfo {author} {\bibfnamefont {H.}~\bibnamefont {Ryu}}, \bibinfo {author} {\bibfnamefont {M.~T.}\ \bibnamefont {Edmonds}}, \bibinfo {author} {\bibfnamefont {H.-Z.}\ \bibnamefont {Tsai}},\ and\ \bibinfo {author} {\bibnamefont {et~al.}},\ }\href {https://doi.org/10.1038/nphys3527} {\bibfield  {journal} {\bibinfo  {journal} {Nature {P}hysics}\ }\textbf {\bibinfo {volume} {12}},\ \bibinfo {pages} {92–97} (\bibinfo {year} {2015})}\BibitemShut {NoStop}%
\bibitem [{\citenamefont {Soumyanarayanan}\ \emph {et~al.}(2013)\citenamefont {Soumyanarayanan}, \citenamefont {Yee}, \citenamefont {He}, \citenamefont {van Wezel}, \citenamefont {Rahn}, \citenamefont {Rossnagel}, \citenamefont {Hudson}, \citenamefont {Norman},\ and\ \citenamefont {Hoffman}}]{Soumyanarayanan_Yee_He_van_wezel_Rahn_Rossnagel_Hudson_Norman_Hoffman_2013}%
  \BibitemOpen
  \bibfield  {author} {\bibinfo {author} {\bibfnamefont {A.}~\bibnamefont {Soumyanarayanan}}, \bibinfo {author} {\bibfnamefont {M.~M.}\ \bibnamefont {Yee}}, \bibinfo {author} {\bibfnamefont {Y.}~\bibnamefont {He}}, \bibinfo {author} {\bibfnamefont {J.}~\bibnamefont {van Wezel}}, \bibinfo {author} {\bibfnamefont {D.~J.}\ \bibnamefont {Rahn}}, \bibinfo {author} {\bibfnamefont {K.}~\bibnamefont {Rossnagel}}, \bibinfo {author} {\bibfnamefont {E.~W.}\ \bibnamefont {Hudson}}, \bibinfo {author} {\bibfnamefont {M.~R.}\ \bibnamefont {Norman}},\ and\ \bibinfo {author} {\bibfnamefont {J.~E.}\ \bibnamefont {Hoffman}},\ }\href {https://doi.org/10.1073/pnas.1211387110} {\bibfield  {journal} {\bibinfo  {journal} {Proceedings of the {N}ational {A}cademy of {S}ciences}\ }\textbf {\bibinfo {volume} {110}},\ \bibinfo {pages} {1623–1627} (\bibinfo {year} {2013})}\BibitemShut {NoStop}%
\bibitem [{\citenamefont {Flicker}\ and\ \citenamefont {van Wezel}(2015)}]{uni_strain_cdw_flicker}%
  \BibitemOpen
  \bibfield  {author} {\bibinfo {author} {\bibfnamefont {F.}~\bibnamefont {Flicker}}\ and\ \bibinfo {author} {\bibfnamefont {J.}~\bibnamefont {van Wezel}},\ }\href {https://doi.org/10.1103/PhysRevB.92.201103} {\bibfield  {journal} {\bibinfo  {journal} {Phys. {R}ev. {B}}\ }\textbf {\bibinfo {volume} {92}},\ \bibinfo {pages} {201103} (\bibinfo {year} {2015})}\BibitemShut {NoStop}%
\bibitem [{\citenamefont {McMillan}(1975)}]{mcmillan-cdw1}%
  \BibitemOpen
  \bibfield  {author} {\bibinfo {author} {\bibfnamefont {W.~L.}\ \bibnamefont {McMillan}},\ }\href {https://doi.org/10.1103/PhysRevB.12.1187} {\bibfield  {journal} {\bibinfo  {journal} {Phys. {R}ev. {B}}\ }\textbf {\bibinfo {volume} {12}},\ \bibinfo {pages} {1187} (\bibinfo {year} {1975})}\BibitemShut {NoStop}%
\bibitem [{\citenamefont {McMillan}(1976)}]{mcmillan-cdw2}%
  \BibitemOpen
  \bibfield  {author} {\bibinfo {author} {\bibfnamefont {W.~L.}\ \bibnamefont {McMillan}},\ }\href {https://doi.org/10.1103/PhysRevB.14.1496} {\bibfield  {journal} {\bibinfo  {journal} {Phys. {R}ev. {B}}\ }\textbf {\bibinfo {volume} {14}},\ \bibinfo {pages} {1496} (\bibinfo {year} {1976})}\BibitemShut {NoStop}%
\bibitem [{\citenamefont {McMillan}(1977)}]{mcmillan-cdw3}%
  \BibitemOpen
  \bibfield  {author} {\bibinfo {author} {\bibfnamefont {W.~L.}\ \bibnamefont {McMillan}},\ }\href {https://doi.org/10.1103/PhysRevB.16.643} {\bibfield  {journal} {\bibinfo  {journal} {Phys. {R}ev. {B}}\ }\textbf {\bibinfo {volume} {16}},\ \bibinfo {pages} {643} (\bibinfo {year} {1977})}\BibitemShut {NoStop}%
\bibitem [{\citenamefont {Jacobs}\ and\ \citenamefont {Walker}(1980)}]{jw-cdw1}%
  \BibitemOpen
  \bibfield  {author} {\bibinfo {author} {\bibfnamefont {A.~E.}\ \bibnamefont {Jacobs}}\ and\ \bibinfo {author} {\bibfnamefont {M.~B.}\ \bibnamefont {Walker}},\ }\href {https://doi.org/10.1103/PhysRevB.21.4132} {\bibfield  {journal} {\bibinfo  {journal} {Phys. {R}ev. {B}}\ }\textbf {\bibinfo {volume} {21}},\ \bibinfo {pages} {4132} (\bibinfo {year} {1980})}\BibitemShut {NoStop}%
\bibitem [{\citenamefont {Walker}\ and\ \citenamefont {Jacobs}(1981)}]{jw-cdw2}%
  \BibitemOpen
  \bibfield  {author} {\bibinfo {author} {\bibfnamefont {M.~B.}\ \bibnamefont {Walker}}\ and\ \bibinfo {author} {\bibfnamefont {A.~E.}\ \bibnamefont {Jacobs}},\ }\href {https://doi.org/10.1103/PhysRevB.24.6770} {\bibfield  {journal} {\bibinfo  {journal} {Phys. {R}ev. {B}}\ }\textbf {\bibinfo {volume} {24}},\ \bibinfo {pages} {6770} (\bibinfo {year} {1981})}\BibitemShut {NoStop}%
\bibitem [{\citenamefont {Walker}\ and\ \citenamefont {Jacobs}(1982)}]{jw-cdw3}%
  \BibitemOpen
  \bibfield  {author} {\bibinfo {author} {\bibfnamefont {M.~B.}\ \bibnamefont {Walker}}\ and\ \bibinfo {author} {\bibfnamefont {A.~E.}\ \bibnamefont {Jacobs}},\ }\href {https://doi.org/10.1103/PhysRevB.25.4856} {\bibfield  {journal} {\bibinfo  {journal} {Phys. {R}ev. {B}}\ }\textbf {\bibinfo {volume} {25}},\ \bibinfo {pages} {4856} (\bibinfo {year} {1982})}\BibitemShut {NoStop}%
\bibitem [{\citenamefont {Cossu}\ \emph {et~al.}(2020)\citenamefont {Cossu}, \citenamefont {Palotás}, \citenamefont {Sarkar}, \citenamefont {Di~Marco},\ and\ \citenamefont {Akbari}}]{strain_stripe_cdw}%
  \BibitemOpen
  \bibfield  {author} {\bibinfo {author} {\bibfnamefont {F.}~\bibnamefont {Cossu}}, \bibinfo {author} {\bibfnamefont {K.}~\bibnamefont {Palotás}}, \bibinfo {author} {\bibfnamefont {S.}~\bibnamefont {Sarkar}}, \bibinfo {author} {\bibfnamefont {I.}~\bibnamefont {Di~Marco}},\ and\ \bibinfo {author} {\bibfnamefont {A.}~\bibnamefont {Akbari}},\ }\bibfield  {journal} {\bibinfo  {journal} {{NPG} {A}sia {M}aterials}\ }\textbf {\bibinfo {volume} {12}},\ \href {https://doi.org/10.1038/s41427-020-0207-x} {10.1038/s41427-020-0207-x} (\bibinfo {year} {2020})\BibitemShut {NoStop}%
\bibitem [{\citenamefont {Goodwin}\ and\ \citenamefont {Fal'ko}(2022)}]{goodwin_falko_cdw_moire}%
  \BibitemOpen
  \bibfield  {author} {\bibinfo {author} {\bibfnamefont {Z.~A.~H.}\ \bibnamefont {Goodwin}}\ and\ \bibinfo {author} {\bibfnamefont {V.~I.}\ \bibnamefont {Fal'ko}},\ }\href {https://doi.org/10.1088/1361-648x/ac99ca} {\bibfield  {journal} {\bibinfo  {journal} {Journal of {P}hysics {C}ondensed {M}atter}\ }\textbf {\bibinfo {volume} {34}},\ \bibinfo {pages} {494001} (\bibinfo {year} {2022})},\ \bibinfo {note} {https://iopscience.iop.org/article/10.1088/1361-648X/ac99ca/pdf}\BibitemShut {NoStop}%
\bibitem [{\citenamefont {McHugh}\ \emph {et~al.}(2023)\citenamefont {McHugh}, \citenamefont {Enaldiev},\ and\ \citenamefont {Fal'ko}}]{falko_tNbSe2}%
  \BibitemOpen
  \bibfield  {author} {\bibinfo {author} {\bibfnamefont {J.~G.}\ \bibnamefont {McHugh}}, \bibinfo {author} {\bibfnamefont {V.~V.}\ \bibnamefont {Enaldiev}},\ and\ \bibinfo {author} {\bibfnamefont {V.~I.}\ \bibnamefont {Fal'ko}},\ }\href {https://doi.org/10.1103/PhysRevB.108.224111} {\bibfield  {journal} {\bibinfo  {journal} {Phys. {R}ev. {B}}\ }\textbf {\bibinfo {volume} {108}},\ \bibinfo {pages} {224111} (\bibinfo {year} {2023})}\BibitemShut {NoStop}%
\bibitem [{\citenamefont {Kohn}\ and\ \citenamefont {Sham}(1965)}]{kohn_dft}%
  \BibitemOpen
  \bibfield  {author} {\bibinfo {author} {\bibfnamefont {W.}~\bibnamefont {Kohn}}\ and\ \bibinfo {author} {\bibfnamefont {L.~J.}\ \bibnamefont {Sham}},\ }\href {https://doi.org/10.1103/PhysRev.140.A1133} {\bibfield  {journal} {\bibinfo  {journal} {Phys. {R}ev.}\ }\textbf {\bibinfo {volume} {140}},\ \bibinfo {pages} {A1133} (\bibinfo {year} {1965})}\BibitemShut {NoStop}%
\bibitem [{\citenamefont {Naik}\ \emph {et~al.}(2019)\citenamefont {Naik}, \citenamefont {Maity}, \citenamefont {Maiti},\ and\ \citenamefont {Jain}}]{kc_ff_tmd}%
  \BibitemOpen
  \bibfield  {author} {\bibinfo {author} {\bibfnamefont {M.~H.}\ \bibnamefont {Naik}}, \bibinfo {author} {\bibfnamefont {I.}~\bibnamefont {Maity}}, \bibinfo {author} {\bibfnamefont {P.~K.}\ \bibnamefont {Maiti}},\ and\ \bibinfo {author} {\bibfnamefont {M.}~\bibnamefont {Jain}},\ }\href {https://doi.org/10.1021/acs.jpcc.8b10392} {\bibfield  {journal} {\bibinfo  {journal} {The {J}ournal of {P}hysical {C}hemistry {C}}\ }\textbf {\bibinfo {volume} {123}},\ \bibinfo {pages} {9770} (\bibinfo {year} {2019})},\ \Eprint {https://arxiv.org/abs/https://doi.org/10.1021/acs.jpcc.8b10392} {https://doi.org/10.1021/acs.jpcc.8b10392} \BibitemShut {NoStop}%
\bibitem [{\citenamefont {Naik}\ and\ \citenamefont {Jain}(2018)}]{Ultraflatbands_solitons_naik}%
  \BibitemOpen
  \bibfield  {author} {\bibinfo {author} {\bibfnamefont {M.~H.}\ \bibnamefont {Naik}}\ and\ \bibinfo {author} {\bibfnamefont {M.}~\bibnamefont {Jain}},\ }\href {https://doi.org/10.1103/PhysRevLett.121.266401} {\bibfield  {journal} {\bibinfo  {journal} {Phys. {R}ev. {L}ett.}\ }\textbf {\bibinfo {volume} {121}},\ \bibinfo {pages} {266401} (\bibinfo {year} {2018})}\BibitemShut {NoStop}%
\bibitem [{\citenamefont {Maity}\ \emph {et~al.}(2021)\citenamefont {Maity}, \citenamefont {Maiti}, \citenamefont {Krishnamurthy},\ and\ \citenamefont {Jain}}]{recon_moire_tmd_maity}%
  \BibitemOpen
  \bibfield  {author} {\bibinfo {author} {\bibfnamefont {I.}~\bibnamefont {Maity}}, \bibinfo {author} {\bibfnamefont {P.~K.}\ \bibnamefont {Maiti}}, \bibinfo {author} {\bibfnamefont {H.~R.}\ \bibnamefont {Krishnamurthy}},\ and\ \bibinfo {author} {\bibfnamefont {M.}~\bibnamefont {Jain}},\ }\href {https://doi.org/10.1103/PhysRevB.103.L121102} {\bibfield  {journal} {\bibinfo  {journal} {Phys. {R}ev. {B}}\ }\textbf {\bibinfo {volume} {103}},\ \bibinfo {pages} {L121102} (\bibinfo {year} {2021})}\BibitemShut {NoStop}%
\bibitem [{\citenamefont {Vitale}\ \emph {et~al.}(2021)\citenamefont {Vitale}, \citenamefont {Atalar}, \citenamefont {Mostofi},\ and\ \citenamefont {Lischner}}]{Vitale_Atalar_Mostofi_Lischner_2021}%
  \BibitemOpen
  \bibfield  {author} {\bibinfo {author} {\bibfnamefont {V.}~\bibnamefont {Vitale}}, \bibinfo {author} {\bibfnamefont {K.}~\bibnamefont {Atalar}}, \bibinfo {author} {\bibfnamefont {A.~A.}\ \bibnamefont {Mostofi}},\ and\ \bibinfo {author} {\bibfnamefont {J.}~\bibnamefont {Lischner}},\ }\href {https://doi.org/10.1088/2053-1583/ac15d9} {\bibfield  {journal} {\bibinfo  {journal} {2{D} {M}aterials}\ }\textbf {\bibinfo {volume} {8}},\ \bibinfo {pages} {045010} (\bibinfo {year} {2021})}\BibitemShut {NoStop}%
\bibitem [{\citenamefont {Weston}\ \emph {et~al.}(2020)\citenamefont {Weston}, \citenamefont {Zou}, \citenamefont {Enaldiev}, \citenamefont {Summerfield}, \citenamefont {Clark}, \citenamefont {Zólyomi}, \citenamefont {Graham}, \citenamefont {Yelgel}, \citenamefont {Magorrian}, \citenamefont {Zhou},\ and\ \citenamefont {et~al.}}]{Weston_Zou_Enaldiev_Summerfield_Clark_Zólyomi_Graham_Yelgel_Magorrian_Zhou_etal}%
  \BibitemOpen
  \bibfield  {author} {\bibinfo {author} {\bibfnamefont {A.}~\bibnamefont {Weston}}, \bibinfo {author} {\bibfnamefont {Y.}~\bibnamefont {Zou}}, \bibinfo {author} {\bibfnamefont {V.}~\bibnamefont {Enaldiev}}, \bibinfo {author} {\bibfnamefont {A.}~\bibnamefont {Summerfield}}, \bibinfo {author} {\bibfnamefont {N.}~\bibnamefont {Clark}}, \bibinfo {author} {\bibfnamefont {V.}~\bibnamefont {Zólyomi}}, \bibinfo {author} {\bibfnamefont {A.}~\bibnamefont {Graham}}, \bibinfo {author} {\bibfnamefont {C.}~\bibnamefont {Yelgel}}, \bibinfo {author} {\bibfnamefont {S.}~\bibnamefont {Magorrian}}, \bibinfo {author} {\bibfnamefont {M.}~\bibnamefont {Zhou}},\ and\ \bibinfo {author} {\bibnamefont {et~al.}},\ }\href {https://doi.org/10.1038/s41565-020-0682-9} {\bibfield  {journal} {\bibinfo  {journal} {Nature {N}anotechnology}\ }\textbf {\bibinfo {volume} {15}},\ \bibinfo {pages} {592–597} (\bibinfo {year} {2020})}\BibitemShut {NoStop}%
\bibitem [{\citenamefont {Enaldiev}\ \emph {et~al.}(2020)\citenamefont {Enaldiev}, \citenamefont {Z\'olyomi}, \citenamefont {Yelgel}, \citenamefont {Magorrian},\ and\ \citenamefont {Fal'ko}}]{stack_domains_enaldiev}%
  \BibitemOpen
  \bibfield  {author} {\bibinfo {author} {\bibfnamefont {V.~V.}\ \bibnamefont {Enaldiev}}, \bibinfo {author} {\bibfnamefont {V.}~\bibnamefont {Z\'olyomi}}, \bibinfo {author} {\bibfnamefont {C.}~\bibnamefont {Yelgel}}, \bibinfo {author} {\bibfnamefont {S.~J.}\ \bibnamefont {Magorrian}},\ and\ \bibinfo {author} {\bibfnamefont {V.~I.}\ \bibnamefont {Fal'ko}},\ }\href {https://doi.org/10.1103/PhysRevLett.124.206101} {\bibfield  {journal} {\bibinfo  {journal} {Phys. {R}ev. {L}ett.}\ }\textbf {\bibinfo {volume} {124}},\ \bibinfo {pages} {206101} (\bibinfo {year} {2020})}\BibitemShut {NoStop}%
\bibitem [{\citenamefont {Bianco}\ \emph {et~al.}(2020)\citenamefont {Bianco}, \citenamefont {Monacelli}, \citenamefont {Calandra}, \citenamefont {Mauri},\ and\ \citenamefont {Errea}}]{weak_dim_dep_cdw}%
  \BibitemOpen
  \bibfield  {author} {\bibinfo {author} {\bibfnamefont {R.}~\bibnamefont {Bianco}}, \bibinfo {author} {\bibfnamefont {L.}~\bibnamefont {Monacelli}}, \bibinfo {author} {\bibfnamefont {M.}~\bibnamefont {Calandra}}, \bibinfo {author} {\bibfnamefont {F.}~\bibnamefont {Mauri}},\ and\ \bibinfo {author} {\bibfnamefont {I.}~\bibnamefont {Errea}},\ }\href {https://doi.org/10.1103/PhysRevLett.125.106101} {\bibfield  {journal} {\bibinfo  {journal} {Phys. {R}ev. {Le}tt.}\ }\textbf {\bibinfo {volume} {125}},\ \bibinfo {pages} {106101} (\bibinfo {year} {2020})}\BibitemShut {NoStop}%
\bibitem [{\citenamefont {Hamill}\ \emph {et~al.}(2021)\citenamefont {Hamill}, \citenamefont {Heischmidt}, \citenamefont {Sohn}, \citenamefont {Shaffer}, \citenamefont {Tsai}, \citenamefont {Zhang}, \citenamefont {Xi}, \citenamefont {Suslov}, \citenamefont {{B}erger}, \citenamefont {Forr{\'o}} \emph {et~al.}}]{hamill2021two}%
  \BibitemOpen
  \bibfield  {author} {\bibinfo {author} {\bibfnamefont {A.}~\bibnamefont {Hamill}}, \bibinfo {author} {\bibfnamefont {B.}~\bibnamefont {Heischmidt}}, \bibinfo {author} {\bibfnamefont {E.}~\bibnamefont {Sohn}}, \bibinfo {author} {\bibfnamefont {D.}~\bibnamefont {Shaffer}}, \bibinfo {author} {\bibfnamefont {K.-T.}\ \bibnamefont {Tsai}}, \bibinfo {author} {\bibfnamefont {X.}~\bibnamefont {Zhang}}, \bibinfo {author} {\bibfnamefont {X.}~\bibnamefont {Xi}}, \bibinfo {author} {\bibfnamefont {A.}~\bibnamefont {Suslov}}, \bibinfo {author} {\bibfnamefont {H.}~\bibnamefont {{B}erger}}, \bibinfo {author} {\bibfnamefont {L.}~\bibnamefont {Forr{\'o}}}, \emph {et~al.},\ }\href@noop {} {\bibfield  {journal} {\bibinfo  {journal} {Nature {P}hysics}\ }\textbf {\bibinfo {volume} {17}},\ \bibinfo {pages} {949} (\bibinfo {year} {2021})}\BibitemShut {NoStop}%
\bibitem [{\citenamefont {Yokoya}\ \emph {et~al.}(2001)\citenamefont {Yokoya}, \citenamefont {Kiss}, \citenamefont {Chainani}, \citenamefont {Shin}, \citenamefont {Nohara},\ and\ \citenamefont {Takagi}}]{yokoya2001fermi}%
  \BibitemOpen
  \bibfield  {author} {\bibinfo {author} {\bibfnamefont {T.}~\bibnamefont {Yokoya}}, \bibinfo {author} {\bibfnamefont {T.}~\bibnamefont {Kiss}}, \bibinfo {author} {\bibfnamefont {A.}~\bibnamefont {Chainani}}, \bibinfo {author} {\bibfnamefont {S.}~\bibnamefont {Shin}}, \bibinfo {author} {\bibfnamefont {M.}~\bibnamefont {Nohara}},\ and\ \bibinfo {author} {\bibfnamefont {H.}~\bibnamefont {Takagi}},\ }\href@noop {} {\bibfield  {journal} {\bibinfo  {journal} {Science}\ }\textbf {\bibinfo {volume} {294}},\ \bibinfo {pages} {2518} (\bibinfo {year} {2001})}\BibitemShut {NoStop}%
\bibitem [{\citenamefont {Dreher}\ \emph {et~al.}(2021)\citenamefont {Dreher}, \citenamefont {Wan}, \citenamefont {Chikina}, \citenamefont {{B}ianchi}, \citenamefont {Guo}, \citenamefont {Harsh}, \citenamefont {Mañas-Valero}, \citenamefont {Coronado}, \citenamefont {Mart{\'\i}nez-Galera}, \citenamefont {Hofmann} \emph {et~al.}}]{dreher2021proximity}%
  \BibitemOpen
  \bibfield  {author} {\bibinfo {author} {\bibfnamefont {P.}~\bibnamefont {Dreher}}, \bibinfo {author} {\bibfnamefont {W.}~\bibnamefont {Wan}}, \bibinfo {author} {\bibfnamefont {A.}~\bibnamefont {Chikina}}, \bibinfo {author} {\bibfnamefont {M.}~\bibnamefont {{B}ianchi}}, \bibinfo {author} {\bibfnamefont {H.}~\bibnamefont {Guo}}, \bibinfo {author} {\bibfnamefont {R.}~\bibnamefont {Harsh}}, \bibinfo {author} {\bibfnamefont {S.}~\bibnamefont {Mañas-Valero}}, \bibinfo {author} {\bibfnamefont {E.}~\bibnamefont {Coronado}}, \bibinfo {author} {\bibfnamefont {A.~J.}\ \bibnamefont {Mart{\'\i}nez-Galera}}, \bibinfo {author} {\bibfnamefont {P.}~\bibnamefont {Hofmann}}, \emph {et~al.},\ }\href@noop {} {\bibfield  {journal} {\bibinfo  {journal} {{ACS} {N}ano}\ }\textbf {\bibinfo {volume} {15}},\ \bibinfo {pages} {19430} (\bibinfo {year} {2021})}\BibitemShut {NoStop}%
\bibitem [{\citenamefont {Campanera}\ \emph {et~al.}(2007)\citenamefont {Campanera}, \citenamefont {Savini}, \citenamefont {Suarez-Martinez},\ and\ \citenamefont {Heggie}}]{moire_def_dft}%
  \BibitemOpen
  \bibfield  {author} {\bibinfo {author} {\bibfnamefont {J.~M.}\ \bibnamefont {Campanera}}, \bibinfo {author} {\bibfnamefont {G.}~\bibnamefont {Savini}}, \bibinfo {author} {\bibfnamefont {I.}~\bibnamefont {Suarez-Martinez}},\ and\ \bibinfo {author} {\bibfnamefont {M.~I.}\ \bibnamefont {Heggie}},\ }\href {https://doi.org/10.1103/PhysRevB.75.235449} {\bibfield  {journal} {\bibinfo  {journal} {Phys. {R}ev. {B}}\ }\textbf {\bibinfo {volume} {75}},\ \bibinfo {pages} {235449} (\bibinfo {year} {2007})}\BibitemShut {NoStop}%
\bibitem [{\citenamefont {Soler}\ \emph {et~al.}(2002)\citenamefont {Soler}, \citenamefont {Artacho}, \citenamefont {Gale}, \citenamefont {García}, \citenamefont {Junquera}, \citenamefont {Ordejón},\ and\ \citenamefont {Sánchez-Portal}}]{Soler_Artacho_Gale_García_Junquera_Ordejón_Sánchez-Portal_2002}%
  \BibitemOpen
  \bibfield  {author} {\bibinfo {author} {\bibfnamefont {J.~M.}\ \bibnamefont {Soler}}, \bibinfo {author} {\bibfnamefont {E.}~\bibnamefont {Artacho}}, \bibinfo {author} {\bibfnamefont {J.~D.}\ \bibnamefont {Gale}}, \bibinfo {author} {\bibfnamefont {A.}~\bibnamefont {García}}, \bibinfo {author} {\bibfnamefont {J.}~\bibnamefont {Junquera}}, \bibinfo {author} {\bibfnamefont {P.}~\bibnamefont {Ordejón}},\ and\ \bibinfo {author} {\bibfnamefont {D.}~\bibnamefont {Sánchez-Portal}},\ }\href {https://doi.org/10.1088/0953-8984/14/11/302} {\bibfield  {journal} {\bibinfo  {journal} {Journal of {P}hysics: {C}ondensed {M}atter}\ }\textbf {\bibinfo {volume} {14}},\ \bibinfo {pages} {2745–2779} (\bibinfo {year} {2002})}\BibitemShut {NoStop}%
\bibitem [{\citenamefont {Cooper}(2010)}]{cooper_xcf}%
  \BibitemOpen
  \bibfield  {author} {\bibinfo {author} {\bibfnamefont {V.~R.}\ \bibnamefont {Cooper}},\ }\href {https://doi.org/10.1103/PhysRevB.81.161104} {\bibfield  {journal} {\bibinfo  {journal} {Phys. {R}ev. {B}}\ }\textbf {\bibinfo {volume} {81}},\ \bibinfo {pages} {161104} (\bibinfo {year} {2010})}\BibitemShut {NoStop}%
\bibitem [{\citenamefont {Troullier}\ and\ \citenamefont {Martins}(1991)}]{TMPP}%
  \BibitemOpen
  \bibfield  {author} {\bibinfo {author} {\bibfnamefont {N.}~\bibnamefont {Troullier}}\ and\ \bibinfo {author} {\bibfnamefont {J.~L.}\ \bibnamefont {Martins}},\ }\href {https://doi.org/10.1103/PhysRevB.43.1993} {\bibfield  {journal} {\bibinfo  {journal} {Phys. {R}ev. {B}}\ }\textbf {\bibinfo {volume} {43}},\ \bibinfo {pages} {1993} (\bibinfo {year} {1991})}\BibitemShut {NoStop}%
\end{thebibliography}%

\end{document}